\documentclass[useAMS,usenatbib]{mn2e}
\usepackage{graphicx}
\usepackage{amssymb}
\usepackage{hyperref}

\newcommand{\msun}{\textrm{M}_{\odot}}
\newcommand{\gcc}{\textrm{g}~\textrm{cm}^{-3}}
\newcommand{\mps}{\msun~\textrm{s}^{-1}}
\newcommand{\cmps}{\textrm{cm}~\textrm{s}^{-1}}
\newcommand{\cmcmps}{\textrm{cm}^2~\textrm{s}^{-1}}
\newcommand{\foeps}{\textrm{foe}~\textrm{s}^{-1}}
\newcommand{\mc}{\multicolumn}
\newcommand{\ye}{Y_{\rm{e}}}
\newcommand{\abar}{\bar{A}}
\newcommand{\qt}{Q_{\rm{T}}}
\newcommand{\enu}{\dot{E}_\nu}

  \title[Disk instabilities in LGRBs]{Long duration gamma-ray bursts:
    hydrodynamic instabilities in collapsar disks}
  \author[P. A. Taylor, J. C. Miller,
    Ph. Podsiadlowski]{P. A. Taylor$^{1,2}$\,\thanks{E-mail:
      pault@aims.ac.za (PT); miller@sissa.it (JCM);
      podsi@astro.ox.ac.uk (PhP).},
    J. C. Miller$^{1,3}$\,\footnotemark[1] and
    Ph. Podsiadlowski$^{1}$\,\footnotemark[1]\\ $^{1}$Department of
    Physics (Astrophysics), University of Oxford, Keble Road, Oxford
    OX1 3RH, UK\\ $^{2}$African Institute for Mathematical Sciences,
    6-8 Melrose Road, Muizenberg 7945, South Africa\\ $^{3}$SISSA,
    International School for Advanced Studies, \& INFN, Via Bonomea
    265, 34136 Trieste, Italy}

\begin{document}

\date{draft version}

\pagerange{\pageref{firstpage}--\pageref{lastpage}} \pubyear{2008}

\maketitle

\label{firstpage}

\begin{abstract}
We present 3D numerical simulations of the early evolution of
long-duration gamma-ray bursts in the collapsar scenario.  Starting
from the core-collapse of a realistic progenitor model, we follow the
formation and evolution of a central black hole and centrifugally
balanced disk.  The dense, hot accretion disk produces freely-escaping
neutrinos and is hydrodynamically unstable to clumping and to forming
non-axisymmetric ($m=1$, 2) modes.  We show that these spiral
structures, which form on dynamical timescales, can efficiently
transfer angular momentum outward and can drive the high required
accretion rates ($\geq0.1-1~\mps$) for producing a jet.  We utilise the
smoothed particle hydrodynamics code, Gadget-2, modified to implement
relevant microphysics, such as cooling by neutrinos, a plausible
treatment approximating the central object and relativistic effects.
Finally, we discuss implications of this scenario as a source of
energy to produce relativistically beamed $\gamma$-ray jets.
\end{abstract}


\section{Introduction}

Gamma-ray bursts (GRBs) were first reported by
\citet*{1973ApJ...182L..85K} using data from the \textit{Vela}
satellites.  The study consisted of 16 bursts lasting $\leq30$~s,
which were detected in an energy range of 0.2-1.5~MeV.  Since then,
the number of observed GRBs has risen into the thousands, using
multiwavelength instruments, with burst durations spanning more than 5
orders of magnitude.  Initial estimates of the total output of
electromagnetic energy have been scaled down to
$\approx10^{50-52}~\rm{erg}$ with the assumption of relativistic
beaming in jets.  Observations to date of temporal duration and
spectral hardness ratios confirm a bimodal distribution for the
population: short duration bursts (SGRBs, $<2$~s) and long duration
bursts (LGRBs, $>2$~s), with distinct progenitor scenarios assumed for
each class.  Here, we examine the latter with the aid of numerical
simulations. In constructing realistic models, we first consider the
observational constraints on progenitors and the microphysical
processes which control the early phases of LGRB evolution and provide
the required conditions for jet production.

A major breakthrough for the understanding of LGRBs came with their
observed association with rather energetic Type Ic (core-collapse)
supernovae (SNe), sometimes called hypernovae (HNe): first, with
GRB980425 and SN1998bw \citep{1998Natur.395..670G}, and then with
GRB030329 and SN2003dh \citep{2003ApJ...591L..17S}.  Properties of
these and subsequent LGRB-SN associations are listed in
Table~\ref{tab:lgrbsn}. Studies of host galaxies have shown that
LGRBs tend to occur in star forming regions
\citep{2003A&A...400..499L}, which may tend to have low-metallicity
($<0.3-0.5~Z_{\odot}$) \citep{2006Natur.441..463F}.  The emergent
picture of the LGRB event is that of a massive (short-lived), stellar
progenitor collapsing in on itself, which led to the adoption of the
`collapsar' model \citep{1993ApJ...405..273W} originally used in the
context of core-collapse SNe.

\begin{table*}
  \begin{center}
    \begin{minipage}{137mm}
      \caption[LGRB - HNe Associations]{\label{tab:lgrbsn} LGRB-HNe
        associations, with quantities calculated from observations
        \citep[][]{2006Natur.442.1011P,2006Natur.442.1018M,2006NCimB.121.1207N,2008MNRAS.383.1485V,2010arXiv1004.2262C,2010arXiv1004.2919S}.}
      \begin{tabular}{cccccccc}
        \hline
        \mc{1}{c}{GRB} & \mc{1}{c}{Distance} & \mc{1}{c}{$\textrm{T}_{90}$} & \mc{1}{c}{$E_{\rm{iso}}$}  & \mc{1}{c}{SN} & \mc{1}{c}{Bol. lum.} & \mc{1}{c}{$M(^{56}$Ni)} & \mc{1}{c}{$v_{\rm{max}}$} \\
        &\mc{1}{c}{($z_{\rm{est}}$)} & \mc{1}{c}{(s) } & \mc{1}{c}{($10^{50}$~erg)}  & &  \mc{1}{c}{($10^{42}$~erg~s$^{-1}$)}& \mc{1}{c}{($\msun$)}& \mc{1}{c}{($10^3$~km~s$^{-1}$)} \\
        \hline
        980425 & 0.0085  & 34.9 & 0.01   &1998bw & 8.3  & 0.5-1.0 & 27 \\
        030329 & 0.169 & 22.8 & 180 &   2003dh &  9.1 & 0.35 &  28     \\
        031203 & 0.105 & 37.0  & 0.26    & ~2003lw$^\dagger$ &  11.0   & 0.55 &  18   \\
       \,~100316D  & 0.059 & $\sim$2300~~  & $\geq0.39$~~    &  2010bh  &  $^*$   & $^*$ &  26   \\

     ~060218$^\ddagger$ & 0.033 &$\sim$2100~~ & 0.62   & 2006aj & 5.8   & 0.2  &  26   \\
        - & - & -  &-  & 2003jd & 5.6    & 0.3  & $>$15~~   \\
        \hline
      \end{tabular}
      \small{$^\dagger$\,first observation 2 weeks after explosion;~~$^\ddagger$\,determined to be XRF;~~$^*$\,values unpublished, to date.}
    \end{minipage}
  \end{center}
\end{table*}

Briefly, the typical scenario begins with a rotating, massive star
($M/\msun>25-30$, possibly smaller if a remnant from a merger)
collapsing to form a black hole (BH) from its core and a
centrifugally-supported accretion disk, with continued infall from the
surrounding stellar interior.  The precise means of converting the
energy of the disk material into high-energy photons and forming a
relativistic jet is uncertain; however, the most studied methods, such
as the Blandford-Znajek (B-Z) mechanism creating Poynting-flux
\citep*{1977MNRAS.179..433B,2000PhR...325...83L,2000PhRvD..61h4016L}
and neutrino annihilation accelerating plasma
\citep*{1990ApJ...363..218P,1999ApJ...518..356P}, require extreme and
broadly similar conditions for the disk-BH system in order to
reproduce observed GRB energies.  In either case, accretion rates must
be quite high, $\dot{M}_{\rm{acc}}\gtrsim0.1-1.0~\mps$.  Structurally,
the disk must be dense ($\rho>10^{8-10}~\gcc$) and hot
($T\gtrsim10^{10}~\rm{K}\sim 1~\rm{MeV}$), placing it in a regime
where various scattering processes lead to large amounts of neutrino
production.  Most regions of these disks, while optically thick to
electromagnetic radiation, are optically thin to neutrinos and
therefore are able to cool quite efficiently; such disks are referred
to as `neutrino-dominated accretion flows' (NDAFs).

It is a great challenge for a physically realistic collapsar model to
form such a disk (see discussion below in \S2).  The progenitor must
possess enough specific angular momentum so that material will balance
centrifugally outside of the BH innermost stable circular orbit
(ISCO), which, for a moderately spinning 2-3~$\msun$ BH, is given by
$j_{\rm{ISCO}}\geq2-4\times10^{16}~\cmcmps$.  It is then difficult to
transfer that momentum outward efficiently to reach the required BH
accretion rates.  We note that some collapsar models have employed a
rapidly rotating magnetar as a central engine
\citep[e.g. ][]{1992Natur.357..472U, 1994MNRAS.270..480T,
  2000ApJ...537..810W, 2009MNRAS.396.2038B}, but we do not consider
these alternative scenarios here.

Early studies treated NDAFs as thin disks using the Shakura-Sunyaev
$\alpha$-viscosity ansatz to drive the accretion, for either steady
state (\citealt{1999ApJ...518..356P}; \citealt*{2001ApJ...557..949N};
\citealt{2002ApJ...577..311K}) or time-dependent
\citep{2004MNRAS.355..950J} solutions, typically assuming that the
magneto-rotational instability (MRI) \citep{1991ApJ...376..214B} could
provide a physical mechanism for angular momentum transfer. A recent
2D study by \citet*{2009ApJ...692..804L} with sophisticated equation
of state (EoS) and cooling prescription found high accretion rates
($>0.1~\mps$) and near-GRB neutrino luminosities ($\lesssim10^{52}$
erg s$^{-1}$) for disks driven solely by the $\alpha$-viscosity,
though (as presented here) the inclusion of azimuthal dynamics seems
to be necessary to understand the full behaviour of collapsar
flows. Ideal magneto-hydrodynamic (MHD) studies
\citep{2003ApJ...599L...5P,2007ApJ...663..437M} have also produced
high accretion rates, as have 2D numerical simulations studying
viscosity-driven convection \citep*[for SGRB
  disks,][]{2005ApJ...632..421L}; the creation of collapsar jets
predominantly by the interplay of magnetic fields and neutrino
annihilation has been investigated using 2D MHD and GRMHD
\citep{2007ApJ...659..512N,2009ApJ...704..937N}. 3D simulations
including general relativity have studied tori related to SGRBs
producing high accretion and neutrino production rates
\citep*{2004MNRAS.352..753S,2007PThPh.118..257S}.  However, the
dominant physical processes during the evolution of a realistic, 3D
collapsar system remains uncertain.

We consider the picture in which the hydrodynamic properties of a
collapsar disk provide a mechanism for the high mass accretion rates
giving rise to the LGRB.  Since a BH formed from the iron core of a
collapsed star has a mass of $\approx2-3~\msun$, for a disk with
sufficient mass to accrete at $0.1~\mps$, the resulting disk/central
object mass ratio, $\gtrsim1/10$, is already quite large compared to
most astrophysical systems.  This ratio, combined with the high
temperatures and densities mentioned earlier, has strong implications
for the evolution of the system.  Such self-gravitating disks are
susceptible locally to clumping and are unstable globally to the
formation of structures such as spiral arms, which are quite efficient
at transferring angular momentum outwards and matter inwards.
Moreover, neutrino production from the cooling material provides a
possible mechanism for creating a jet from accelerated plasma, in
addition to further destabilising such disks
\citep{2003MNRAS.339.1025R}.

It is important to note that LGRBs exhibit a wide variety of
characteristics in lightcurve shapes and afterglow spectra, as well as
in associations with HNe \citep[e.g., see discussion
  by][]{2007arXiv0707.2219N}.  To date, only four LGRBs have been
linked with HNe (Table~\ref{tab:lgrbsn}), while many partnerless ones
have been seen at relatively close distances.  Broadlined SNe Ic
bright enough to be classified as HNe have been observed without LGRBs
\citep[][and within]{2008AJ....135.1136M}; and at least one `HN' has
been associated with a less-energetic X-ray flash (XRF), XRF 060218-SN
2006aj \citep{2006Natur.442.1008C}.  Certainly, viewing effects due to
jet collimation play a role in some of the differences between events
\citep{2005Sci...308.1284M}.  For example, XRFs and unaccompanied HNe
may be explained as LGRBs viewed slightly and greatly off-angle,
respectively \citep[the latter interpretation is supported by observed
  rates, e.g. ][]{2004ApJ...607L..17P}.

It may be possible to explain much of this heterogeneity by
considering the variation of a few physical factors among
(qualitatively) similar progenitors: different values of rotation,
core size or metallicity; varied circumstellar environments; etc.
Certainly, LGRBs and related events have been observed across a wide
range of epochs and environments. Assuming that (most) characteristics
have reasonable ranges for permitting bursts, one would expect to have
not only significant variation in burst behaviour but also the
occurence of closely related, non-GRB events.  For instance, systems
with most (but not all) of the appropriate progenitor factors would
likely yield `near-' or imperfect LGRBs, i.e., producing weak or
unobservable jets, or yielding sub-critical amounts of nucleosynthetic
material for powering HNe.  Complete understanding of LGRBs would
include both mapping out the phase space of these factors, as well as
determining properties of related phenomena by altering factors of
successful events.
  
Here, we present a self-consistent model for powering LGRB engines via
the collapsar mechanism using 3D numerical simulations.  Starting with
a physically realistic progenitor (from a stellar evolution code) soon
after collapse, we follow the subsequent formation and evolution of an
accretion disk around a compact object (a proto-neutron star (PNS)
which accretes matter and becomes a BH). The aim of this study is to
include as many as possible of the important microphysical and
dynamical features of the system.  Therefore, it has been
necessary to use several approximations such as a simplified equation
of state, appropriate for a radiation dominated polytrope; a
time-independent estimation of electron fraction; and assumptions of
nuclear statistical equilibrium (NSE), where appropriate, in
determining constituent matter.  To include effects of general
relativity (GR) tractably, we first turn to standard pseudo-potentials
with kinematic rescaling.  We discuss the individual validity of these
simplifications below.  We note that, even while including several
different types of physics, we generally find that the average
estimated error for each approximation is $\lesssim$10-20\%.

In \S2, observational constraints on LGRBs are discussed, leading
to our choice of progenitor and requirements of disk behaviour; in
\S3, we briefly introduce the numerical method used here, smoothed
particle hydrodynamics (SPH); in \S4, the further physics added to the
SPH code are outlined, including a central object, relativistic
effects, the calculation of electron fraction and constituent nuclei
in the material, and cooling from neutrino production; in \S5, the
initial conditions of the various models are described; in \S6-7,
results of simulations with the shellular and cylindrical angular
momentum profiles, respectively, are presented, with those of low
temperature (shellular) models in \S8; in \S9, a test scenario with a
simple jet is given; in \S10, comparisons are made between models
utilizing Newtonian and pseudo-Newtonian potentials; and finally,
general trends and conclusions are discussed in \S11.

\section{Progenitor and Engine Requirements}

In this section we review the broad constraints which LGRB
observations and previously understood physics require from any
physical collapsar progenitor.  We also discuss how the success of any
model is determined here, by judging the results of the formation and
evolution of the disk-compact object system.

\subsection{Observational and physical constraints}

First, the progenitor must be massive, $M/\msun\geq 25-30$ for a
single-star \citep{1999ApJ...522..413F}, in order to form a BH from
mainly core material.  Also, a large supply of matter must be present,
assuming that rapid accretion continues for most of the LGRB duration
and that the event may be accompanied by a SN; e.g., an average
accretion rate of $\dot{M}=0.5~\mps$ for a 100~s burst would require
50~M$_{\odot}$, plus the mass of the SN ejecta as well.  This
progenitor range is in accordance with that of the large remnant mass
($\approx3~\msun$) observed for SN 1998bw and the associated SN
kinetic energies (related to the amount of radioactive nickel
produced), $2-5\times10^{52}~\rm{erg}$ \citep{2001ApJ...550..991N}.
To date, all HNe associated with LGRBs have possessed broad-line
spectral features, suggesting expansion velocities
$\geq~20,000~\rm{km~s}^{-1}$ (e.g., see \citet{2006Natur.442.1011P}
and references within).

Second, the LGRB-associated SNe were classified as Type Ic, due to the
observed spectra containing neither H nor He lines.  Therefore, it
seems that, in general, the progenitor should possess no envelope at
the time of collapse or, at least, one with only a very small amount
of H and He.  Furthermore, this lack of an envelope assists jet
propagation near breakout from the star, greatly reducing the amount
of radiation energy through dissipation into kinetic energy (avoiding
the so-called `baryon mass-loading problem').

In order to form a disk, a progenitor must be rotating rapidly, so
that the entire star does not collapse directly into a BH.  Instead, a
large fraction becomes centrifugally balanced around a PNS/BH formed
by the core, which in turn accretes material.  Disk-BH systems are
common in astrophysics (quasars, X-ray binaries, etc.) and tend to
produce collimated outflows aligned along or near the rotation axis.
Such systems are also typically very efficient in converting the
energy of infalling matter into outgoing, beamed radiation.  Any
centrifugally supported disk must have enough specific angular
momentum to remain outside of the BH $R_{\rm{ISCO}}$, noted before as
$j\approx \rm{a~few}\,\times10^{16}~\rm{cm^2~s^{-1}}$.

Finally, any collapsar disk must transport angular momentum outwards
and mass inwards with great enough efficiency to produce the accretion
rates required to power a GRB jet. In a study by
\citet{1999ApJ...518..356P}, most models required accretion rates of
$\dot{M}>0.1~\mps$ to produce a jet by B-Z or of
$\dot{M}>0.1-1~\mps$ for $\nu$-production and annihilation to
produce luminosities $\gtrsim10^{51}\textrm{erg}~\textrm{s}^{-1}$,
although this depended on several factors such as BH spin, disk
viscosity, etc.  In the collapsar, a disk will be maintained by the
continually infalling, high-$j$ stellar material which reaches
centrifugal balance.

These few requirements are perhaps surprisingly difficult to combine
within a self-consistent, robust model.  As mentioned previously,
during the evolution of the centrifugally formed disk, it becomes a
formidable challenge to transfer angular momentum outwards efficiently
enough to meet the necessary accretion rates.  Similarly, difficulties
arise in satisfying the characteristics of the progenitor itself; a
massive star can easily lose its H/He envelope, but the typical
mechanism of radiatively driven winds then causes the star to lose
angular momentum efficiently as well
\citep*{1998A&A...329..551L,1998RvMA...11...57L}.

Here, we address the latter difficulty by utilising a pre-collapse
progenitor evolved from the merger of two helium stars (more details
below in \S\ref{sec:progenmodel}).  The remnant object remains
massive, while the merging process is able to eject outer envelope
material from the system during a common-envelope phase
\citep*{1992ApJ...391..246P}.  The polar direction is particularly
cleared of H/He, benefitting jet propagation, and the remnant can
easily retain sufficient angular momentum to form a centrifugally
balanced disk.  This is neither a very exotic nor overly restrictive
requirement for the progenitor population, as at least half of all
massive stars are estimated to be in binary and multi-star systems
\citep*{1980ApJ...242.1063G,2007ApJ...670..747K}.

\subsection{Disk evolution}
We discuss here the dynamics controlling the evolution of the
collapsar disk in different phases.  First, after the initial
collapse, viscous processes (along with neutrino cooling) determine
transport and energy dissipation.  Then, given the large self-gravity
and the rapid cooling rate of the disk, we expect spiral arms to
form on a dynamical timescale due to hydrodynamic instability.  While
viscous dissipation is still present in this latter phase, the spiral
structure dominates the global transfer of angular momentum and leads
to rapid accretion.

\subsubsection{Disk viscosity}
The evolution of thin and slim disks has been  widely studied,
typically in a low density regime but with an increasing interest in
high density disks.  Viscous processes in accretion disks convert
kinetic energy to heat and transfer angular momentum outwards, though
generally not at rates nearly as high as those required to drive LGRB
production.  Typically, this viscosity is attributed to the formation
of convective currents, due to some combination of Maxwell (magnetic)
and Reynolds (turbulent) stresses.  Often in studies, these processes
are parametrised as the $\alpha$ of \citet{1973A&A....24..337S},
similar to an efficiency, which relates the overall viscosity to the
local sound speed, $c_{\rm{s}}$, and scale height, $H$:
\begin{equation}
\nu = \alpha c_{\rm{s}} H\,.
\end{equation}
Though a topic of some dispute, it is reasonable to assume that
$\alpha\leq1$, corresponding to a convection cell with size less
than the disk height and with subsonic turnover. 

Studies of Newtonian disks have shown that the presence of magnetic
fields can lead to a magnetorotational instability (MRI), which
induces efficient angular momentum transfer and dissipation
\citep{1991ApJ...376..214B}, dominating disk dynamics with a maximal
growth rate on the order of the dynamical timescale.  This analysis
assumes a Boussinesq (incompressible) flow, where the rotation profile
is monotonically decreasing with radius, and that the plasma
$\beta\geq 3$ (ratio of hydrostatic to magnetic pressure).
Simulations using ideal (infinite conductivity) MHD have shown both
axisymmetric and non-axisymmetric perturbations, leading to large
accretion rates (\citealt*{2004ApJ...616..357F};
\citealt{2004ApJ...616..364F}).  In a physical disk, however, both the
finite conductivity and resistivity of material can be important in
damping the growth of the MRI, as discussed in Appendix A.  We do not
include any magnetic effects here, except by proxy through the
influence of the $\alpha$-viscosity.

It has been shown that the numerical viscosity terms in SPH behave
very similarly to a shearing viscosity in rotating systems.  In the
case of thin disks, (2D) studies have shown that the effective
viscosity, $\nu_{\rm{SPH}}$, is proportional to the $\alpha$-viscosity
(\citealp{1996MNRAS.279..402M}; Taylor \& Miller, in preparation).
Therefore, the physical viscous process are essentially parametrised
by the SPH viscosity.

\subsubsection{Hydrodynamic instability}
\label{sec:INTtoomr}
The Toomre parameter, $\qt$,
\citep{1960AnAp...23..979S,1964ApJ...139.1217T,1965MNRAS.130...97G}
has been used to characterise stability in several gaseous, stellar
and hydrodynamic self-gravitating disk systems 
\citep[e.g.,][]{2001ApJ...553..174G,2003MNRAS.339.1025R}. Moreover,
it has been applied to describe both local stability against clumping
and fragmentation, and the more global formation of spirals and
non-axisymmetric modes.  These structures can dramatically change the
evolution of a disk.  The Toomre parameter is given by
\begin{eqnarray}\label{toomreq}
    \qt&=& \frac{\kappa c_{\rm{s}} }{\pi G \Sigma}\,, \\
    \kappa^2 &\equiv& \frac{1}{R^3}\frac{d(\rm{a_R}\textit{R}^3)}{dR}\,, \nonumber
\end{eqnarray}
where $\kappa$ is the epicyclic frequency ($=\Omega$, orbital
velocity, for Newtonian cases, and then often the stability parameter
is approximated by $Q'_{\rm{T}}=\Omega c_{\rm{s}} / \pi G
\Sigma_{\rm{D}}$); $\rm{a_R}$, radial acceleration; $c_{\rm{s}}$,
local sound speed; $G$, gravitational constant; and $\Sigma$, surface
density of the disk.  The exact critical value below which a disk is
unstable to non-axisymmetric perturbations varies with the specific
system but is usually $\approx1$.  For example, a finite-thickness,
isothermal fluid disk has a critical value, $\qt\approx0.676$
\citep{1965MNRAS.130...97G,2001ApJ...553..174G}.

Physically, the Toomre parameter describes a competition between the
locally stabilising influences of pressure, $P$ ($c_{\rm{s}}^2\propto
T\propto P/\rho$), and epicyclic motion, and the destabilising
influence of self gravity, which tends to enhance local clumpiness and
global non-axisymmetry. From a global perspective,
\citet*{1987ApJ...323..592H} have shown that (for cloud fragmentation)
a parameter similar to $\qt$ can be derived by comparing relevant
timescales in the system: the free-fall time, $\tau_{\rm{ff}}=(3\pi
/32{\it G}\rho)^{\rm{1/2}}$; the sound crossing time,
$\tau_{\rm{sd}}=2\pi {\it R}/c_{\rm{s}}$, for a cylinder of
characteristic radius, $R$; and the epicyclic period,
$\tau_{\rm{ep}}=2\pi /\kappa$.  Besides identifying two stability
criteria from the ratios $\tau_{\rm{ff}}/\tau_{\rm{sd}}$ and
$\tau_{\rm{ff}}/\tau_{\rm{ep}}$, the product of these ratios is
directly proportional to the Toomre parameter as given by
Eq.~\ref{toomreq} above, $\qt\propto
\tau_{\rm{ff}}^2/\tau_{\rm{sd}}\tau_{\rm{ep}}$.

Studies of rotating polytropic disk models have suggested that $m=1$
and $m=2$ instabilities can develop on dynamical timescales
\citep*{2000ApJ...528..946I}.  There, angular momentum was transported
outward in an initial pulse, followed by continued mass drain through
the bar/spiral's interaction with outer material.

Additionally, cooling plays a large role in the disk dynamics and
evolution.  Besides causing vertical contraction, the rate of cooling,
$\dot{Q}^-$, has an important effect on the dynamical formation of
non-axisymmetric structure.  Simulations of both local and global
stability
\citep[][respectively]{2001ApJ...553..174G,2003MNRAS.339.1025R} showed
that when the cooling timescale for a fluid element of specific
internal energy, $\epsilon$, is of the order of the disk orbital
period, i.e.
\begin{equation}\label{coolvdyn}
  t_{\rm{cool}} \equiv
  \frac{\epsilon}{\dot{Q}^-}\approx\rm{a~few}\,\Omega^{-1},
\end{equation}
then massive disks can rapidly become unstable to fragmentation and/or
to non-axisymmetric structure formation on dynamical timescales.

One can think of this cooling effect entering into the Toomre
criterion via thermal pressure and the $1^{\rm{st}}$~Law of
Thermodynamics.  If a fluid element in equilibrium is compressed
without a cooling mechanism, then $P$ ($\propto\epsilon\rho$)
increases, resisting compression and returning the volume to
equilibrium.  If the same element is compressed in the presence of a
cooling mechanism, the internal energy necessarily increases less (or
decreases), and $P$ cannot restore the same equilibrium.  In some
cases, these local overdensities propagate as spirals, becoming global
modes.

While useful predictions of disk stability can be made from these
analytic arguments, the resulting phenomena are highly
nonlinear. Furthermore, in the collapsar case the further effects of
relativistic dynamics (on both the epicyclic frequency and the disk
structure) are uncertain.  Therefore, any discussion of disk behaviour
and evolution in the presence of non-axisymmetric perturbations
requires 3D numerical modelling for a range of initial conditions.

\subsection{Energetic requirements}
\label{energreq}
The full details of the process of producing a relativistically beamed
GRB jet are uncertain, and in this study we do not select one of the
several proposed mechanisms.  Here, we discuss the necessary
conditions given by analytic arguments for disk-BH evolution in both
the Blandford-Znajek (B-Z) and neutrino annihilation scenarios, which
currently are the leading jet mechanisms and have been widely studied.
The very complicated LGRB central engine and process of jet formation
are thus characterised in terms of a few measurable, `physical'
quantities of the evolving system, which is obviously a great
simplification and relies on many assumptions for each model.  In the
current field of LGRB studies, these approximations serve as the best
evaluation for a collapsar system to produce a burst, but they can
neither preclude new mechanisms nor definitively describe the full
behaviour of an individual one.

In the B-Z mechanism, rotational energy of the disk-BH system is
converted into Poynting-flux jets along the rotation axis.  In
addition to the BH spin and mass accretion rates, the flux of the
outflow depends on the strength of a central magnetic (dipole) field
which reaches outward from the BH.  For a $3~\msun$ BH with a
$10^{15}$~G magnetic field, the jet energy is given as a function of
BH spin, $a$ ($\equiv Jc/GM^2$, the ratio of total BH specific angular
momentum to mass), and accretion rate, $\dot{M}$, as
\citep*{1986bhmp.book.....T}:
\begin{equation}\label{edot_bz}
  \dot{E}_{\rm{B-Z}} \approx 3 \times
  10^{52}\,a^2\,\left(\frac{\dot{M}}{\mps}\right)~~~\rm{erg~s^{-1}}.
\end{equation}

In the case of neutrino annihilation, disk models studied by
\citet{1999ApJ...518..356P} only reached GRB-like luminosities of
$\gtrsim10^{51}~\rm{erg~s^{-1}}$ with accretion rates of
$\approx0.1~\mps$ and a few$\,\times\,0.1~\mps$ for $a=0.95$ and
0.5, respectively. Neutrino production varied greatly with BH spin and
disk viscosity, and in general, estimations to translate $L_{\nu}$
directly into jet energy remain quite uncertain.  The efficiency of
neutrino annihilation in plasma is unknown, particularly in the
presence of other physical factors such as magnetic fields, baryon
pollution, etc.  Analysing idealised disks around Kerr BHs using
ray-tracing techniques, \citet{2007A&A...463...51B} estimated the
effects of both GR and disk geometry on neutrino annihilation rates.
In general the effects were small (most less than a factor of 2), with
more efficient annihilation associated with thinner disks and higher
BH spins.  For this study, we note in particular the important
influence of the latter, as the smaller $R_{\rm{ISCO}}$ with increased
$a$ leads to greater $L_{\nu}$ in the inner regions as well.

The neutrino case can be generalised more broadly with the simple
assumption \citep[e.g.,][]{1999ApJ...527L..39J} that the useable
neutrino energy rate for producing a LGRB jet, $\enu$, scales directly with
the total disk $L_{\nu}$:
\begin{equation}
\enu = \eta_{\nu\bar{\nu}} L_{\nu}\,,
\end{equation}
where $\eta_{\nu\bar{\nu}}$ now represents the total, unknown
efficiency for the physical process of neutrino annihilation.  This is
a very uncertain quantity, and studies have shown that this efficiency
may not be constant.  Here, we will consider it to have a reasonably
high value, $\eta_{\nu\bar{\nu}}=0.01$, as found by
\citet{1999ApJ...527L..39J} \citep[see also,][]{1998A&A...338..535R},
accepting that this is a very rough approximation.

In this study $\dot{E}_{\rm{GRB}}=10^{51}~{\rm erg~s^{-1}}
\equiv1~\foeps$ is defined as our fiducial LGRB-jet energy flux, the
threshold value which a successful central engine must produce.  This
value is generally consistent with observed LGRB energetics and would
then require, for a BH with $a=0.5$, for example, an accretion rate of
$\dot{M}>0.13~\mps$ in the case of the B-Z mechanism.  However, we do
not intend this to preclude the possibility of lower-luminosity bursts
which, as of this time, have not been detected \citep*[see discussion
  in][]{2009MNRAS.395.1515C}. In fact, such objects or `failed' LGRBs
may produce other observed phenomena, such as polarised SNe, XRFs,
etc., particularly in the presence of a large central $B$-field.

\section{SPH}
For this study we use smoothed particle hydrodynamics (SPH)
\citep{1977MNRAS.181..375G,1977AJ.....82.1013L}, which has found wide
usage in modeling various astrophysical phenomena.  SPH has advantages
over other numerical techniques in that it naturally adapts to
following dynamic flows and arbitrary geometries without the need for
mesh refinement or other readjustment techniques required in
Eulerian-based codes.  The Lagrangian method allows natural
definitions for the energetics of fluid elements and also efficiently
maps structures whose density spans several orders of magnitude.  We
utilise the public release version of the Gadget-2 code
\citep{2005MNRAS.364.1105S}, which, furthermore, strictly conserves
momentum (linear and angular), energy and entropy.  We add additional
relevant physics to the code, such as cooling, relativistic effects
and the presence of a central object.

Briefly, in SPH a continuous fluid is sampled at a finite number of
points, and discretised versions of the three hydrodynamic equations
(mass, momentum and energy) are calculated.  Gadget-2 utilises a
polytropic EoS and, rather than energy, evolves an
entropic function, $K(s)=P/\rho^{\gamma}$, where $\gamma$ is the usual
adiabatic index; $P$ and $\rho$, the hydrodynamic pressure and
density, respectively, related to specific internal energy,
$\epsilon=(\gamma-1)P/\rho$; and $s$ is specific entropy.  The
smoothed kernels determine a set of nearest neighbours which interact
with symmetric hydrodynamic forces.  Entropy is created from
discontinuities and shocks, via a viscosity formulation which
possesses two terms.  The first provides the bulk and shear viscosity
(linear in converging velocity), and the second, an artificial von
Neumann-Richtmyer (quadratic) viscosity, mimics naturally dissipative
processes in addition to preventing interparticle penetration
\citep{1997JCoPh.136..298M}. The strength of the Gadget-2 viscosity is
parametrised by a single coefficient of order unity (here,
$\alpha_{\rm v}=0.8$), and the Balsara prescription to reduce
numerical viscosity introduced in rotating flows was utilized
\citep{1995JCoPh.121..357B,1996MNRAS.278.1005S}. Density is not
explicitly evolved but instead is recalculated at each timestep from a
weighted average of the total mass within a smoothed kernel.
Gravitational forces are calculated efficiently with a hierarchical
multipole expansion, using standard tree algorithms
\citep[e.g.][]{1986Natur.324..446B}.

\section{Additional Physics}
To these general hydrodynamic and gravitational components, we include
physical features specifically relevant to the collapsar evolution.
Unless otherwise stated, all temperatures, $T$, are given in units
of K, and both mass and proton density, $\rho$ and $\ye\rho$,
respectively, in units of $\gcc$.

\subsection{Central object}
\label{sec:PHYScentob}
After collapse, the iron core of the initial progenitor first forms a
(rotating) PNS, which subsequently accretes material to become a BH.
Since our main concern is studying the behaviour of the accretion
flow, we focus on simulating only the main effects of each distinct
phase on the hydro-flow. The two most important influences of the
central object are to provide an inner accretion boundary for the
inflowing material and to exert gravitational accelerations (see
\S\ref{sec:grsr} below, regarding relativity).  Here, we discuss how
the PNS and BH differ importantly in the presence and absence,
respectively, of a surface pressure.

In the first few seconds after formation, the central object is quite
hot and is best described as a PNS, with slightly larger radius than
the typical 10-12~km of a NS, as well as a `softer' surface.  As the
PNS accretes matter, its radius decreases, and we approximate its
mass-radius (M-R) progression using the recent calculations of
\citet{2006nucl.th...7055N}.  Examples (using baryonic mass) of
($M_{\rm{PNS}}/\msun$, $R_{\rm{PNS}}/{\rm{km}}$) from his model T30Y04
are: (1.6, 35), (1.7, 24) and (1.9, 16), beyond which a BH is formed.

To approximate the surface of a (spherical) PNS, we include a
radially-outward acceleration at the given $R_{\rm{PNS}}$, which
interacts with SPH particles that have approached to within a fraction
of their smoothing-length.  For inflowing material, we add into the
particle's Euler equation an acceleration term which mimics the steep
pressure gradient of the outer PNS, $a_{\rm{R}}=-\nabla P/\rho$,
approximated from the Tolman-Oppenheimer-Volkoff (TOV) equation for
our model.  However, fluid elements with density greater than that at
neutron drip ($\rho > \rho_{\rm{drip}}=4\times10^{11}\gcc$) accrete
directly.  The important aspect in these specifics is to provide a
reasonable representation of the PNS effects on the evolving disk
material, and we do not expect a strong model-dependence from using
these relations.

A fluid particle accretes onto the central object in two ways.  Most
basically, when it reaches $R_{\rm{PNS}}$ with specific angular
momentum, $j$, then the particle's mass, $m$, and total angular
momentum about the axis of rotation, $J=mj$, are added to that of the
PNS (calculating $j$ is further discussed in \S\ref{sec:grsr}).  Also,
since SPH particles represent a density distributed through a kernel,
we `accrete' a fraction of the mass, $m_{\rm{f}}$, and total angular
momentum, $J=m_{\rm{f}}j$, as the particle approaches $R_{\rm{PNS}}$
to within a fraction of the smoothing length.  This action also
requires that any accreting particle's smoothing length must be much
less than $R_{\rm{PNS}}$, effectively establishing a resolution
requirement for the simulation.

Above its maximum mass, the PNS becomes a BH, whose properties are
determined, in this case, solely by its mass and spin.  During this
stage, the flow boundary possesses no surface pressure and is
approximated as the surface of a sphere with radius
$r_{\rm{in}}=R_{\rm{ISCO}}$ of a Kerr BH in the equatorial plane.  All
matter is co-rotating with the BH, and $r_{\rm{in}}$ can be calculated
exactly \citep[][and see Eq.~(\ref{bhparam}),
  below]{1989pbh..book.....N}.  The
numerical procedure of the accretion is the same as in the PNS case,
and we note that the final $R_{\rm{PNS}}$ is fortuitously similar to
$r_{\rm{in}}$ for a newly-formed BH, which prevents the introduction
of numerical errors at the conversion.

\subsection{General and special relativity}
\label{sec:grsr}

In order to include effects of general relativity in an approximate
manner, we add the acceleration of a pseudo-Newtonian potential for
the central object to the gravity calculated between the SPH
particles, in both the BH and PNS phases; though, the effects are
larger for the BH case due to its larger mass and greater spin, and
also $R_{\rm{PNS}}>r_{\rm{in}}$ for all masses, so that the innermost
flow is at a larger radius, further decreasing PNS-GR effects.  In
this section we utilise relativistic units with $c=G=1$, and, while
the same treatment of acceleration is used for both types of compact
object, below we refer to them collectively as `BH'.

Due to the large angular momentum of the accreted material, one must
utilise approximations appropriate for a rotating central object.
From the many pseudo-potenials which have been developed in different
studies, we have chosen to use the form of inward radial free-fall
acceleration, $\rm{a_R}(\textit{R})$, for a Kerr BH with mass, $M$,
and specific angular momentum, $a$, given by
\citet{2003ApJ...582..347M}:
\begin{equation} \label{SEPequ}
\rm{a_{SEP}}(\textit{R})=\frac{\textit{M}}{\textit{R}^2}\left[1 - \left(\frac{\textit{r}_{\rm{in}}}{\textit{R}}\right)+  \left(\frac{\textit{r}_{\rm{in}}}{\textit{R}}\right)^2   \right],
\end{equation}
derived from what they call the `Second-order Expansion Potential'
(SEP), where all radii have been scaled to gravitational units,
$r_{\rm{g}}=GM/c^2$, and $r_{\rm{in}}$ is the location of the ISCO,
given by:
\begin{eqnarray}\label{bhparam}
 r_{\rm{in}} &= & 3 + Z_2 - [(3 - Z_1)(3 + Z_1 + 2Z_2)]^{1/2}\,, \\[0.2cm]
 Z_1 &= & 1 + (1-a^2)^{1/3}[(1+a)^{1/3} + (1-a)^{1/3}]\,,\nonumber \\[0.2cm]
 Z_2 &= & (3a^2 + Z_1^2)^{1/2}\,. \nonumber
\end{eqnarray}
We discuss this choice of pseudo-potential and compare its
characteristics to others, as well as to full GR solutions, in
Appendix B.

In order to keep track of the BH spin parameter, $a$, we add to it the
total angular momentum of all accreted material, $J=mj$.  For material
accreting from the equatorial plane, we approximate the specific
angular momentum, $j$, as that of a corotating particle in circular
orbit at $r_{\rm{in}}$, given as \citep{1983bhwd.book.....S}:
\begin{equation}
j_{\rm{K}} = \sqrt{\frac{M}{r_{\rm{in}}}}  \frac{(r_{\rm{in}}^2 - 2a\sqrt{Mr_{\rm{in}}}+a^2)}
{(r_{\rm{in}}^2 - 3Mr_{\rm{in}} +2a\sqrt{Mr_{\rm{in}}})^{1/2}}\,, 
\end{equation}
which is a several times larger than the Newtonian equivalent in the
vicinity of the accretion boundary.  In the case of polar accretion
(defined as when the distance to the equatorial plane is greater than
that to the rotation axis), we simply use the Newtonian value,
$j=xv_{\rm{y}}-yv_{\rm{x}}$.  Necessarily, this material will tend to
have much less angular momentum anyway, so that this is a reasonable
approximation.

For all velocities discussed here, we utilise a simple
special-relativistic rescaling of the values occuring within the
pseudo-potential, $v_{\rm{ps}}$:
\begin{equation}
v_{\rm{SR}} = \frac{v_{\rm{ps}}}{\sqrt{1 + v^2_{\rm{ps}}}}\,.
\end{equation}
This has been shown by \citet{1996A&A...313..334A} to produce
velocities within 5\% of relativistic calculations around non-rotating
BHs when used in conjunction with the Paczy\'{n}ski-Wiita potential
\citep{1980A&A....88...23P}.

We recognize that using these approximations to the Kerr metric in
order to represent the gravitational contribution of the central
object is a rough way of proceeding, but we believe that it
constitutes a legitimate approach at the present stage of the work
(see further discussion in Appendix B).  In the case where the central
object is a BH, it is clearly very approximate to treat the
gravitational acceleration on fluid elements as the sum of
contributions of a vacuum BH pseudo-potential and the Newtonian
self-gravity of surrounding matter.  In the case where the central
object is a PNS, the surrounding spacetime differs from that of the
Kerr metric anyway, with the former having quadrupole moments
considerably larger than those for equivalent BHs. (However, it must
be noted that the effect of the quadrupole falls off as $R^3$, so the
influence of this difference for the disk flow may not be very great.)
Improving on this treatment remains a topic for future work, but we
believe that the approach being used here is a reasonable one at
present.

\subsection{Neutrino cooling}
\label{sec:PHYSneutc}

\begin{figure}
  \includegraphics[width=84mm]{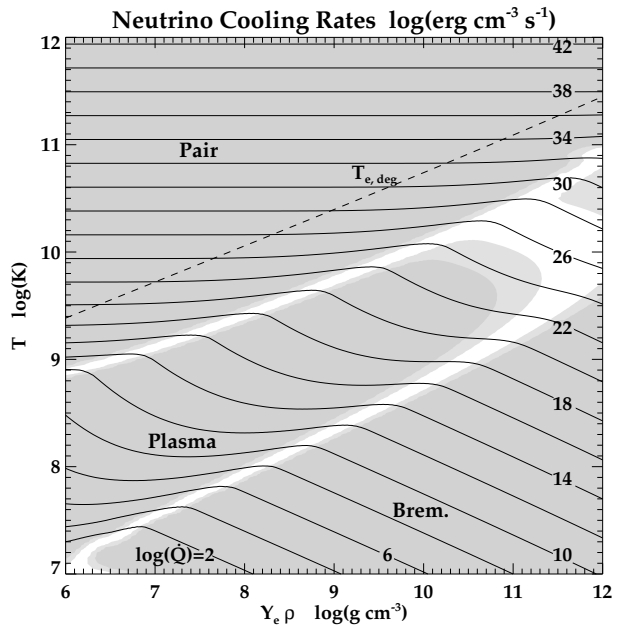}
  \caption{\label{fig_qdot} Solid lines show contours of the total
    combined cooling rate, $\dot{Q}_\nu$.  The dark shaded regions
    show where a single production process dominates $>90\%$ of the
    total cooling rate, and the lighter, $>75\%$.  Included for
    reference is the electron degeneracy temperature,
    T$_{\rm{e,deg}}$.}
\end{figure}

In the inner regions of the collapsing star, the flow is optically
thick to electromagnetic radiation, so that radiative transfer is very
inefficient and the energy carried away by photons is negligible.
Instead, much of the material, particularly in the disk itself,
occupies a region of ($\ye\rho,~T$) phase space where cooling via
neutrino emission is important (calculating $\ye$ is discussed in the
next section).  In the collapsar system local neutrino production
greatly influences disk stability, and, summed globally, it may
provide a mechanism for relativistic jet production.

For the regions which are optically thin to neutrinos, energy from
processes such as pair annihilation, plasmon decay and
(non-degenerate) bremsstrahlung is carried away freely.  Very dense
regions with $\rho > 5\times10^{12}$ are considered to be optically
thick to neutrinos, and using the common `lightbulb approximation',
cooling shuts off.

For the ranges $7\leq\log({\textit T})\leq12$ and
$6\leq\log(\ye\rho)\leq12$, we utilise tables of data from detailed
calculations by \citet*{1994ApJ...425..222H} for the plasma processes
and by \citet{1996ApJS..102..411I} for pair annihilation; for
non-degenerate bremsstrahlung scattering, we use the analytic
approximation given by \citet{1998ApJ...507..339H},
$\dot{Q}_{\rm{Br}}=1.5\times10^{33}\,T_{11}^{5.5}\,\rho_{13}^2~\rm{erg}~\rm{cm}^{-3}~\rm{s}^{-1}$,
with the notation $T_{11}=T/10^{11}$, etc.  Fig.~\ref{fig_qdot} shows
contours of the combined rates of neutrino cooling, $\dot{Q}_{\nu}$,
with shaded regions where each process dominates total cooling.
During the evolution of the collapsar disk, we note that nearly all
material for which cooling is relevant occupies ($\ye\rho,~T$) states
dominated by plasma processes and pair production.

\begin{figure}
    \includegraphics[width=84mm]{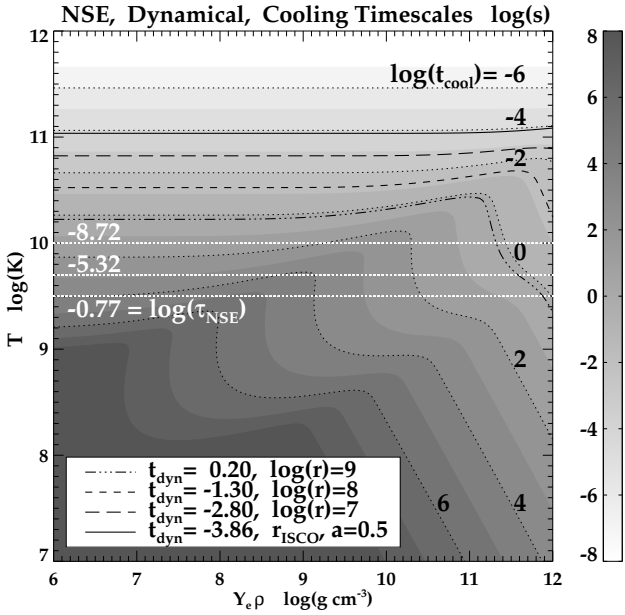}
    \caption{\label{fig_tcool} A comparison of timescales: the black
      (white) dotted lines show contours of $t_{\rm{cool}}$
      ($\tau_{\rm{NSE}}$), from Eq.~(\ref{coolvdyn}).  In most
      regions $\tau_{\rm{NSE}} \ll t_{\rm{cool}}$.  Values of
      dynamical timescale, $t_{\rm{dyn}}=\Omega^{-1}$, are given and
      overplotted on the $t_{\rm{cool}}$ contours for direct
      comparison regarding local stability.}
\end{figure}

The importance of the relations between various timescales for the
stability of the collapsar disk has already been introduced in
\S\ref{sec:INTtoomr} (and $\tau_{\rm{NSE}}$ is discussed in
\S\ref{sec:PHYSnse}).  Fig.~\ref{fig_tcool} shows the relevant
cooling timescales ($t_{\rm{cool}}=\epsilon/\dot{Q}_{\nu}$) for
neutrino processes in this phase space.  Typical values of the
dynamical timescale ($t_{\rm{dyn}}=\Omega^{-1}$, orbital period)
around a 3~M$_{\odot}$, $a=0.5$~BH at different radii have also been
calculated.  For direct comparison, the dynamical timescales are
overlaid on the phase space, with reference to the cooling-stability
criterion of Eq.~(\ref{coolvdyn}).  Fluid at a given radius is stable
against clumping for the various ($\ye\rho,~T$) states in the
region below the $t_{\rm{dyn}}({\it r})$ contour, and unstable for
those a few times above.

\subsection{Electron fraction}
\label{sec:elecfr}
\begin{figure}
  \includegraphics[width=84mm]{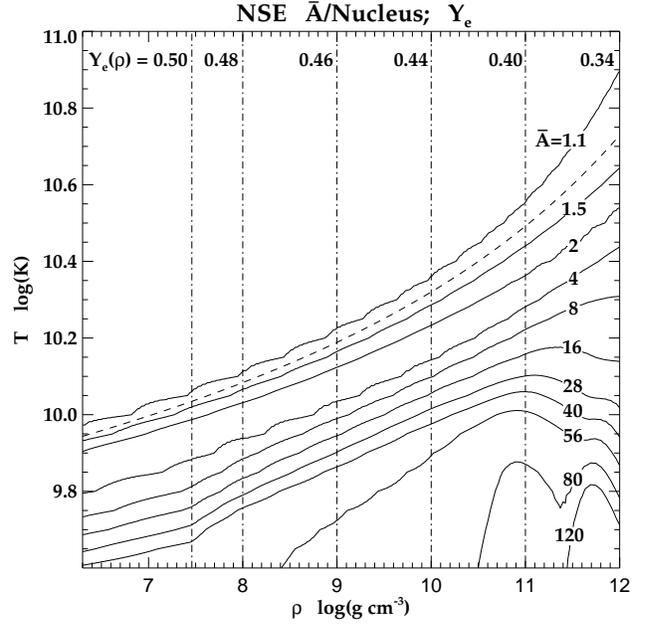}
  \caption{Contours show the average nucleon number, $\abar$, in
    NSE for a given temperature and density, as calculated by F. Timmes
    (private communication).  The dashed curve shows the commonly used
    onset of free nucleon matter.  Vertical dot-dashed lines show
    density-dependent electron fraction, $\ye$.}
  \label{fig_nse}
\end{figure}

As the density of matter increases, so does the rate of capture of
electrons by nuclei. This neutronisation decreases the electron
fraction, $\ye$, and necessarily the proton/neutron ratio, below
$0.5$.  This change has important consequences for nuclear processes
and cooling via neutrino production, which are dependent on the proton
density, $\ye\rho$. For densities $\rho > 10^{7}$, we follow the
simple prescription of \citet{2005ApJ...633.1042L}, who noted in
core-collapse simulations that $\ye$ remains nearly constant in time
and temperature for a given density.  Using their `G15' set of
parameters, the approximate electron fraction, $\ye(\rho)$ is given
by:
\begin{eqnarray}
  \ye(\textit{x})&= & 0.354 - 0.111x \nonumber \\
  &~& + \, 0.035|x|\,[1 + 4\,(|x| - 0.5)(|x| - 1)]\,,\nonumber\\[0.1cm]
  x(\rho) &= &\max \left[ -1 , \min \left( 1 , 0.3434\,\log(\rho) - 3.5677 \right) \right]\,.
\end{eqnarray}
The electron fraction for a range of densities is shown by the
vertical lines in Fig.~\ref{fig_nse}.  Thus, roughly-speaking, the
electron fraction remains both temperature- and time-independent, with
the restriction that, for a given $i$th fluid element,
$Y_{\rm{e},{\it i}}$ monotonically decreases over time, i.e.,
upwards fluctuations are not permitted.  Matter with density $\rho <
10^{7}$ retains its initial value of $Y_{\rm{e},{\it i}}$.

\subsection{NSE}
\label{sec:PHYSnse}

In regions of high temperature and density, nuclear reactions
involving strong interactions equilibrate with their inverse
reactions, and nuclear statistical equilibrium (NSE) is reached
\citep[e.g.,][and within]{2006NuPhA.777..188H}.  Assuming that matter
remains transparent to neutrinos, element/isotope abundances may then
be calculated using statistical relations which involve a large
network of relevant isotopes, for material with $T\geq T_{\rm{NSE}}$
($\equiv4\times10^9~\rm{K}\approx0.34$~MeV).  Importantly, in these
regions Fe-group elements are formed, predominantly from Si-group
elements created during a lower $(\rho, T)$, quasi-static equilibrium
phase \citep*[QSE, e.g.][]{1968PhRvL..20..161B,1998ApJ...498..808M}.

For calculating the average nucleon number, $\abar_{\rm{NSE}}$, of a
fluid element during the simulation, we consider matter with both
$T>T_{\rm{NSE}}$ and $\rho>10^6$ to be in NSE (for a discussion of
$\tau_{\rm{NSE}}$, see below). Computationally, a fluid element below
$T_{\rm{NSE}}$ retains its pre-collapse $\abar$ value.  Knowing
($\rho,~\ye,~T$), $\abar_{\rm{NSE}}$ is determined from a table of
values calculated by F. Timmes \citep*[private communication, and
  see][]{1999ApJS..124..241T} and shown as a function of temperature
and density (as above, $\ye=\ye(\rho)$) in Fig.~\ref{fig_nse}.  Also
shown as a guide is the often-used approximation for where the mass
fraction of free neutrons and protons in material is unity,
$X_{\rm{nuc}}=1$ \citep{1992ApJ...391..228W}.

We calculate the abundances of specific $\alpha$- and heavy-elements
produced by NSE conditions in the inner regions of the collapsar using
results of the same reaction network calculations by F. Timmes (see
Appendix C).  Due to the observational association of LGRBs with
HNe/SNe, we investigate in particular the production of $^{56}$Ni, the
radioactive isotope which is the dominant energy source of SN
lightcurves via decay to $^{56}$Co.  A typical core collapse SN
contains $^{56}\rm{Ni}(\approx0.1)$ at time of explosion, and
estimates for the brighter HN may require up to $0.5-0.7~\msun$ of
$^{56}$Ni; some SNe may possibly require up to $\sim5~\msun$
(\citet{2008ApJ...673.1014U} and references within), but this latter
value remains uncertain.  The masses of competing end products,
non-radioactive $^{54}$Fe and elemental Co, are also approximated;
free nuclei and light $\alpha$-elements are expected to dominate much
of the hot, dense collapsar disk.

Elemental abundances provide further means for understanding the
observed association of asymmetric and polarised SNe with LGRBs.
\citet{2005Sci...308.1284M} showed that, among other features, a HN
viewed along the axis of the GRB jet would possess a strong,
single-peaked [Fe II] emission line, while one seen in the equatorial
plane of the system would possess a double-peaked [O I] emission line.
It was reasoned that Fe-group elements were produced during the
collapse near the jet, while the lighter elements, remaining from the
progenitor's evolution, settled into the equatorial plane and were
ejected from the evolving disk.  Their analysis matched the Fe-line
observed in SN 1998bw (associated with GRB 980425), and the O-feature
in SN 2003jd (no GRB).

Therefore, we examine the amounts and locations of Fe and O produced
in the collapsar from NSE, noting the difference in origin for the
latter in this case; at larger radii, a significant mass of O created
during the stellar evolution of the progenitor is expected to remain
in roughly a spherical distribution.  The timescale for reactions to
reach NSE varies strongly with temperature and very weakly with
density.  Here, we utilise the approximation
$\tau_{\rm{NSE}}=\exp(149.7/T_9 - 46.054)$~s
\citep*{2005ApJ...623..337G}. This value for fluid elements is
compared with other relevant timescales in the disk (shown in
Fig.~\ref{fig_tcool}), particularly the dynamical and viscous
timescales in the inner and outer parts, respectively.  In regions
where the nuclear timescale is significantly less than either of
these, one can calculate the nuclear abundances of material with
relative ease \citep[again, see e.g.,][]{2006NuPhA.777..188H}.

\subsection{The role of non-NSE nucleosynthesis}
\label{sec:nonnse}
The $^{56}$Ni required to power HN lightcurves may also be produced by
non-NSE nucleosynthetic tracks, for example during the expansion and
cooling of hot material. Such time-dependent nuclear processes (e.g.,
the $r$-process and explosive nucleosynthesis) have not been included
in this study. However, the feasibility of producing a significant
amount of $^{56}$Ni by non-NSE processes can be gauged through an
examination of the properties of the material comprising candidate
regions for rapid expansion, namely in outflows
\begin{enumerate}
\renewcommand{\labelenumii}{\arabic{enumii}}
 \item from the outer edge of disk: equatorial outflows due to
   angular momentum transfer;
 \item from equatorial plane: vertical neutrino or thermally driven
   winds from hot disk material;
 \item from the inner edge of the accretion disk: hot `bubbles' coming
   from non-accreted flow or created from jet-interaction, ejected in
   the polar direction.
\end{enumerate}

In typical cases for SNe and outflows from compact objects, Fe-group
and heavier elements are manufactured by high entropy material (of
order $10~k_{\rm{B}}/A$ and higher, as a minimum, where $k_{\rm{B}}$
is Boltzmann's constant) which is expanding rapidly ($v\sim 0.1c$).
Furthermore, the electron fraction of material ($\ye\lesssim0.5$
initially for most regions of interest here) is important in
determining nucleosynthetic results; outflow with an extremely low
electron fraction permits $r$-processing (yielding neutron-rich
elements), while that which is more in the neighbourhood of
$0.44\lesssim\ye<0.5$ is a candidate to evolve to $\ye\approx0.5$, at
which point significant $^{56}$Ni may be produced (though the precise
evolution depends on multiple timescales, on the presence of
neutrinos, etc.; see, e.g. \citealt*{2003ApJ...586.1254P};
\citealt{2004ApJ...617L.131J}; \citealt*{2004ApJ...606.1006P};
\citealt*{2006ApJ...643.1057S}; \citealt{2008ApJ...685L.129S}).

\section{Progenitor model}
\label{sec:progenmodel}

\begin{table*}
  \begin{center}
    \begin{minipage}{170mm}
      \caption{\label{tab:models} Summary of collapsar models, mapped
        from \citet{2005ApJ...623..302F} pre-collapse binary merger
        (details in text).  Rotation and $E_{\rm{th}}$ profiles have
        been scaled in cases to test a wide array of scenarios.
        Values of quantities are given for the initial $M_0=0.5~\msun$
        of free material in the models, which utilise the SEP
        pseudo-potential unless otherwise stated.  At $t=0$~s, the
        total potential energy for the SEP and Newtonian potentials is
        $E_{\rm{V,SEP}}=-11.9\times10^{50}$~erg and
        $E_{\rm{V,New}}=-12.0\times10^{50}$~erg, respectively; all
        models have the same total radial kinetic energy,
        $E_{\rm{kin},R}=4.3\times10^{50}$~erg, and the integrated
        (inward) radial mass-flux, $\dot{M}_{\rm{R}}=3.44~\mps$.}
    \begin{tabular}{lcccccccc}
      \hline
      \mc{1}{c}{Model}& \mc{1}{c}{Velocity}& \mc{1}{c}{$v_{\phi}$}   & \mc{1}{c}{ $E_{\rm{th}}$} & \mc{1}{c}{$J_0$} & \mc{1}{c}{$y_j$} & \mc{1}{c}{$E_{\rm{kin}}$~[$\textit{E}_{\rm{kin},\phi}\,^{\dag}$]} & \mc{1}{c}{$E_{\rm{th}}$}& \mc{1}{c}{Further}  \\
      &\mc{1}{c}{Profile}& \mc{1}{c}{Scaling }& \mc{1}{c}{Scaling}    & \mc{1}{c}{($10^{50}$~erg~s)} & \mc{1}{c}{($\bar{j}_0/j_{\rm ISCO}$)} & \mc{1}{c}{($10^{50}$~erg)}& \mc{1}{c}{($10^{50}$~erg)}&  \mc{1}{c}{ Notes }\\
\hline
PSm2		& shell		& $\times$2	& full  &15.64&65.5	&7.78\,[3.48]	& 3.83		& -- \\
PSm1		& shell		& full		& full 	&~7.82 &32.8	&6.04\,[1.74]	&3.83		&  -- \\
PSdsq2		& shell		& /$\sqrt{2}$	& full  &~5.53& 23.1	&5.17\,[0.87]	&3.83		& -- \\
PSdsq2J$^{\ddag}$& shell	& /$\sqrt{2}$	& full  &~5.53&23.1 	&5.17\,[0.87]	&3.83		& jet added at late time  \\
PSd2		& shell		& /2		& full  & ~3.91 &16.4	&4.74\,[0.44]	&3.83		&  -- \\
PSd5		& shell		& /5		& full 	&~1.56 &~6.5	&4.37\,[0.07]	&3.83		& -- \\
PCm1		& cyl.		& full		& full  &~8.68 &36.3 	&5.50\,[1.20]	& 3.83		& -- \\
PCdsq2		& cyl.		& /$\sqrt{2}$	& full  &~6.14 &25.7	&4.90\,[0.60]	&3.83		& -- \\
PCd2		& cyl.		& /2		& full  &~4.34 &18.2  	&4.60\,[0.30]	& 3.83		& -- \\
PSm1Kd5		& shell		& full		& /5   	&~7.82&32.8	&6.04\,[1.74]	& 0.77		& -- \\
PSd2Kd5		& shell		& /2		& /5    &~3.91 &16.4 	&4.74\,[0.44]	&0.77		& -- \\
NSm1		& shell		& full		& full 	&~7.82& 32.8	&6.04\,[1.74]	&3.83		& Newtonian potential \\
NSd2		& shell		& /2		& full  &~3.91 &16.4 	&4.74\,[0.44]	&3.83		& Newtonian potential \\
\hline
\end{tabular}
\medskip
{$^{\dag}$}\small{The azimuthal contribution to
  $E_{\rm{kin}}$.~~~~$^{\ddag}$Exactly the same evolution as PSdsq2
  until $t=1.10$~s.}
\end{minipage}
\end{center}
\end{table*}

The `physical' progenitor used in these simulations comes from the
resultant of the merger of 2 He stars (8+8 M$_{\odot}$), evolved to
the point of core collapse by \citet{2005ApJ...623..302F}.  We utilise
the results of their model `M88c', which was predicted to form a BH
instead of a NS and which also possessed a large amount of angular
momentum (the maximum of their several scenarios of the merger
process).  We now desecribe the mapping of the 1D, pre-collapse
(polytropic, $\gamma=5/3$) model into our 3D axially symmetric,
post-collapse ($\gamma=4/3$) starting point, and also the range of
related models which we analyse here.

The collapsar simulation begins $\approx2$~s after core collapse,
after the stellar core has formed a $1.75~\msun$ PNS
($>\rm{M}_{Chandra}\equiv1.4~\msun$, but at this stage, the PNS is
quite `oversized' before neutrinos have diffused out).  A rarefaction
wave moving radially outward at the local sound-speed leads to
continued infall.  Following \citet*{1993PhR...227..157C} and
\citet*{1996ApJ...460..801F}, we assume that an isentropic, convective
envelope in pressure equilibrium has formed around the PNS.  This
material possesses a high specific entropy per baryon
($s_{\rm{env}}=15~\textit{k}_{\rm{B}}/A$), and at its outer
radius ($\approx 10^7$~cm) the hydrostatic pressure balances the ram
pressure (=$\rho v^2$) of the infalling stellar matter.

From the 1D \citet{2005ApJ...623..302F} profile, we first generate a
3D pre-collapse model which is spherically symmetric in density and
other EoS variables.  Similarly, we assume that $\ye$ and $\abar$ (not
necessarily in NSE) are spherically symmetric.  We implement two
general classes of rotation profile: first, shellular, where the
angular velocity, $\dot{\phi}$, is constant for a given radius from
the centre; and second, cylindrical, which is expected for a state of
permanent rotation with aligned surfaces of constant density and
pressure, as corollaries of the Poincar\'{e}-Wavre Theorem
\cite[e.g.,][]{1978trs..book.....T}.  Shellular rotation is expected
in systems with strong, anisotropic turbulence oriented perpendicular
to equipotential surfaces \citep{1992A&A...265..115Z}.

From the above, there are three distinct regions to be translated from
the pre-collapse progenitor: PNS, envelope and infall.  The inner
1.75~M$_{\odot}$ simply becomes the PNS, whose outer radius and
numerical properties have been described in \S\ref{sec:PHYScentob}.
The envelope is then built up in spherically symmetric layers of
hydrostatic equilibrium, with radial density profile:
\begin{equation}
\rho(R) =
\left[ 
  \frac{G M_{\rm{PNS}}}{4 K_{\rm{env}}}
  \left( \frac{1}{{\it R}} - \frac{1}{R_{\rm{PNS}}} \right)
  + \left( \frac{P_{\rm{PNS}}}{K_{\rm{env}}} \right)^{1/4}
  \right]^3
\end{equation}
where $P_{\rm{PNS}}$ is the pressure at the inner edge of the
envelope.  The entropic function in the envelope,
$K_{\rm{env}}=P/\rho^{\gamma}$, defines the thermal profile and is,
for a radiation-dominated gas, simply related to specific entropy per
baryon, $s$, as $K= (3c/4\sigma)^{1/3}
(k_{\rm{B}}\,s/4\rm{m}_{p})^{4/3}$, where $\sigma$ is the
Stefan-Boltzmann constant, and m$_{\rm{p}}$, the proton mass.

To determine the structure of the infalling material, we use an
approximate rate of fallback, $\dot{M}\lesssim 1~\mps$, across a
spherical shell at $2\times10^{7}$~cm, appropriate for collapsing
massive stars \citep[e.g.,][]{2006NewAR..50..492F}.  The radial
velocity is taken to be that of freefall, $v_{\rm{ff}}$, from the
original radius of the shell, and therefore, the infalling density is
given by $\rho_{\rm{inf}}(R)=\dot{M}/(4\pi R^2v_{\rm{ff}})$.  For the
azimuthal velocity, $v_\phi$, we preserve the specific angular
momentum of the fluid elements in collapse.  Finally, the internal
energy of the fluid elements is calculated by assuming nearly steady
flow with negligible cooling, so that the Bernoulli equation applies.

In both the envelope and the infalling region, for material with $\rho
>10^6$ and $T > 10^7$ ($T>4\times10^9$), $\ye$ ($\abar_{\rm{NSE}}$) is
calculated explicitly, as described in \S\ref{sec:elecfr}.  Otherwise,
those quantities are unchanged in a fluid element from their
pre-collapse values.  The PNS (and later, BH) surface defines the
inner boundary condition.  At the outer boundary mass is regularly
added to the system as continued infall, ensuring that the outermost
radius at any time is much larger than the accretion disk/inner region
of interest.  At $t=0$~s, the simulation contains $M_0 = 0.5~\msun$ of
free material modelled using $2.25\times10^5$ SPH particles of equal
mass (which remains the same for all added infall particles).

\begin{figure}
  \begin{center}
    \includegraphics[width=74mm]{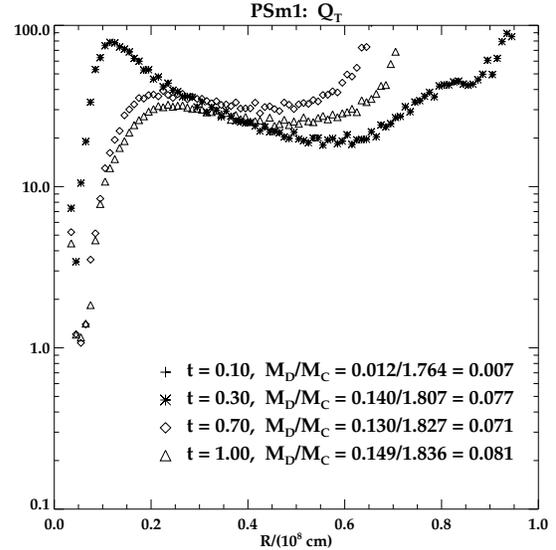}
    \caption{\label{fig:PSm1_qts} Model PSm1: evolution of the Toomre
      parameter, $\qt$. The disk remains globally stable, as
      $\qt>1$.  One local ring (where $\qt\approx1$)
      develops non-axisymmetry which eventually is damped globally
      (see Fig.~\ref{fig:PSm1_eqplden}).}
    \end{center}
\end{figure}

Table~\ref{tab:models} lists the properties of models tested in this
study, calculated only from the initially present $M_0$.  They do not
include properties of the subsequent infalling layers, and as such,
the bulk values are relevant measures of relative energies and angular
momentum, but they are not absolute quantities for the entire
collapsar.  In addition, we define the dimensionless ratio, $y_j$, of
the inflow average specific angular momentum, $\bar{j}_0=J_0/M_0$, to
the characteristic quantity, $j_{\rm ISCO}$ of the central object
(even for the Newtonian potentials, as an approximation), and we
discuss further application of this in \S\ref{sec:discusssim}.  In
order to further explore a wider range of possible LGRB progenitors,
various physical properties of the pre-collapse progenitor are
rescaled. In particular, we focus on quantities which appear in $\qt$,
affecting the stability of the forming accretion disk.  Scalings of
both the shellular and cylindrical rotation profiles are tested, which
affect $\Sigma$ by determining the location of centrifugal balance for
a fluid element.  Similarly, the specific internal energy ($\propto
c_{\rm{s}}^2$) is decreased, to examine the effects on $\dot{M}$ and
the neutrino cooling rates by changing the initial location of a fluid
element in the ($\rho, T$) phase space.  For comparison, a Newtonian
potential is used in two models to gauge the effects of relativity on
the collapsar system.  Finally, in one model a `jet' is inserted at
late times after the formation of a spiral, in order to examine the
production of outflows.

\begin{figure*}
  \vspace{-.4cm}
  \begin{center}
    \begin{tabular}{cc}
      \includegraphics[width=78mm]{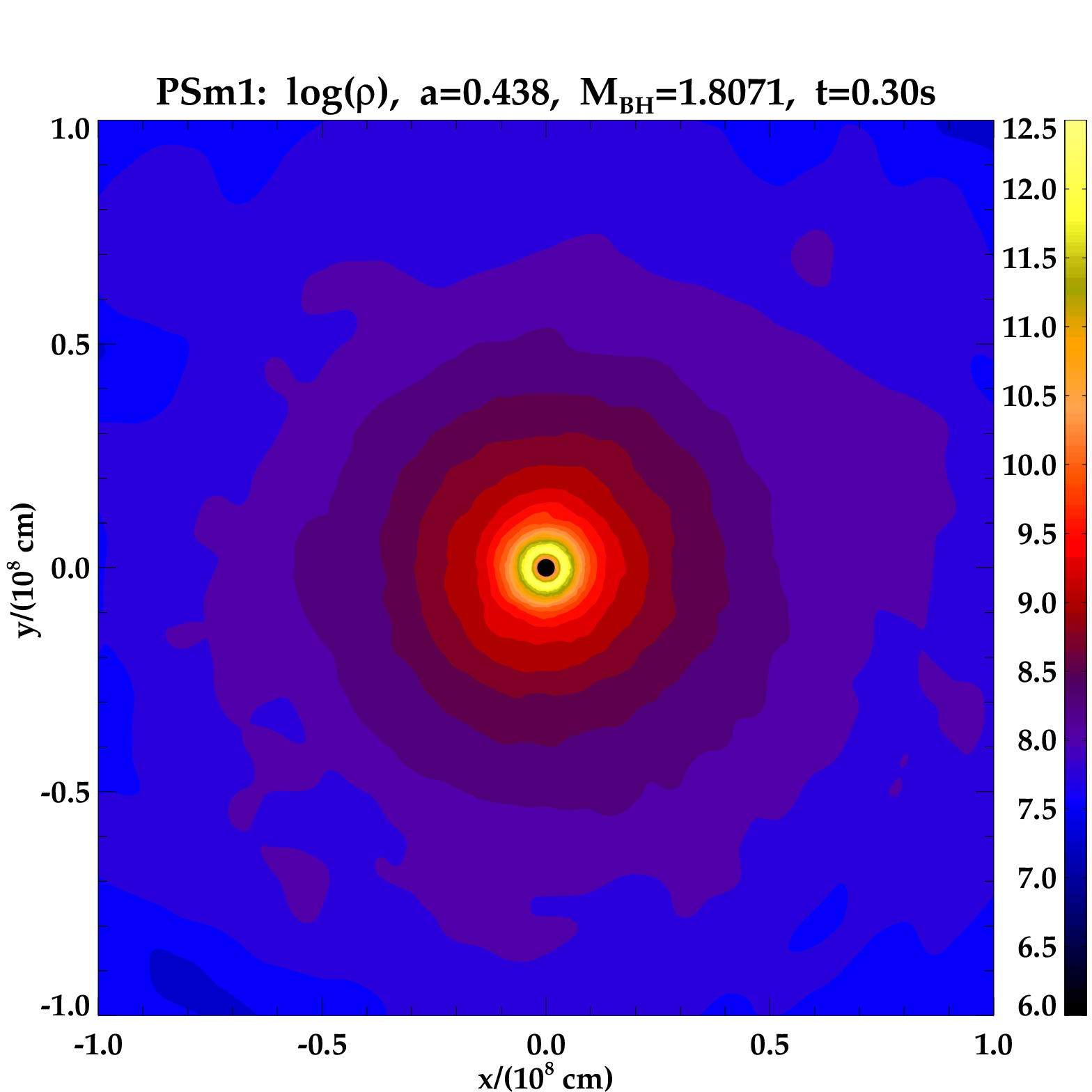}&
      \includegraphics[width=78mm]{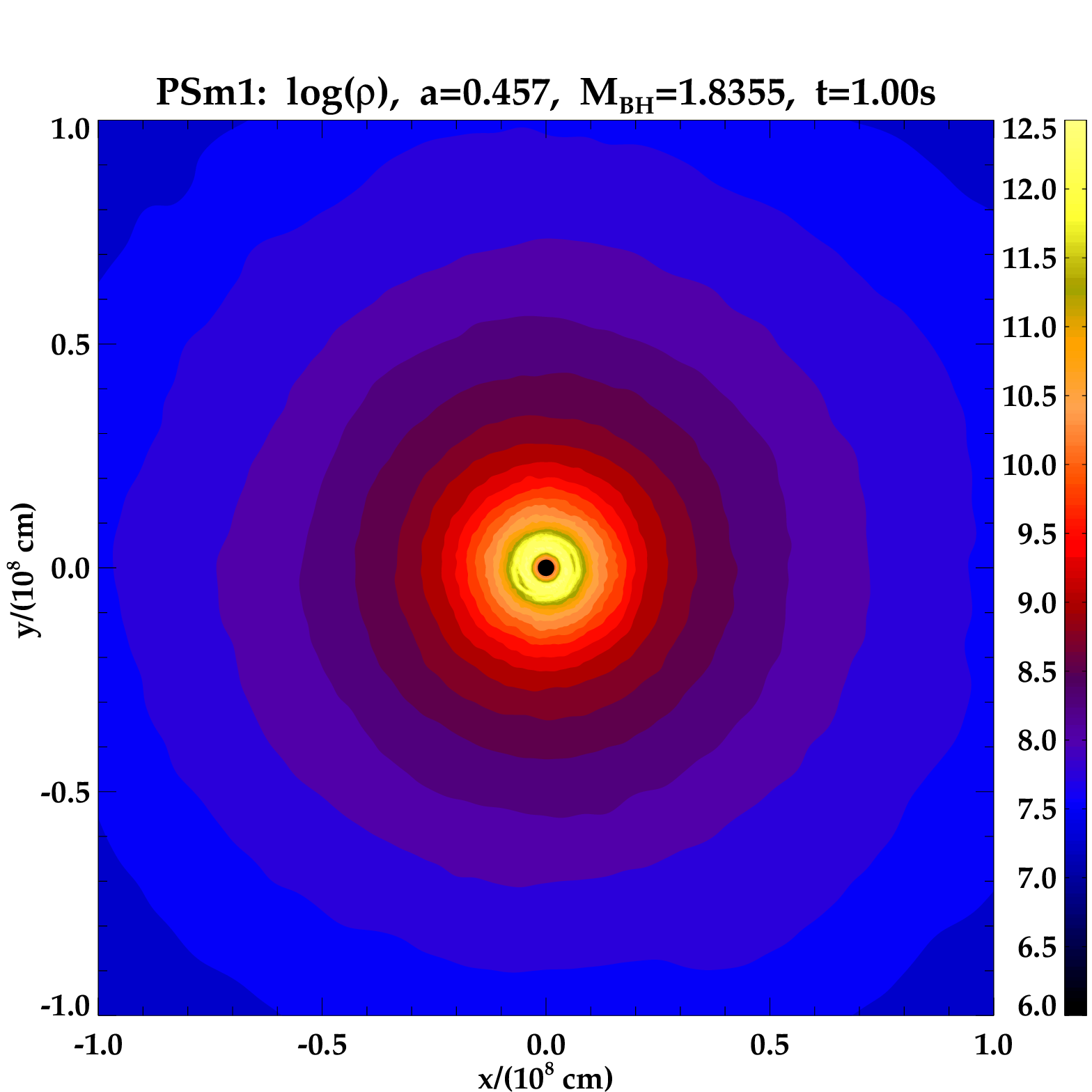}\\[-.4cm]
      \includegraphics[width=78mm]{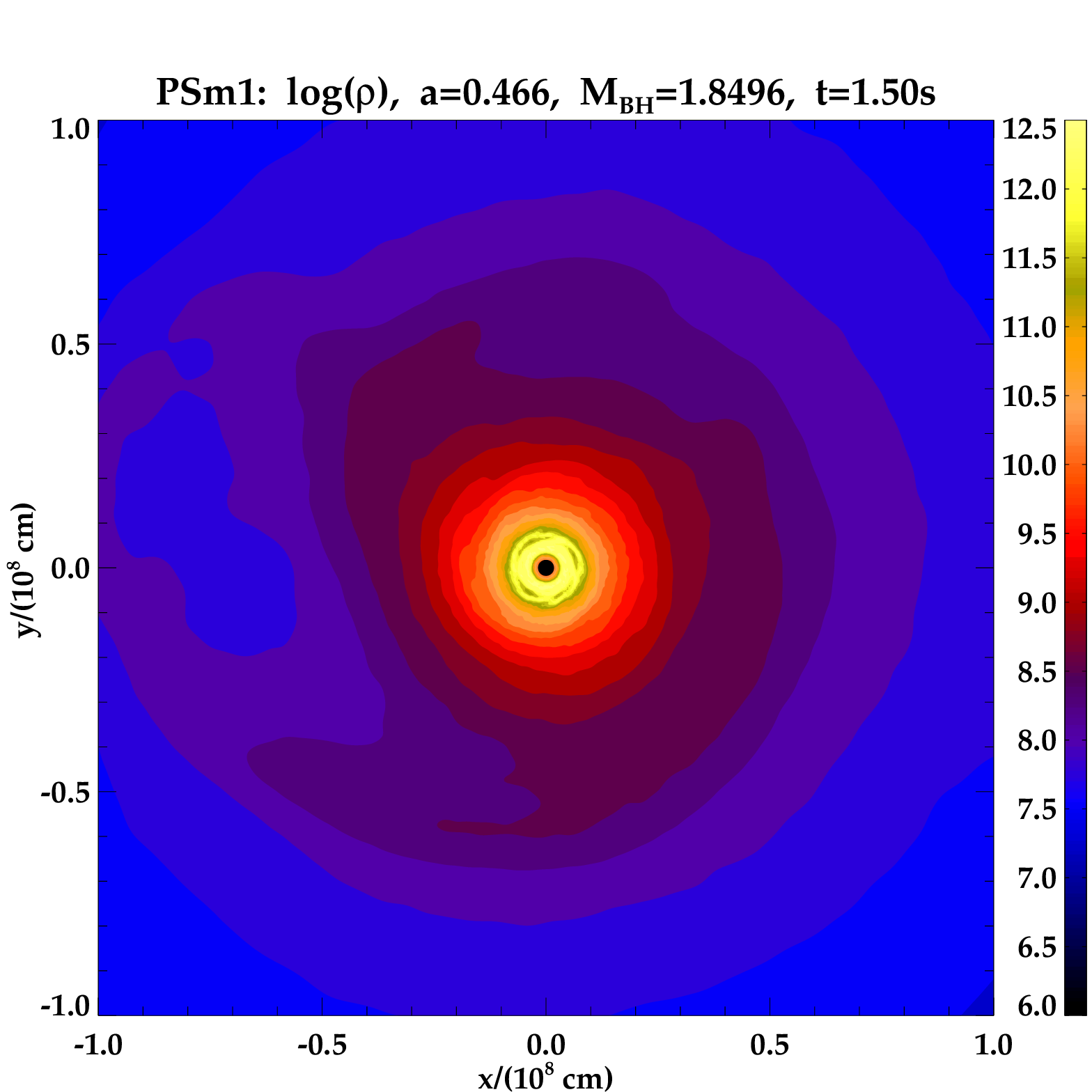}&
      \includegraphics[width=78mm]{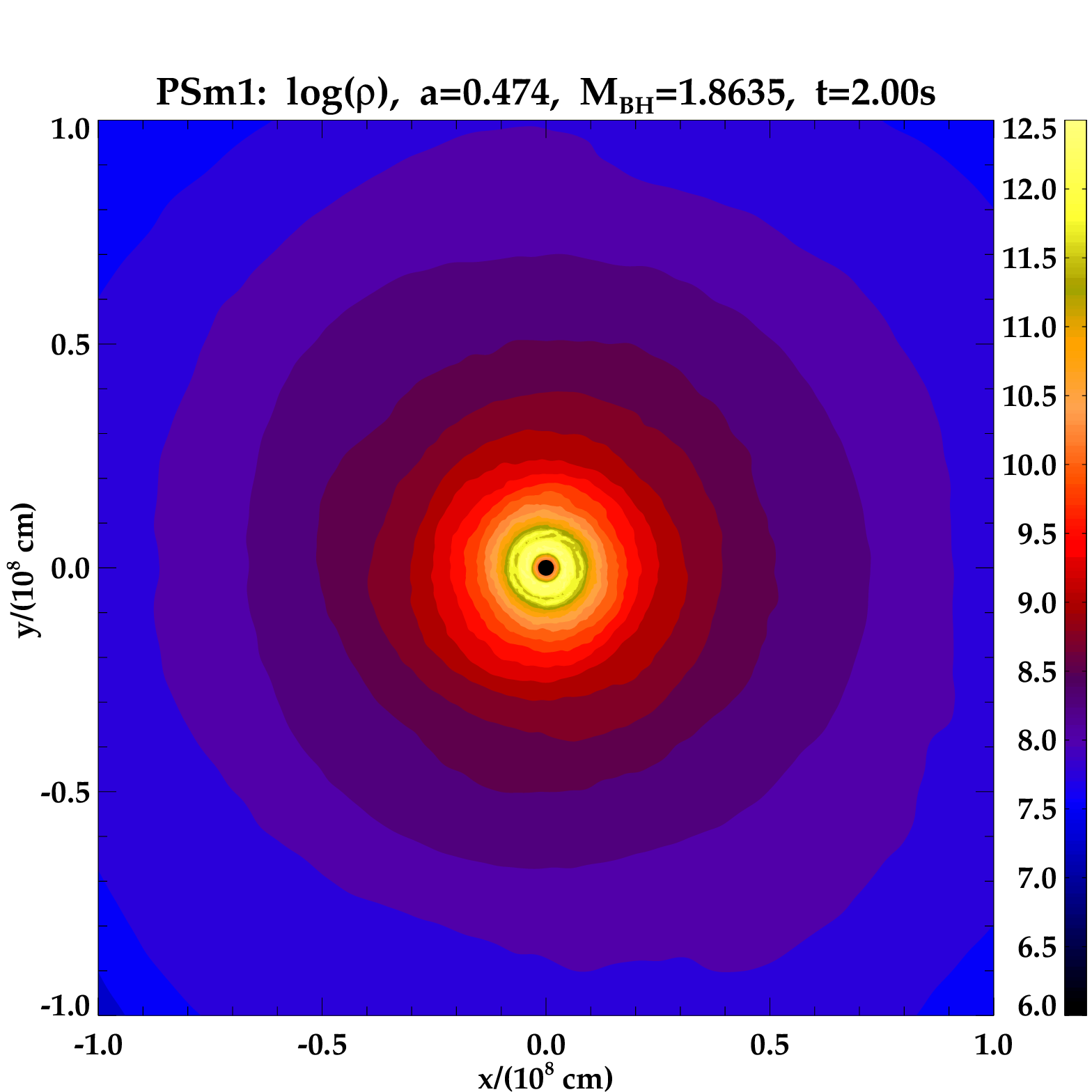}\\[-.2cm]
      \end{tabular}
      \caption{\label{fig:PSm1_eqplden} Model PSm1 (equatorial density)
        closely follows the evolution of $\qt$, as the disk is
        globally stable to hydrodynamic perturbations present in the
        early phases. A locally unstable region leads to the growth of
        temporary non-axisymmetric structure, which damps out due to
        global stability}
   \end{center}
\end{figure*}

\section{Shellular rotation models}
\label{sec:shell}

We now describe the evolutions of various models with shellular
profiles. In order to appreciate the evolution of a number of the more
dynamic models, we also make reference to animations\footnote{Located
  at:
  \href{http://www-astro.physics.ox.ac.uk/\string~ptaylor}{http://www-astro.physics.ox.ac.uk/\string~ptaylor}. All animations show the evolution of a thin slice of the
  equatorial plane, excepting PSdsq2\_invden\_mov.gif (vertical plane
  slice).} showing the evolution of density ($\rho$), specific entropy
per baryon ($s/A$) and entropic function ($K$).

\subsection{PSm1}
\label{sec:models_sphere1}

Fig.~\ref{fig:PSm1_qts} shows the evolution of $\qt$, for the
evolving PSm1 accretion disk.  Also shown are estimates of the mass of
the forming disk compared to central object mass ($\rm{M_D/M_C}$).  As
the disk grows, $\qt$ approaches the region of instability in
one localised region.  At late times, the minimum remains just greater
than unity throughout the disk, which is therefore predicted to remain
globally stable against the development of non-axisymmetric structures
(though locally, one region remains near instability).  In testing the
often-used approximation for classifying stability, it was found that
the shape of the $Q'_{\rm{T}}$ curve was very similar to that in
Fig.~\ref{fig:PSm1_qts} but with systematically smaller values, in
roughly constant ratio, $Q'_{\rm{T}}/\qt\approx 70-80\%$.  This
behaviour remains in the other pseudo-potential systems as well, and
only $\qt$ has been used in analysis.

\begin{figure*}
  \vspace{-.4cm}
  \begin{center}
    \begin{tabular}{cc}
      \includegraphics[width=78mm]{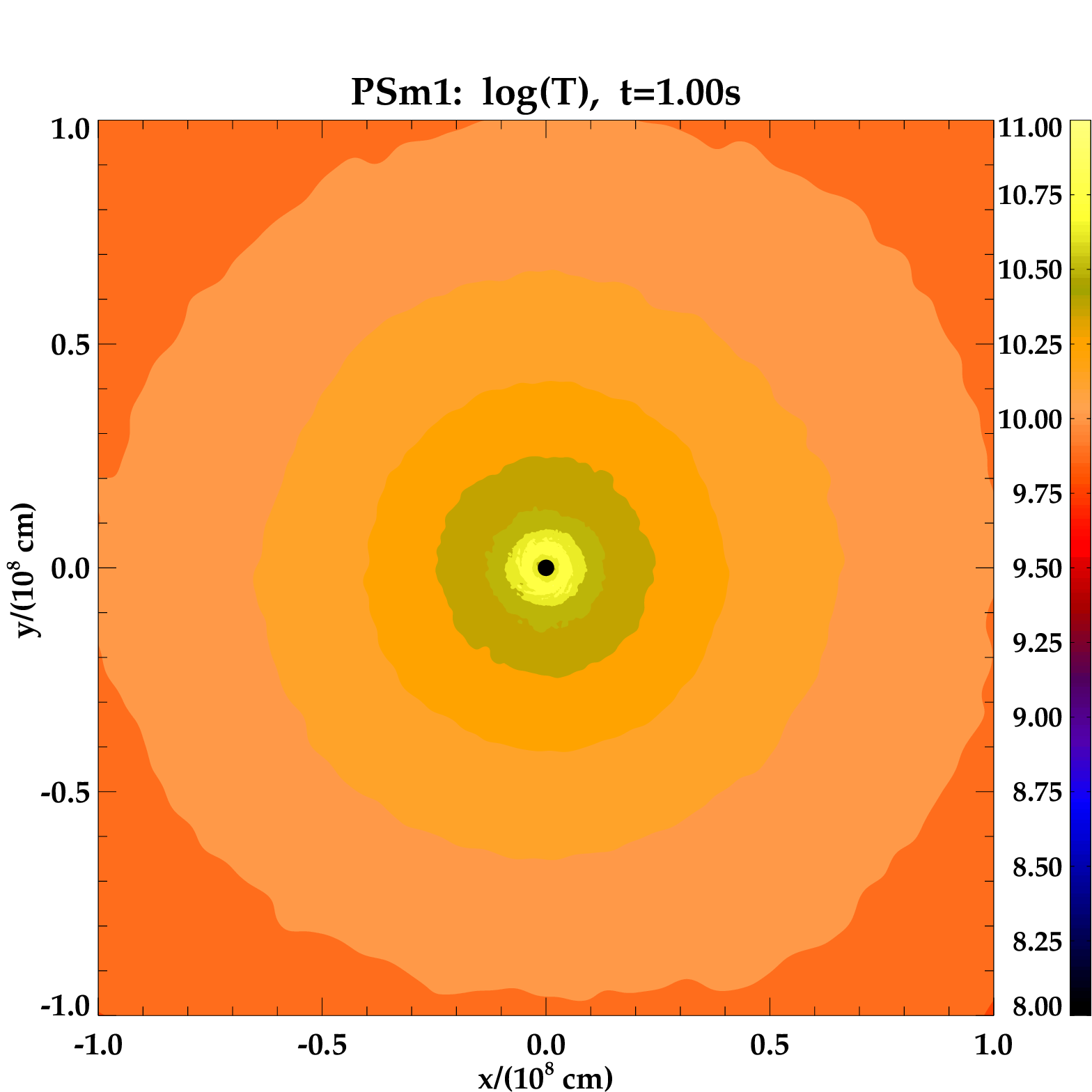}&
      \includegraphics[width=78mm]{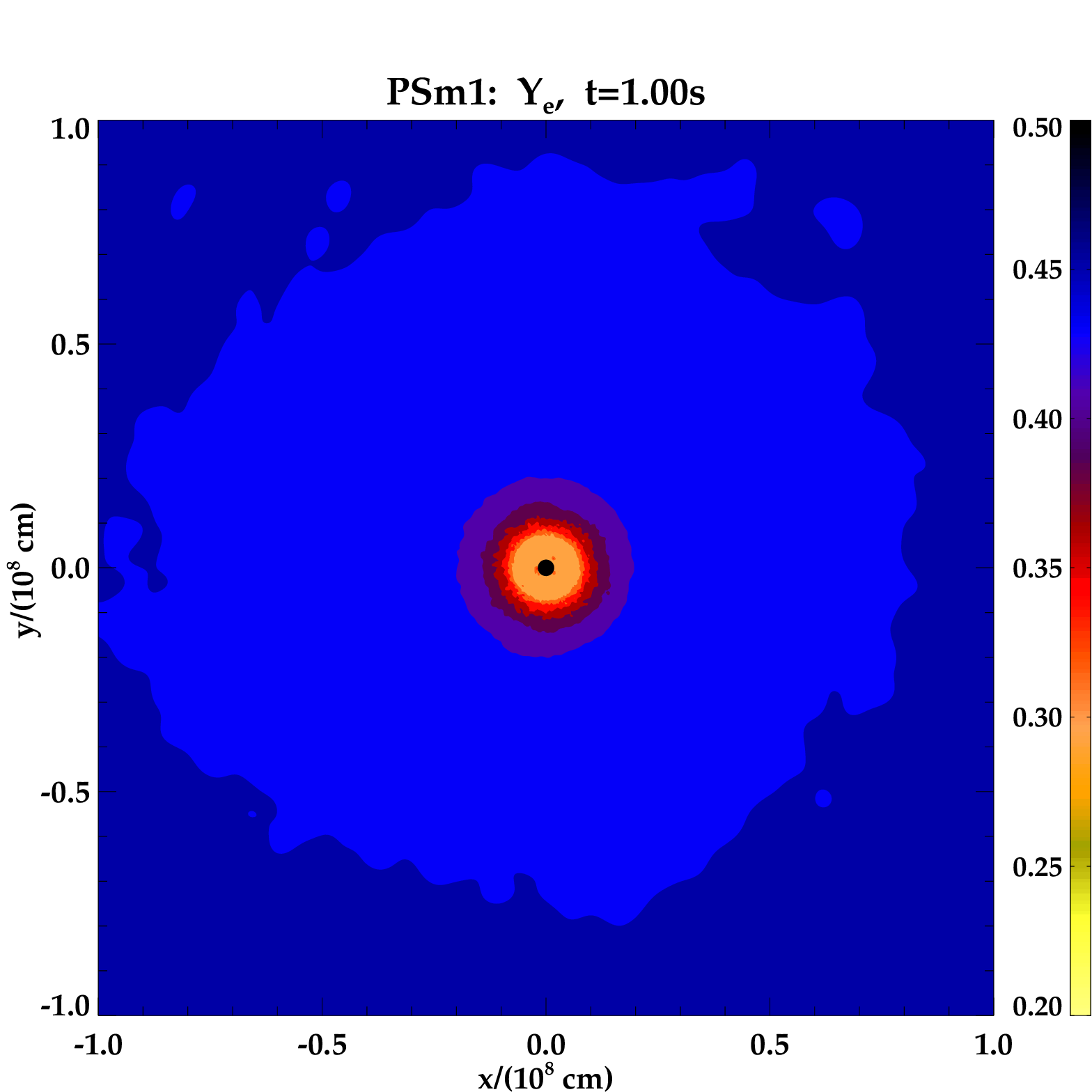}\\[-.4cm]
      \includegraphics[width=78mm]{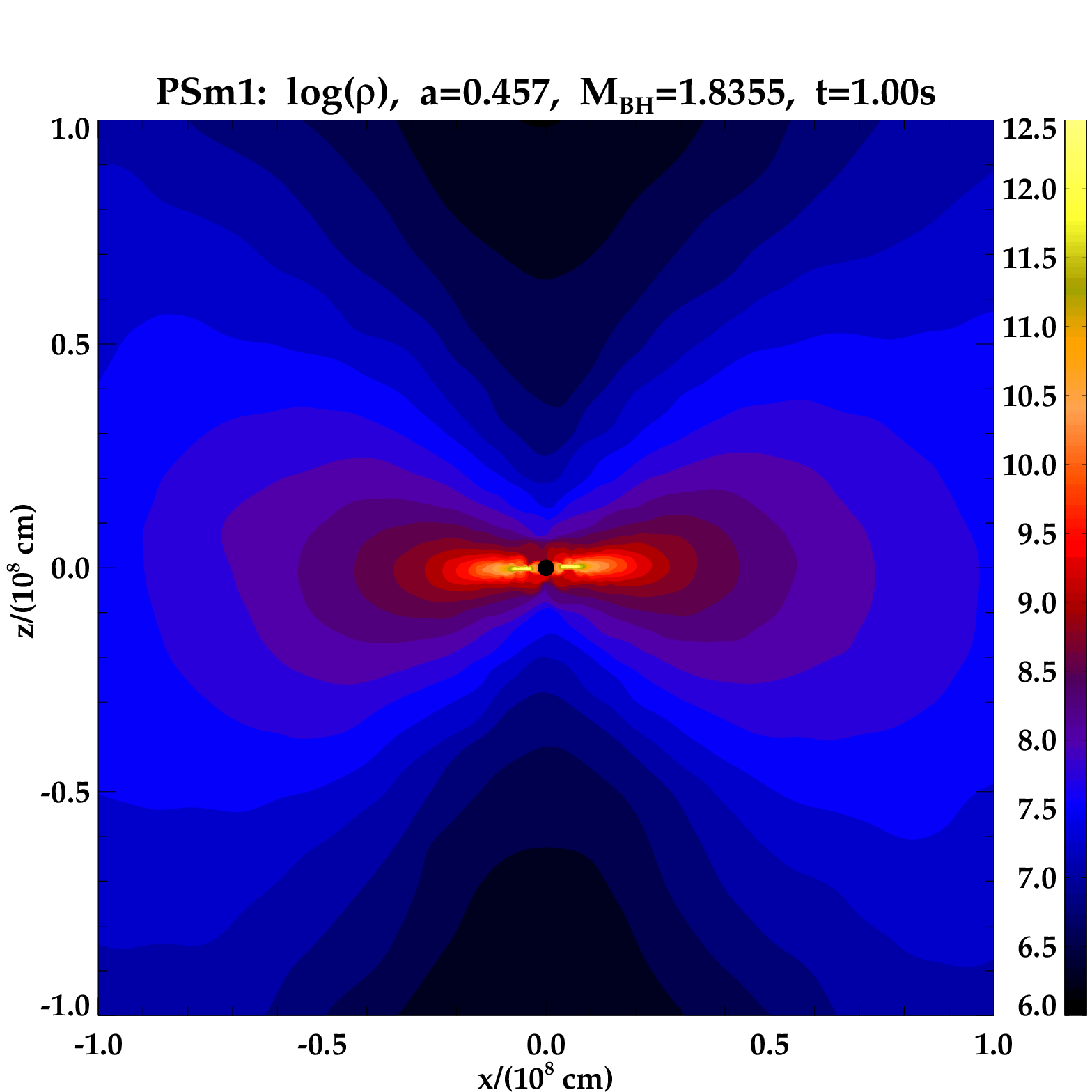}&
      \includegraphics[width=78mm]{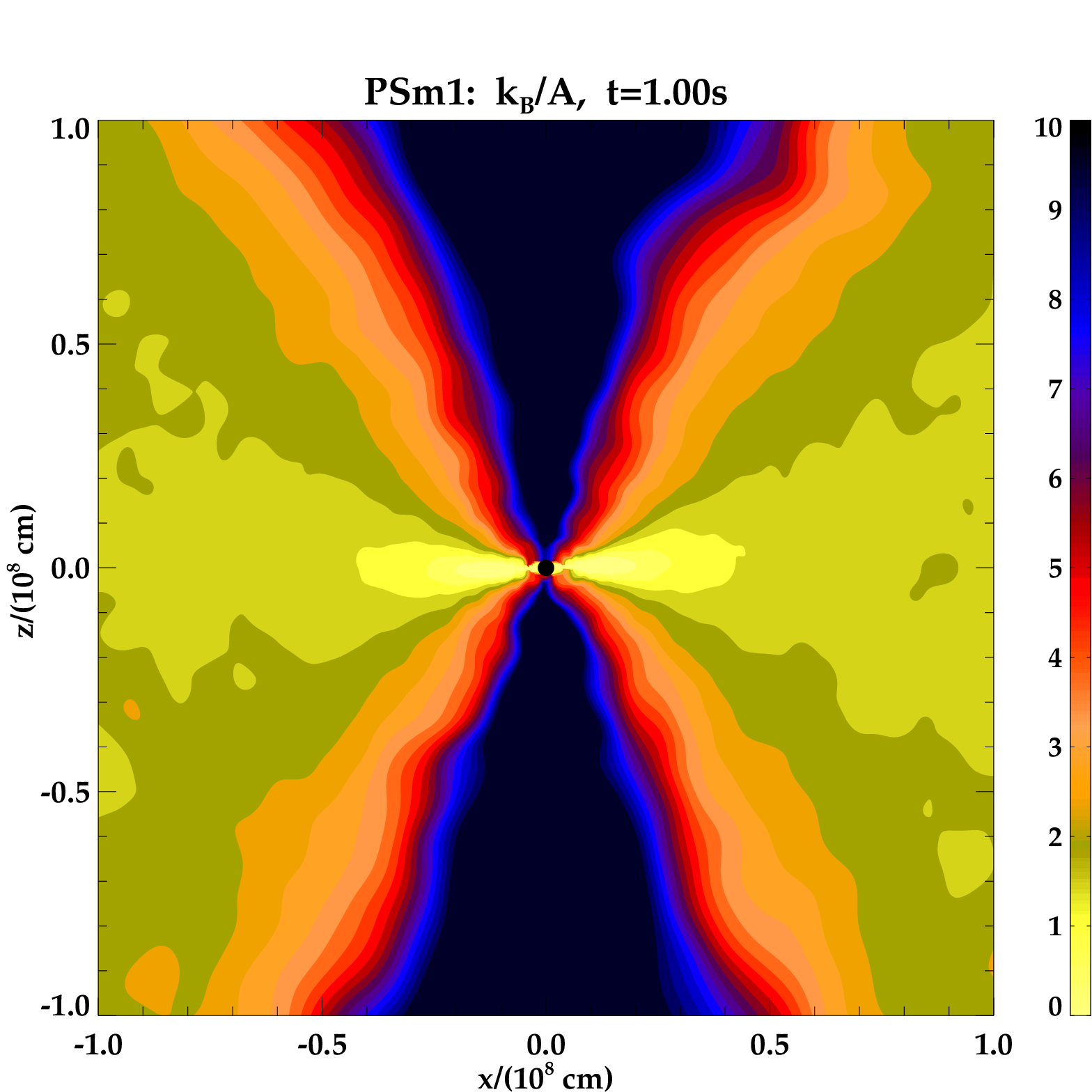}\\[-.4cm]
      \includegraphics[width=78mm]{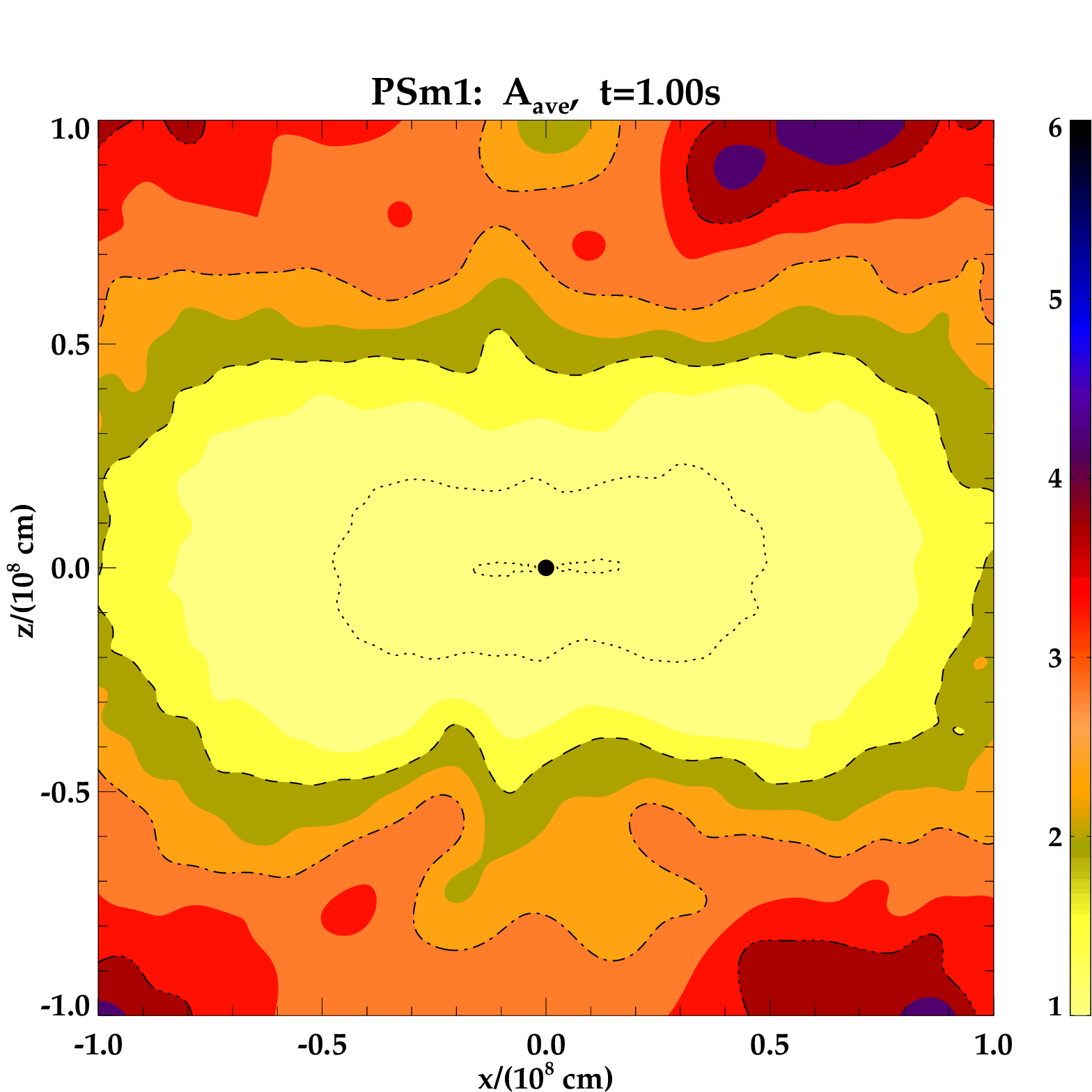}&
      \includegraphics[width=78mm]{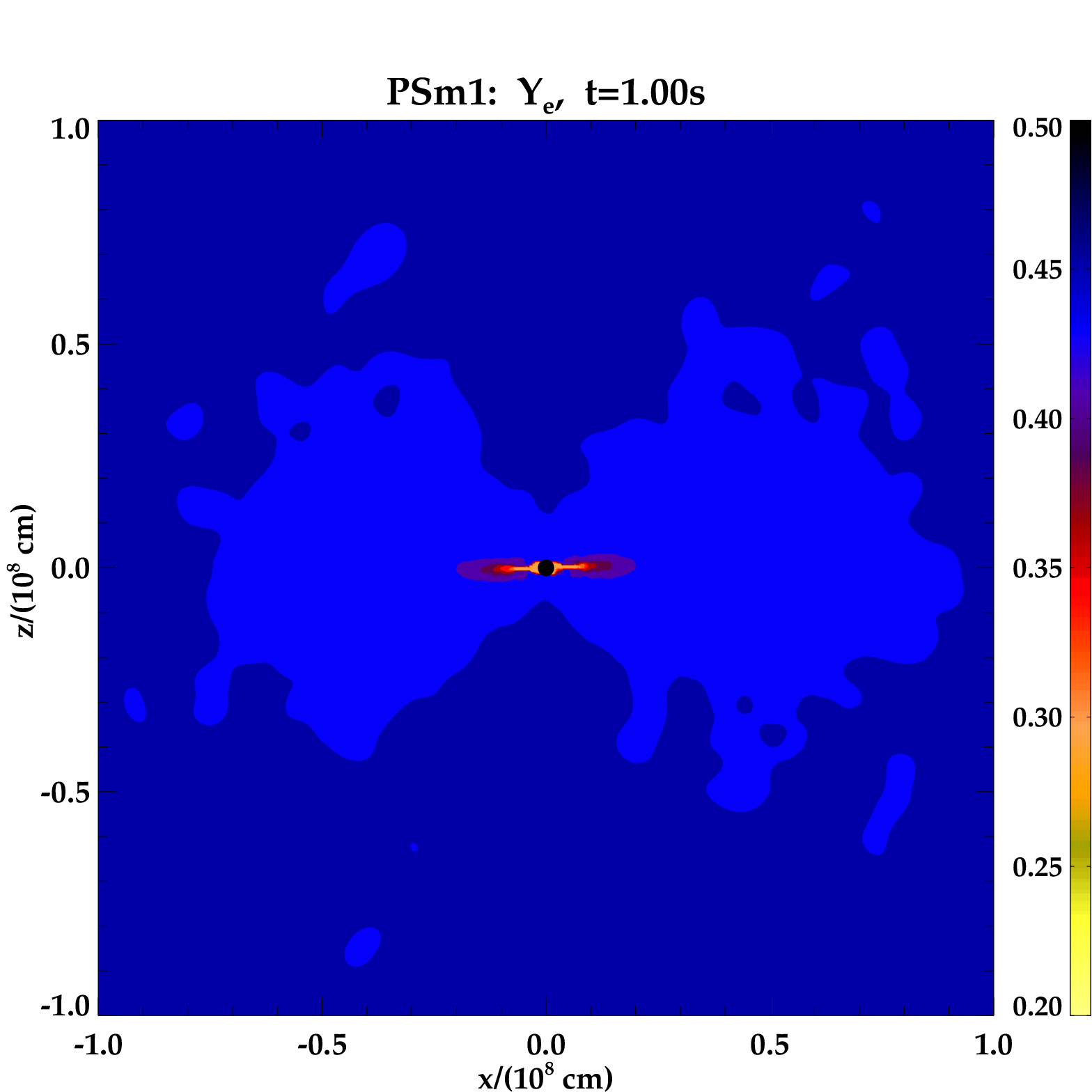}\\[-.2cm]
      \end{tabular}
      \caption{\label{fig:PSm1_eqplt} Model PSm1 properties at
        $t=1.00$~s: $T$ and $\ye$ in the equatorial plane (top); in
        the vertical plane, $\rho$ and $s$ (middle), and $\ye$ and
        $\abar$ (bottom), for which in the latter are included lines
        of $\abar=1$ (dotted), 2 (dashed), 3 (dot-dashed), 4
        (dot-dot-dashed).}
   \end{center}
\end{figure*}

The disk behaviour during evolution, as shown by the equatorial
density contours in Fig.~\ref{fig:PSm1_eqplden}, closely follows the
Toomre stability analysis.  Large non-axisymmetric perturbations which
appear at early times ($t\sim0.30$~s) have become globally
axisymmetric by $t=1.00$~s.  The small, $m=2$ mode in density which is
beginning to form at $R\lesssim10^7$~cm (at $t=1.00$~s), where
$\qt\approx 1$ predicted a locally unstable region, develops into a
higher order mode by $t=1.50$~s.  While outer regions become
temporarily disturbed, the disk remains globally stable and returns to
a quasi-steady state\footnote{The disk is not an isolated system; the
  collapsing star provides continued infall (at a slowly decreasing
  rate), and the properties of successive shells vary according to the
  pre-collapse progenitor structure.  Therefore, `quasi-steady' refers
  to an axisymmetric flow in which profiles of density, velocity and
  other hydrodynamic variables change slowly with time.} by
$t=2.00$~s.  Thus, the instability remains local and does not lead to
the formation of large spirals and to enduring, non-axisymmetric
modes. The disk is dense ($\rho>10^{8-12}$) and hot ($T>$ a few MeV),
with inner regions reaching electron fractions $\ye\lesssim 0.3$, as
shown in the top panel of Fig.~\ref{fig:PSm1_eqplt}.

\begin{figure} \vspace{-0.4cm}
  \begin{center} 
    \includegraphics[width=78mm]{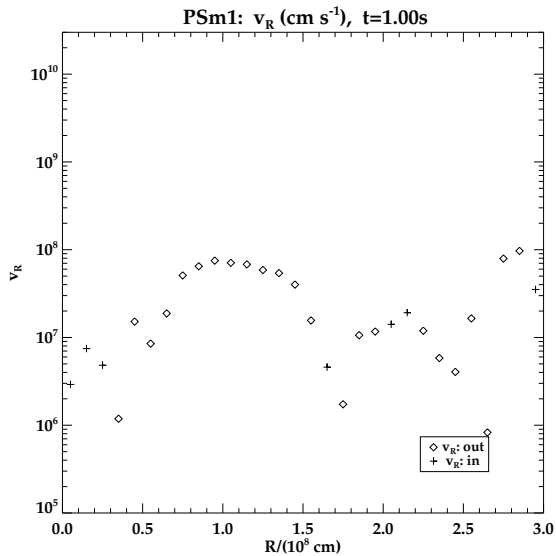}
    \caption{\label{fig:PSm1_jspan} Radial velocity profile of PSm1
      at $t=1.00$~s, for which `diamonds' and `+' show outflows and
      inflows, respectively. }
    \end{center}
\end{figure}

The vertical disk structure is shown in the lower panels of
Fig.~\ref{fig:PSm1_eqplt}.  The density and specific entropy show
the inner `thin disk' region extending out to $R_{\rm{T}}\lesssim
2\times 10^7$~cm, with an extended quasi-thin disk out to
$\approx2~R_{\rm{T}}$.  A large `corona' of hot ($T\sim1$~MeV) and
fairly dense material extending from the equatorial plane is in
vertical near-equilibrium (i.e., not in an outward-driven wind, but
instead, with relatively small velocities) and balances the continued
infall, with uniform $\ye$ and dominated by material with $\abar\le4$.
At irregular intervals, radial outflows are produced in the equatorial
plane (Fig.~\ref{fig:PSm1_jspan}, bottom), with velocities up to $\sim
1000~\rm{km~s}^{-1}$, which carry the low-$\abar$,-$\ye$ material
outwards.

\begin{figure}
  \begin{center}
    \begin{tabular}{c}
      \includegraphics[width=78mm]{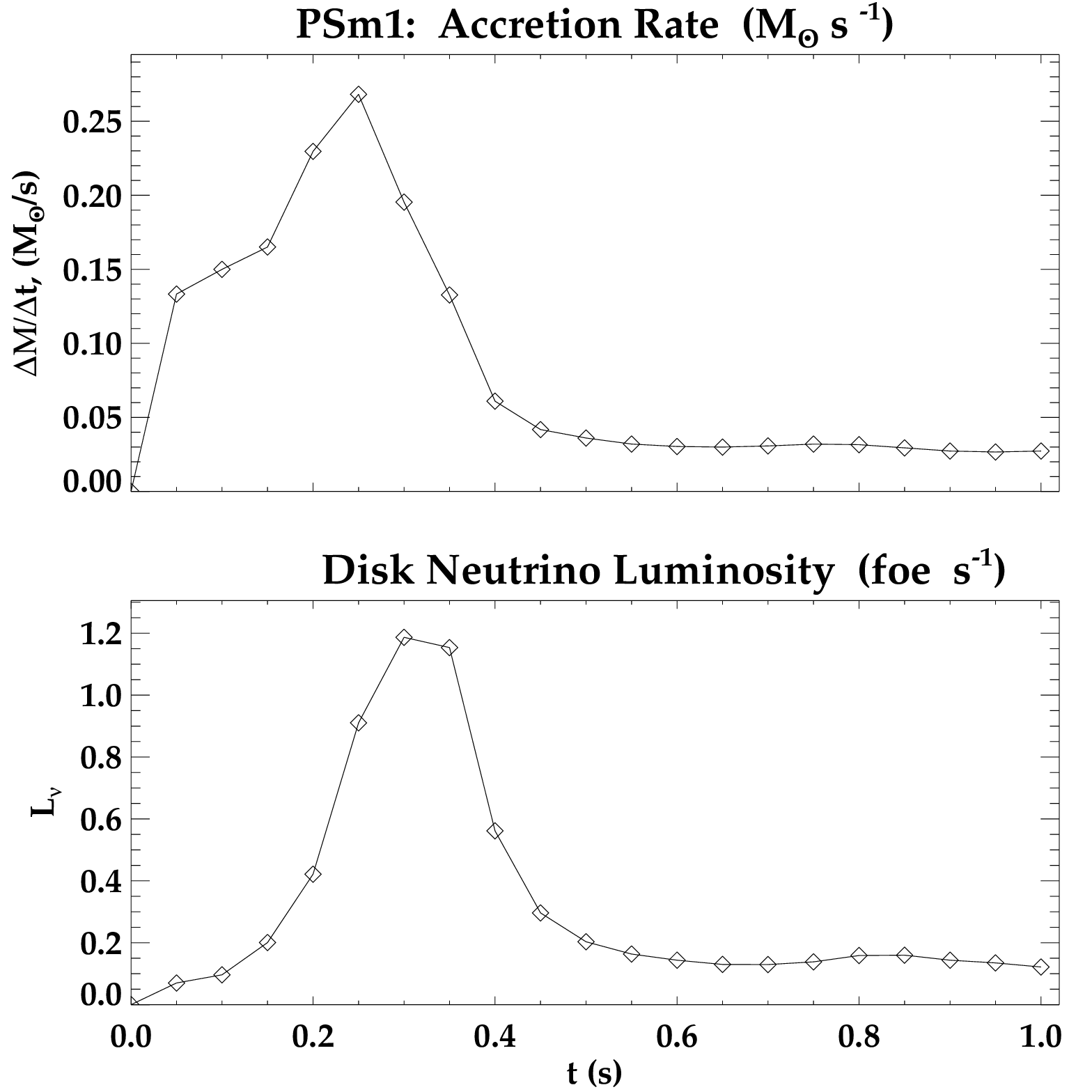}
    \end{tabular}
    \caption{\label{fig:PSm1_mdot} Model PSm1: evolution of mass
      accretion rates and neutrino luminosity.  After a brief period of
      rapid accretion of low-$j$ material, $\dot{M}$ decreases rapidly
      as the disk reaches a steady, axisymmetric state.  The neutrino
      luminosity behaves similarly; after an initial period of rapid
      cooling of dense, shock-heated material, $L_{\nu}$ decreases below
      1 foe~s$^{-1}$.}
  \end{center}
\end{figure}

The rate of accretion, $\dot{M}$, onto the central object is shown in
Fig.~\ref{fig:PSm1_mdot}, along with the disk neutrino luminosity,
$L_{\nu}$.  During early times, $\dot{M}$ is large (while low angular
momentum material accretes quasi-spherically), as is $L_{\nu}$ since
infalling material shocks around the PNS surface and begins to cool
rapidly, with a maximum just as $\dot{M}$ begins to decline.  By late
times, a stable disk has formed with steady (predominantly equatorial)
accretion, and neutrino production decreases.  By
Eq.~(\ref{edot_bz}), these values of $\dot{M}$ are at least a factor
of 3-4 too low to produce $\dot{E}_{\rm{B-Z}}\approx\dot{{\it
    E}}_{\rm{GRB}}$.

For neutrinos, even utilising a reasonably optimistic annihilation
efficiency, $\eta_{\nu\bar{\nu}} L_{\nu} \ll \dot{E}_{\rm{GRB}}$.
Therefore, this model will not produce a successful LGRB by either
mechanism, as accretion rates from the usual viscous processes by
which the thin $\alpha$-disk evolves are too low.  The collapsing
material possesses too much angular momentum and thermal support to be
globally unstable, even with large amounts of neutrinos being produced
from the inner regions.

\begin{figure}\vspace{-0.4cm}
  \begin{center}
    \begin{tabular}{c}
      \includegraphics[width=78mm]{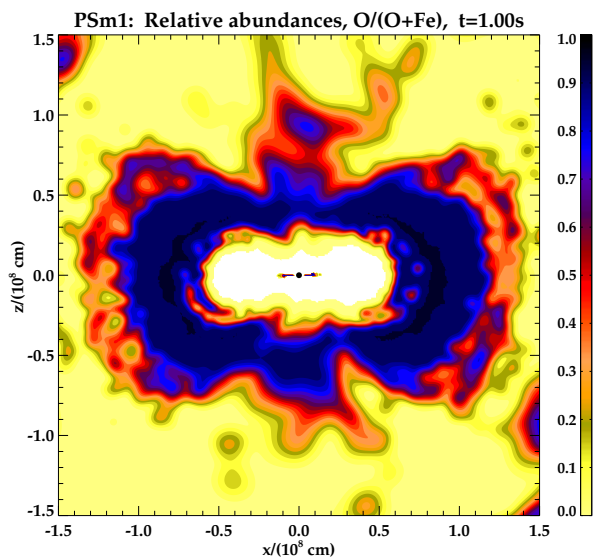}
    \end{tabular}
    \caption{\label{PSm1_smooth3_100} For model PSm1, contours compare
      the relative locations of O and Fe produced by the disk and
      inner region in NSE (vertical slice above the equatorial plane).
      In the central (empty) region, amounts of both elements are
      negligible.}
  \end{center}
\end{figure}

To analyse the collapsar from the point of view of producing a HN, we
have calculated the quantities of various elements produced in the
disk and inner regions in NSE, which are given for PSm1 (and for other
models) in Table~\ref{tab:nsenucl}.  As noted above, light elements
dominate the disk material, and most of the heavy metals are produced
in the corona and shocked infall regions.  For PSm1, the NSE mass of
$^{56}$Ni remains fairly constant at $0.02~\msun$, which is nearly
double the mass of $^{54}$Fe ($0.012~\msun$) and roughly equal to
total elemental Fe ($0.021~\msun$, though the iron content slowly
increases with time).  Therefore, NSE-nucleosynthesis in and around
the (steady) collapsar disk falls short of the amount of $^{56}$Ni
required to power the bright HNe by roughly an order of
magnitude\footnote{However, this simulation encompasses only a single
  accretion event. It is likely that continued infall would lead to
  disk reformation, and subsequent repetitions of accretion events may
  then iteratively increase the total $^{56}$Ni yield.}.

A comparison of the relative production of elemental Fe and O in the
fluid elements is shown in Fig.~\ref{PSm1_smooth3_100}.  As the disk
settles into a steady flow, there seems to be only a slight preference
for Fe being produced in the polar direction and O equatorially in the
coronae around the dense disk.  It should be noted that what is
plotted is only the relative, local abundances, and that the total
mass of O produced in this case is several orders of magnitude less
than the Fe.  However, this mass difference may be unsurprising given
that Fe, unlike O, is an NSE end product which dominates the
mass-fraction for a large area of relevant ($\rho, T$) phase space
(cf. Appendix C).

\begin{table*}
\begin{center}
  \begin{minipage}{155mm}
    \caption{\label{tab:nsenucl} Representative NSE-nucleosynthetic
      yields (in $\msun$) at late times for each model.  The largest
      fluctuations in elemental abundance ($>$ order of magnitude)
      occured in PSdsq2, for which the values are given at two
      times. }
    \begin{tabular}{lccccccccccc}
      \hline
      \mc{1}{c}{Model~[$t$ (s)]}& \mc{1}{c}{$^{56}$Ni}& \mc{1}{c}{Co}   & \mc{1}{c}{Fe} & \mc{1}{c}{$^{54}$Fe}& \mc{1}{c}{Cr} & \mc{1}{c}{Ca}& \mc{1}{c}{Si}& \mc{1}{c}{O} & \mc{1}{c}{$^4$He} & \mc{1}{c}{$n$} & \mc{1}{c}{$p$}\\
\hline
PSm2~[1.0]&$10^{-4.0}$ &$10^{-2.7}$ & $10^{-2.0}$  &$10^{-2.5}$   &$10^{-2.5}$   &$10^{-4.4}$  & $10^{-4.0}$  & $10^{-5.9}$ & 0.17 &0.03 &0.02 \\
PSm1~[1.0]&0.02  & 0.01 & 0.03 & 0.12 & $10^{-2.2}$  & $10^{-3.5}$  &$10^{-3.5}$  &$10^{-5.4}$  & 0.38 & 0.19 & 0.14	\\
PSdsq2~[1.0]& $10^{-2.2}$   & $10^{-2.2}$ & 0.02 & 0.01 & $10^{-2.7}$ & $10^{-3.7}$ & $10^{-3.7}$ & $10^{-5.5}$ & 0.25 & 0.52 & 0.35\\
PSdsq2~[1.25] & $10^{-8.3}$ & $10^{-5.4}$ & $10^{-4.8}$ & $10^{-6.0}$ & $10^{-5.0}$ &    $10^{-6.2}$ & $10^{-6.4}$ &  $10^{-6.6}$  & 0.09 & 0.15 & 0.10\\
PSd2~[0.42]& 0.04 & 0.03 & 0.07 & 0.05 & $10^{-2.2}$ & $10^{-3.0}$ & $10^{-3.1}$ &$10^{-6.0}$ & 0.14 & 0.43 & 0.25 \\
PCm1~[1.0] &$10^{-2.3}$ & 0.01  & 0.04 & 0.02 & $10^{-2.1}$ & $10^{-3.7}$ & $10^{-3.4}$ & $10^{-5.7}$ & 0.54 & 0.35 & 0.31 \\
PCdsq2~[0.58]& 0.01 & 0.02 & 0.05 & 0.03 & $10^{-2.2}$  & $10^{-3.4}$ & $10^{-3.3}$ &  $10^{-5.7}$ & 0.27  & 0.50 & 0.35 \\
PCd2~[0.44]	&0.03 &0.03 &0.06 & 0.04  &$10^{-2.2}$  &$10^{-3.1}$ &$10^{-3.2}$ & $10^{-5.6}$ & 0.19 &0.45  &0.26 \\
PSm1Kd5~[1.0]	&$10^{-3.0}$  &$10^{-2.4}$  &0.02  &$10^{-2.2}$  &$10^{-2.4}$ &$10^{-4.1}$ &$10^{-3.7}$ &$10^{-5.6}$ &0.40  &0.46 &0.36 \\
PSd2Kd5~[0.40]&0.10 & 0.02  &0.03 &0.02  &$10^{-2.5}$  &$10^{-3.2}$ &$10^{-3.5}$ &$10^{-6.0}$& 0.10 & 0.48 &0.24 \\
NSm1~[1.0] &$10^{-4.2}$ & $10^{-2.7}$ & $10^{-2.1}$ &$10^{-2.7}$  &$10^{-2.5}$  &$10^{-4.7}$ &$10^{-4.2}$ &$10^{-6.1}$  &0.18  &0.01 & 0.07 \\
NSd2~[0.35] &0.03  & 0.03 &0.07 &0.05 &$10^{-2.1}$ & $10^{-3.0}$ &$10^{-3.0}$ &$10^{-5.6}$  &0.16 &0.42 &0.23  \\
PSdsq2J~[1.25] &$10^{-5.1}$ &$10^{-7.1}$ & $10^{-3.3}$& $10^{-4.0}$& $10^{-3.7}$ &$10^{-5.9}$  &$10^{-5.5}$&$10^{-6.5}$ &0.17  &0.16  &0.10  \\
\hline
\end{tabular}
\end{minipage}
\end{center}
\end{table*}

We now investigate the three regions of potential non-NSE
nucleosynthesis as mentioned in \S\ref{sec:nonnse}.  For region (i),
some equatorial outflow material reached velocities of order
$10^3~\rm{km~s}^{-1}$ at $R\sim10^8$~cm, where $\ye\approx0.43-0.45$
(Fig.~\ref{fig:PSm1_eqplt}).  The entropy through much of the disk is
quite low, but in the outermost regions and corona reaches
$\sim10~k_{\rm{B}}/A$.  However, efficient $^{56}$Ni-production would
probably require significantly greater ejection velocities as well as
higher entropy.

There is not a great deal of outflow vertically from the disk for
region (ii).  The coronal material is optically thin to neutrinos and
possesses a quasi-steady vertical structure; the velocities (not
shown) of fluid elements are relatively small ($\ll0.1~c$) and
non-uniform, except in the azimuthal direction (for near-Keplerian
rotation).  Therefore, these coronal regions do not appear to be a
possible site for $r$-process nucleosynthesis, and these
characteristics are similar throughout the disks analysed here.

Finally, for region (iii) near the inner edge of the disk, some
material has $\ye\approx0.3$, which is beneficial for $r$-processing,
though the specific entropy in these regions is quite low ($s/A\sim1$)
due to efficient neutrino cooling.  However, as seen in
Fig.~\ref{fig:PSm1_eqplt}, high-$s$ regions surround the inner
disk edge, where material is still less neutron rich
$\ye\approx0.43-0.45$.  Furthermore, unlike in region (ii) where
vertical motion of material is inhibited by continued infall, after
the initial core-collapse the cone around the rotation axis has a
relatively low density.  One might expect vertical trajectories in
this region either from material shock-heated to form a high entropy
`bubble' which rises, or from material swept up in the collimated
outflow which creates the actual LGRB (discussed briefly in
\S\ref{sec:psdsq2j}).

\subsection{PSdsq2}
\label{sec:psdsq2}
\begin{figure}
  \begin{center}
    \includegraphics[width=74mm]{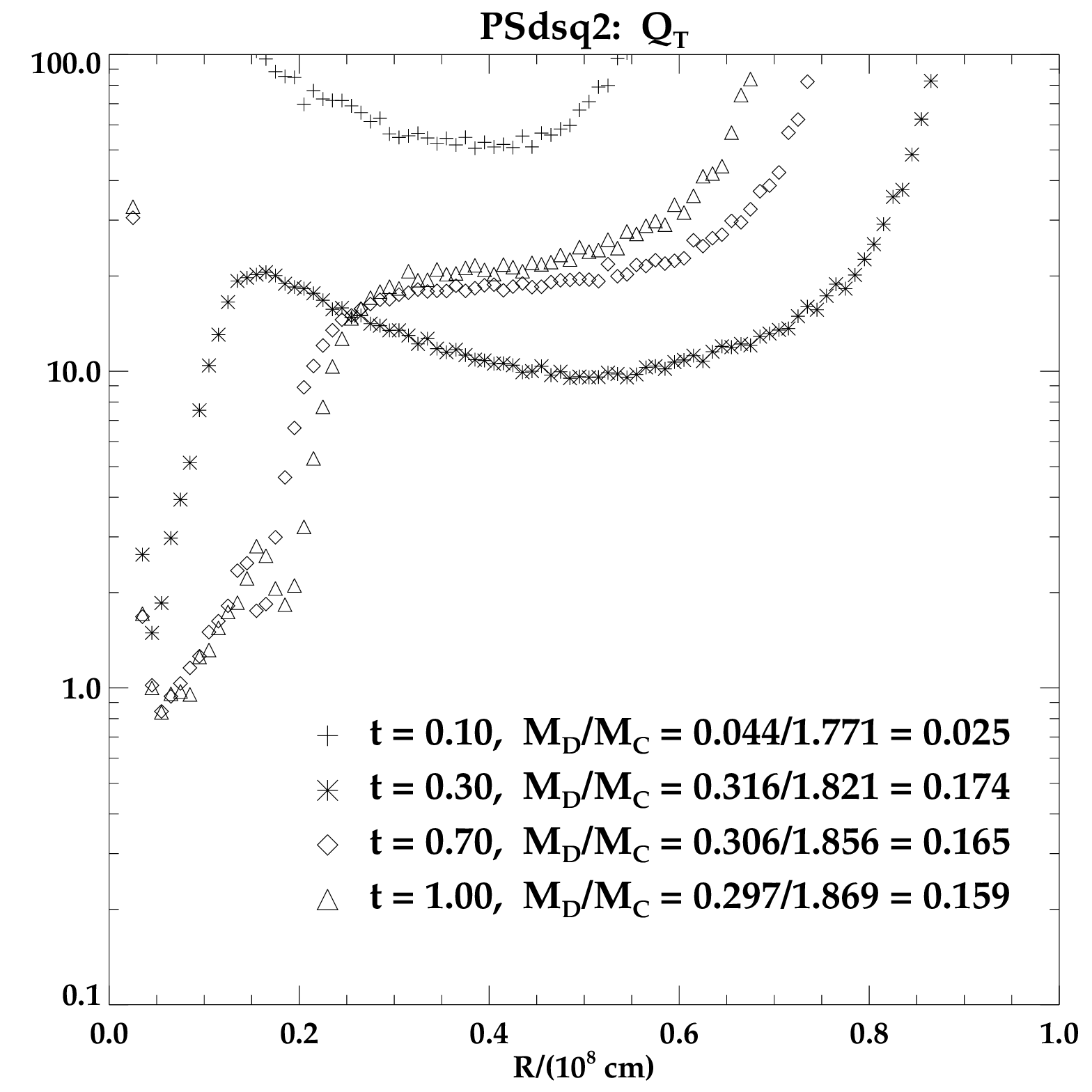}
    \caption{\label{fig:PSdsq2_qt} Model PSdsq2: evolution of the
      Toomre parameter, $\qt$.  By $t=0.70$~s, a significant
      fraction of the disk is unstable ($\qt<1$), and spiral
      structure develops dynamically.}
    \end{center}
\end{figure}

\begin{figure*}\vspace{-0.4cm}
  \begin{center}
    \begin{tabular}{cc}
      \includegraphics[width=78mm]{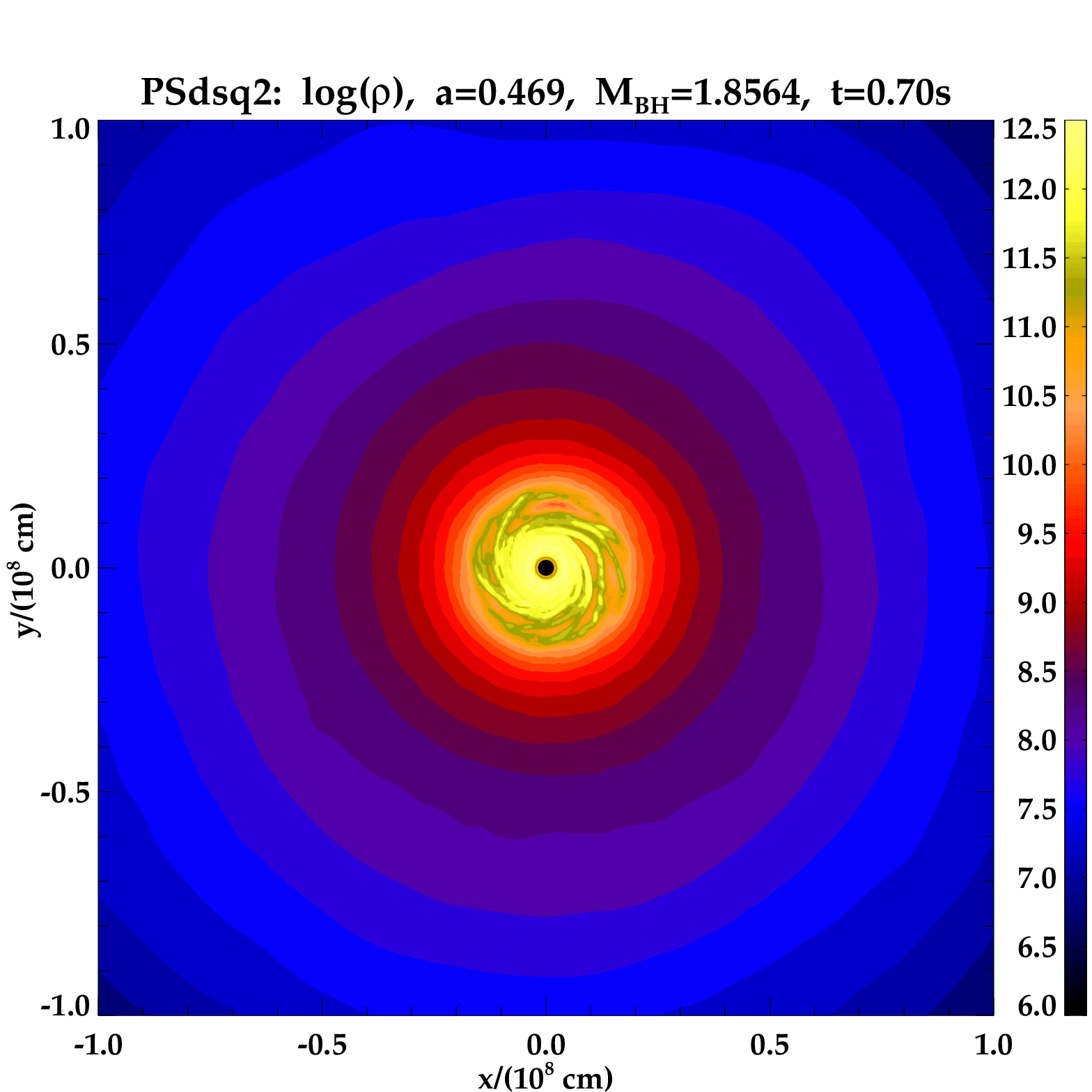}&
      \includegraphics[width=78mm]{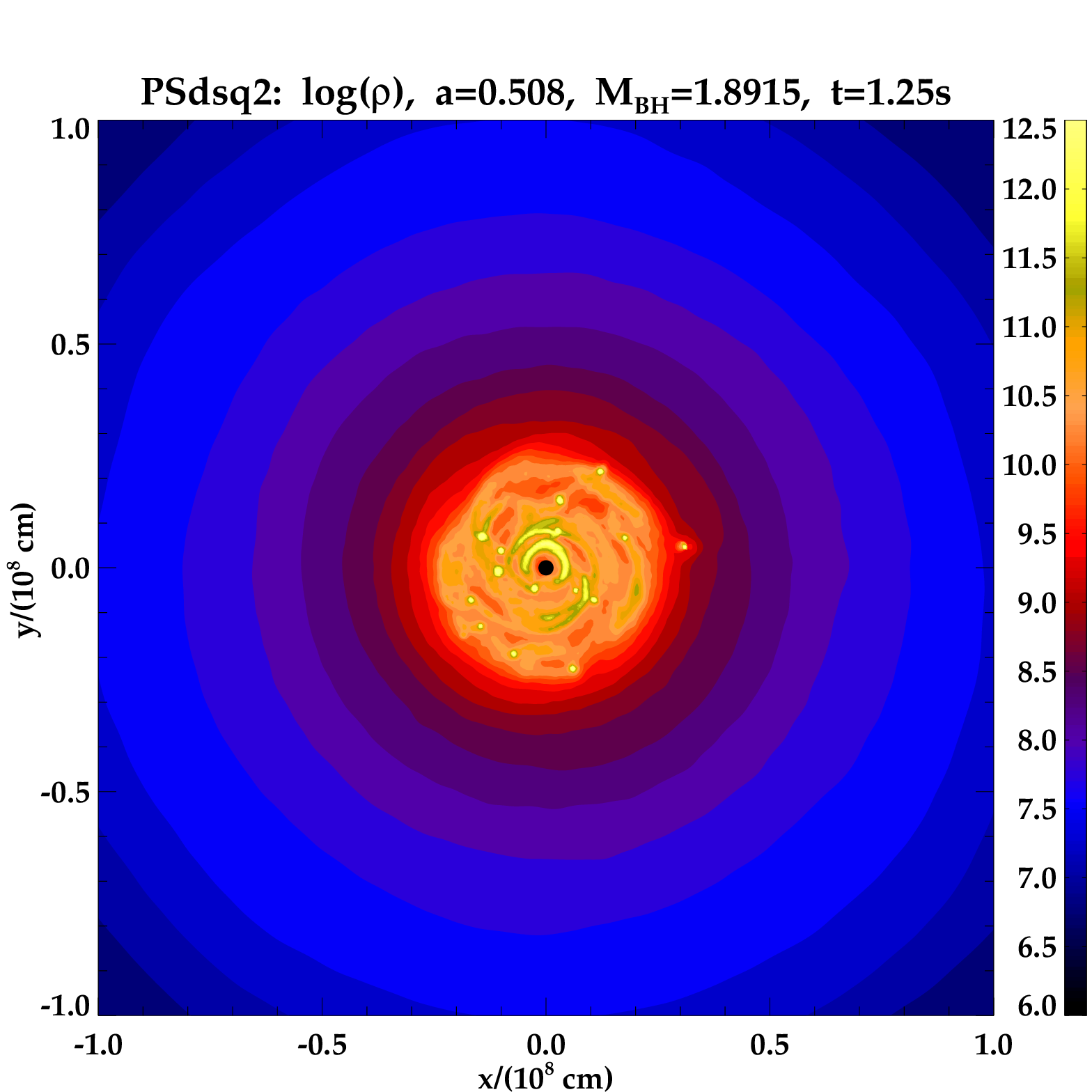}\\[-.4cm]
      \includegraphics[width=78mm]{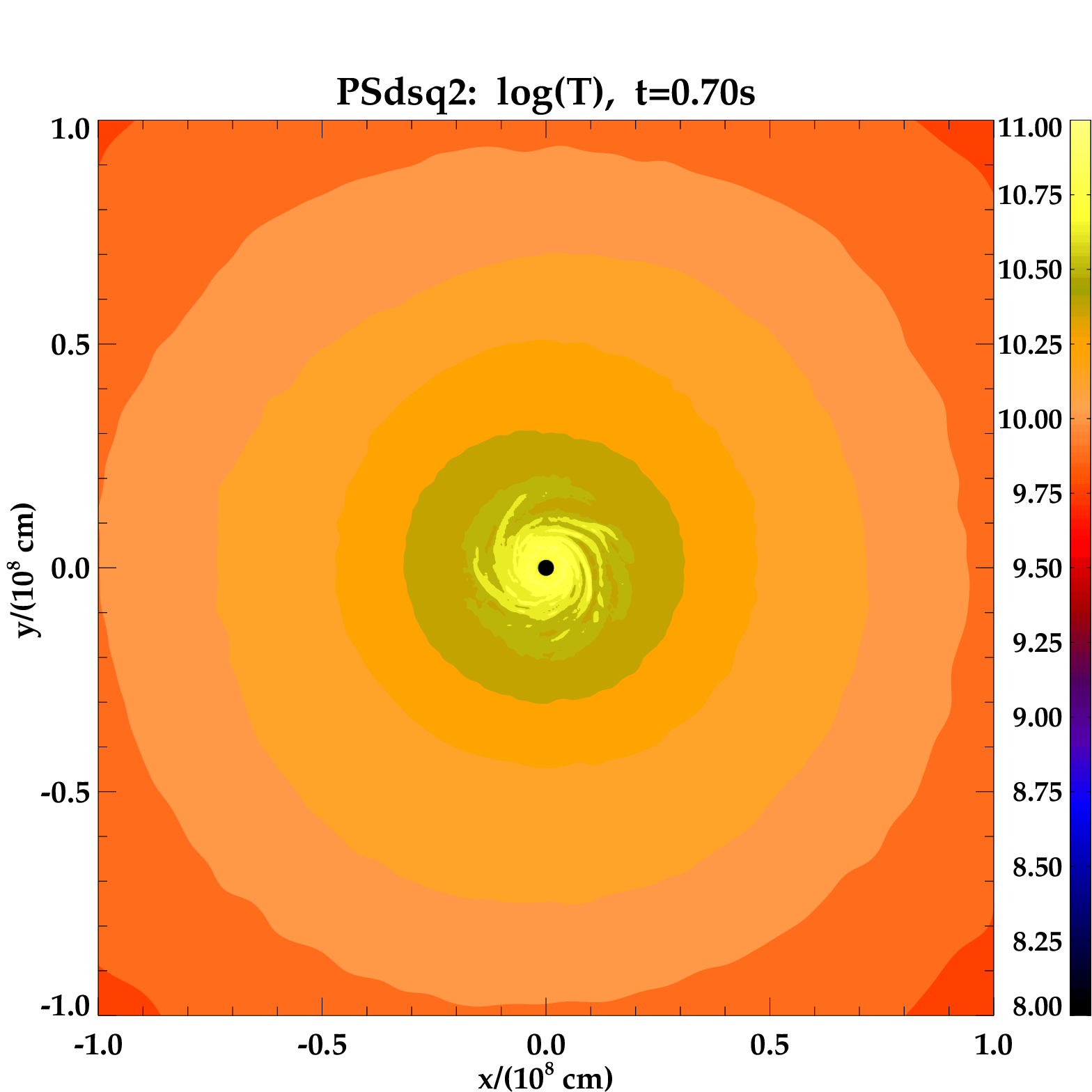}&
      \includegraphics[width=78mm]{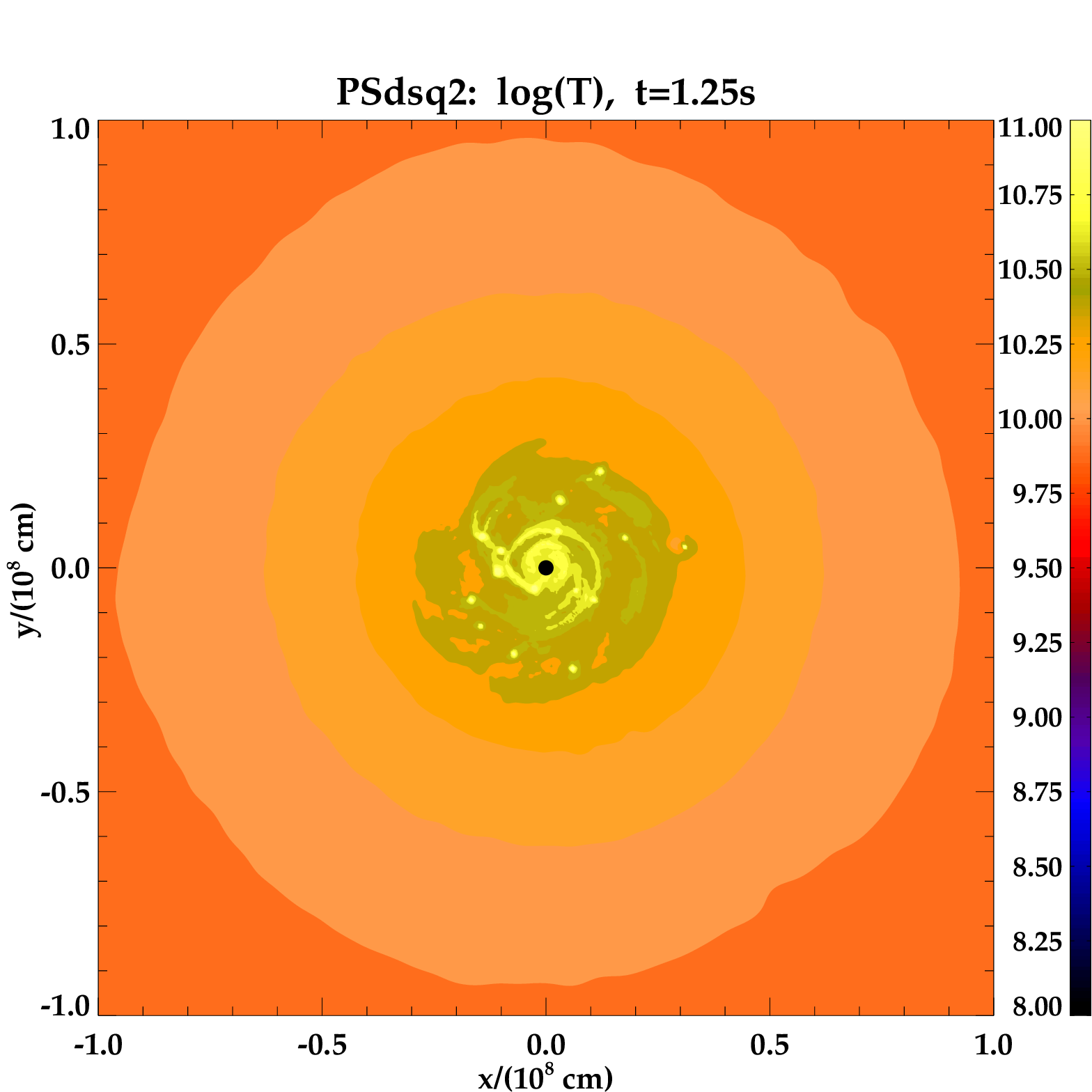}\\[-.4cm]
      \includegraphics[width=78mm]{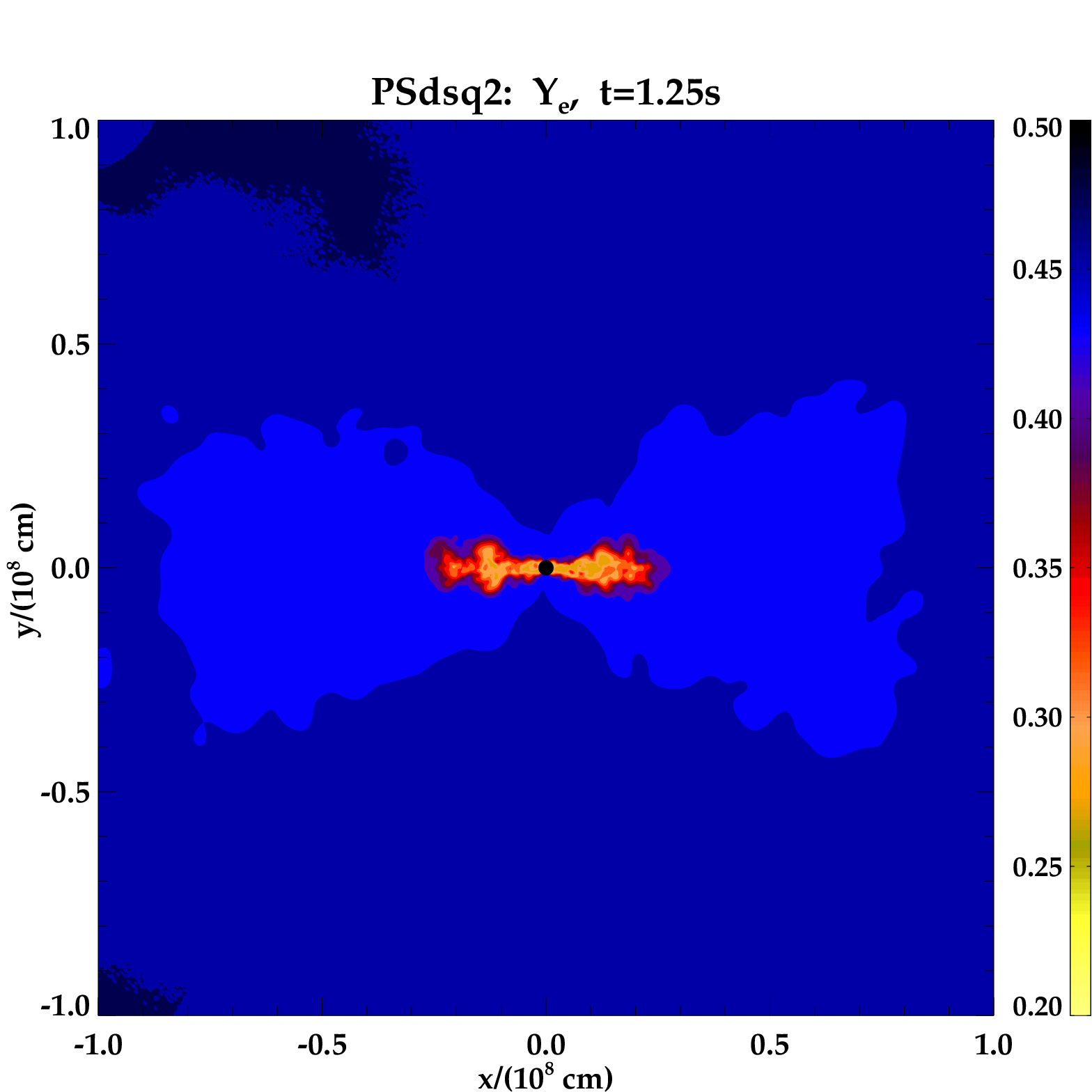}&
      \includegraphics[width=78mm]{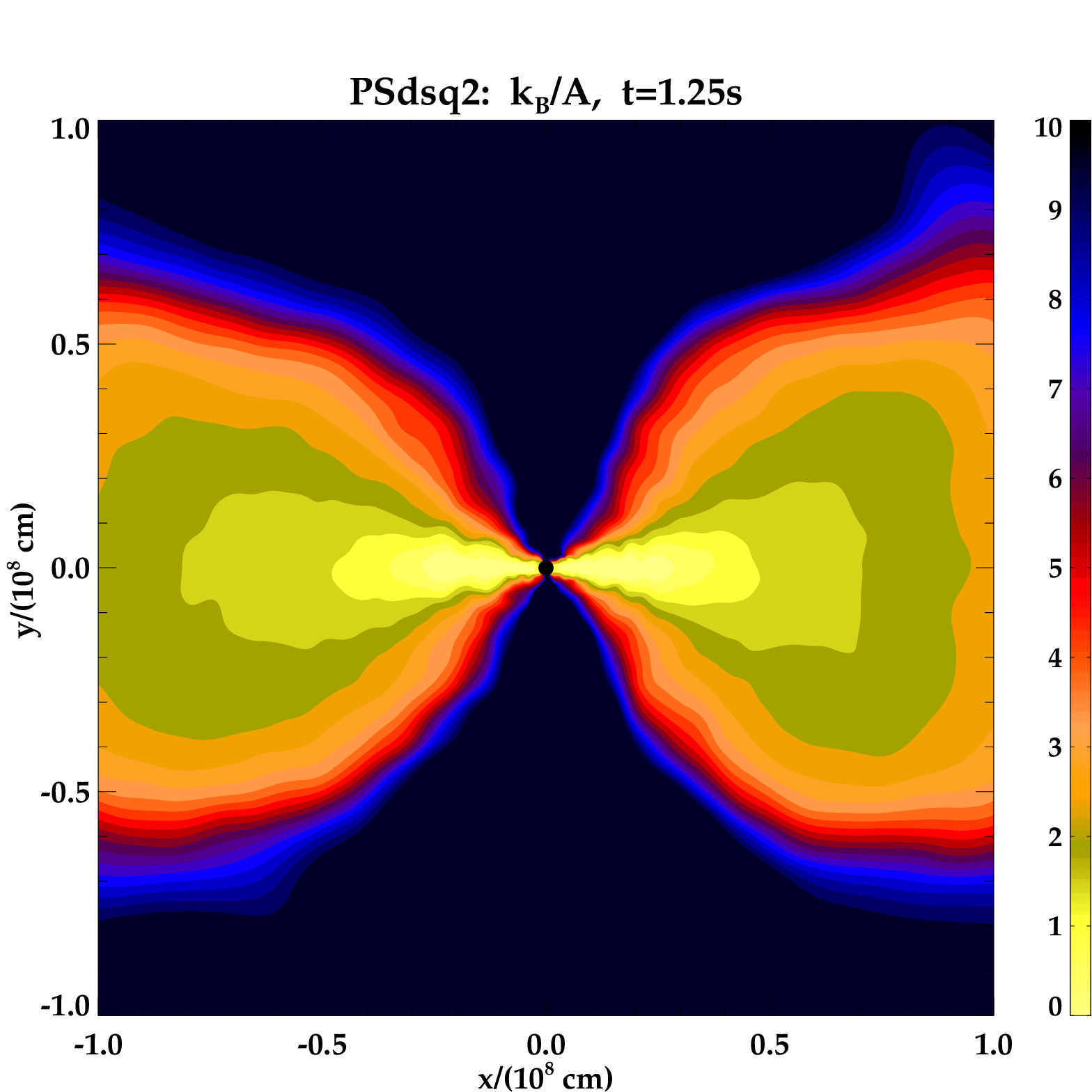}\\[-.2cm]
      \end{tabular}
      \caption{\label{fig:PSdsq2_rhot} Model PSdsq2 in the equatorial
        plane, non-axisymmetric evolution of $\rho$ and $T$ at
        $t=0.7$~s (top) and $t=1.25$~s (middle); spiral structure
        develops and leads to shocks and angular momentum
        transfer. (bottom) Electron fraction and specific entropy in
        the vertical plane at $t=1.25$~s.}
   \end{center}
\end{figure*}

We now investigate a model ($v_{\phi}$ globally decreased by a factor
of $\sqrt{2}$) with one half of the $E_{\rm{kin},\phi}$ of the
marginally unstable PSm1.  The evolution of $\qt$ is shown in
Fig.~\ref{fig:PSdsq2_qt}, as well as the disk/central object mass
ratios, which are significantly larger than in the previous model.  A
large part of the inner disk has $\rm{Q_T}<1$ and is therefore
predicted to be unstable to density perturbations.  Indeed,
non-axisymmetric structures form by $t=0.50$~s and
Fig.~\ref{fig:PSdsq2_rhot} (top) shows spirals in the density
evolution as well as localised density clumps.  The same figure shows
the effect of shocks in transferring kinetic energy into thermal,
easily seen in the $T$-profiles.  The (online) animations
of density and specific entropy illustrate the full dynamics of the
global behaviour of the disk, in particular the creation of
oscillating, $m=0$ flows in the equatorial plane.

\begin{figure}
  \begin{center}
    \includegraphics[width=78mm]{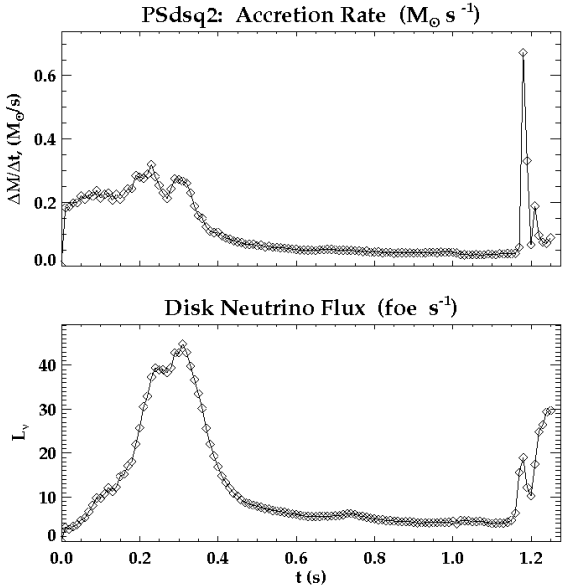}
    \caption{\label{fig:PSdsq2_mdot} Model PSdsq2: mass accretion rates
      and neutrino luminosity.}
    \end{center}
\end{figure}

The mass accretion and neutrino production rates are shown in
Fig.~\ref{fig:PSdsq2_mdot}, with the same initial phase of low-$j$
accretion. Even after the initial onset of spirals at $t>0.50$~s,
$\dot{M}$ remains low until $t>1.10$~s, at which time spikes of rapid
accretion develop\footnote{Due to the combined effects of the high
  accretion rates, densities and rapidly changing dynamics, the
  timesteps of the SPH particles decreased rapidly in these late
  times, causing the simulation to halt.  These numerical
  considerations are discussed further in \S\ref{sec:discusssim}.}.
In this case accretion rates with $\dot{M}>0.6~\mps$ give rise to
$\dot{E}_{\rm{B-Z}}\approx 5~\foeps> \dot{\it{E}}_{\rm{GRB}}$.
$L_{\nu}$ increases by an order of magnitude during the rapid
accretion to $\approx 30~\foeps$. Therefore, while the system
provides conditions for a central engine of sufficient power to create
a GRB jet via the B-Z mechanism, the process of neutrino annihilation
is probably too inefficient to produce a `successful' jet, as
$\enu\approx 0.3~\textrm{foe~s}^{-1}<\dot{E}_{\rm{GRB}}$.

The size of the $T>1$~MeV inner region of this collapsar is larger
than that of PSm1, and a much smaller mass of Fe-group elements is
present, particularly by $t=1.25$ (Table~\ref{tab:nsenucl}).  As a
consequence, only a very small amount of $^{56}$Ni is created, and
there is no preferential production of Fe in the polar direction at
late times.  The region of $\abar<4$ material is less flattened along
the equatorial plane, and likewise the distribution of O vs Fe is less
polarised (particularly due to the extremely low iron content at late
times).

\begin{figure} \vspace{-0.4cm}
  \begin{center}
    \includegraphics[width=78mm]{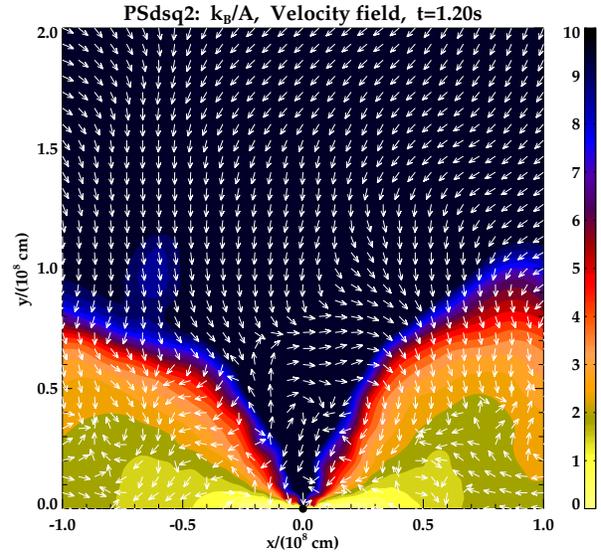}
    \caption{\label{fig:PSdsq2_vel} Specific entropy contours with
      normalised velocity vector field showing direction of flow in
      vertical slice half-plane (above equatorial plane) during
      accretion phase. The flow directly above the poles shows small
      turbulent eddies.}
    \end{center}
\end{figure}

The coronal structure (Fig.~\ref{fig:PSdsq2_rhot}, bottom panel) is
smaller than that in PSm1.  It remains fairly steady, even during
periods of outflow, and the non-aligned nature of the (non-azimuthal)
components of velocity vectors is shown in Fig.~\ref{fig:PSdsq2_vel},
ruling out case (ii) outflows for nucleosynthesis.  Peak equatorial
outflows reach a factor of a few higher than for PSm1, but are still
probably too low for case (i).  The low-entropy region has a smaller
vertical height from the equatorial plane, and material with higher
$s/A$ and moderate $\ye$ lies quite close to the central object in
this model (Fig.~\ref{fig:PSdsq2_rhot}).  Minimum values of the
electron fraction in the disk have decreased due to high $T$ and
$\rho$ of the shocked regions to roughly $\ye=0.25$, but the specific
entropy remains low.  Again, case (iii) outflows in the presence of a
jet flow remain the most likely location for significant
$^{56}$Ni-production.

\subsection{PSd2}
\begin{figure}
  \begin{center}
    \includegraphics[width=74mm]{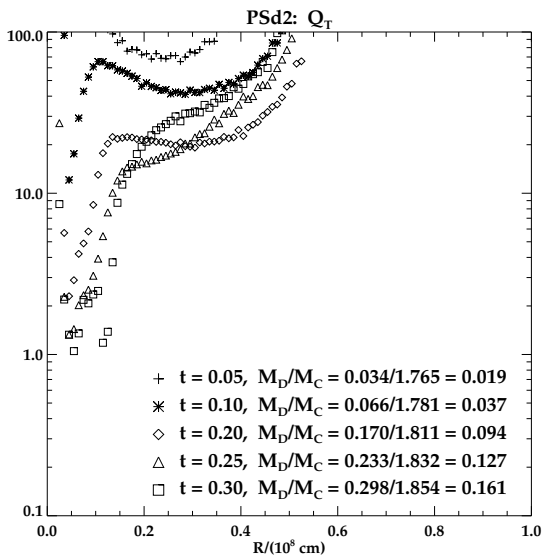}
    \caption{\label{fig:PSd2_qt} By $t=0.25$~s, part of the disk has
      reached an unstable state ($\qt\approx1$), and a larger
      extent of the disk has become unstable by $t=0.30$~s.}
    \end{center}
\end{figure}

\begin{figure*}\vspace{-0.4cm}
  \begin{center}
    \begin{tabular}{cc}
      \includegraphics[width=78mm]{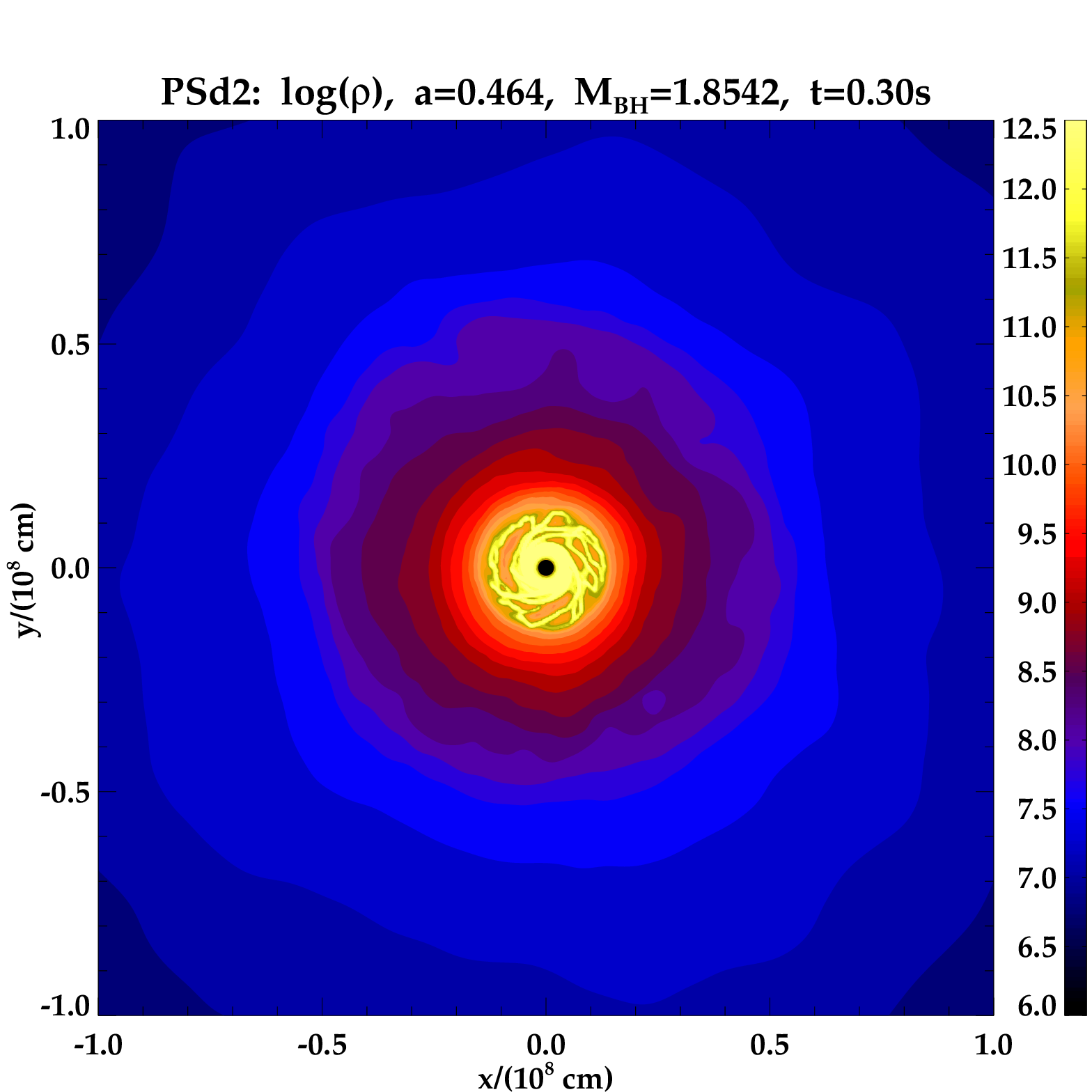}&
      \includegraphics[width=78mm]{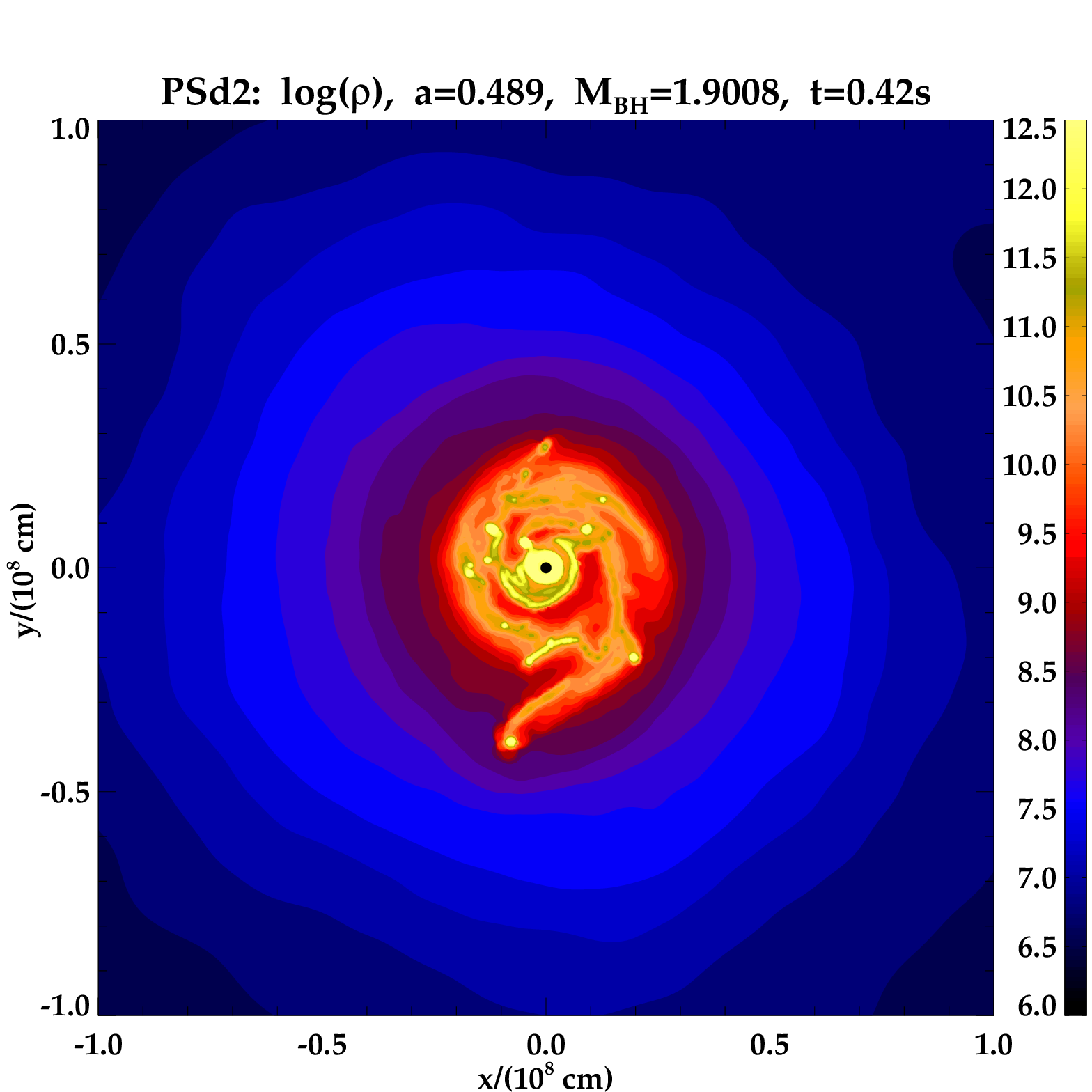}\\[-.2cm]
      \end{tabular}
      \caption{\label{fig:PSd2_rho} Model PSd2 in the equatorial
        plane, non-axisymmetric evolution of $\rho$ at
        $t=0.30, 0.42$~s. Small structures develop quickly into
        global modes and angular momentum transfer.}
   \end{center}
\end{figure*}

The model PSd2 ($v_{\phi}$ decreased by a factor of 2) possesses 25\%
of the $E_{\rm{kin},\phi}$ of PSm1 and continues the trend of showing
greater instability at earlier times after collapse
($t\approx0.20-0.25$~s).  Fig.~\ref{fig:PSd2_qt} shows the evolution
of $\qt$ and the increasing disk/central object mass ratio.
Fig.~\ref{fig:PSd2_rho} shows the perturbations evolving as a complex
configuration: tightly wound spirals develop into filaments within a
global $m=2$ mode (full evolution in the online animations).  The
inner region is comprised of alternating dense `fingers' and shocked
flow with higher entropy.  Though intially a smaller disk forms than
in previous models, the non-axisymmetric structure extends to larger
radii.

\begin{figure}
  \begin{center}
    \includegraphics[width=78mm]{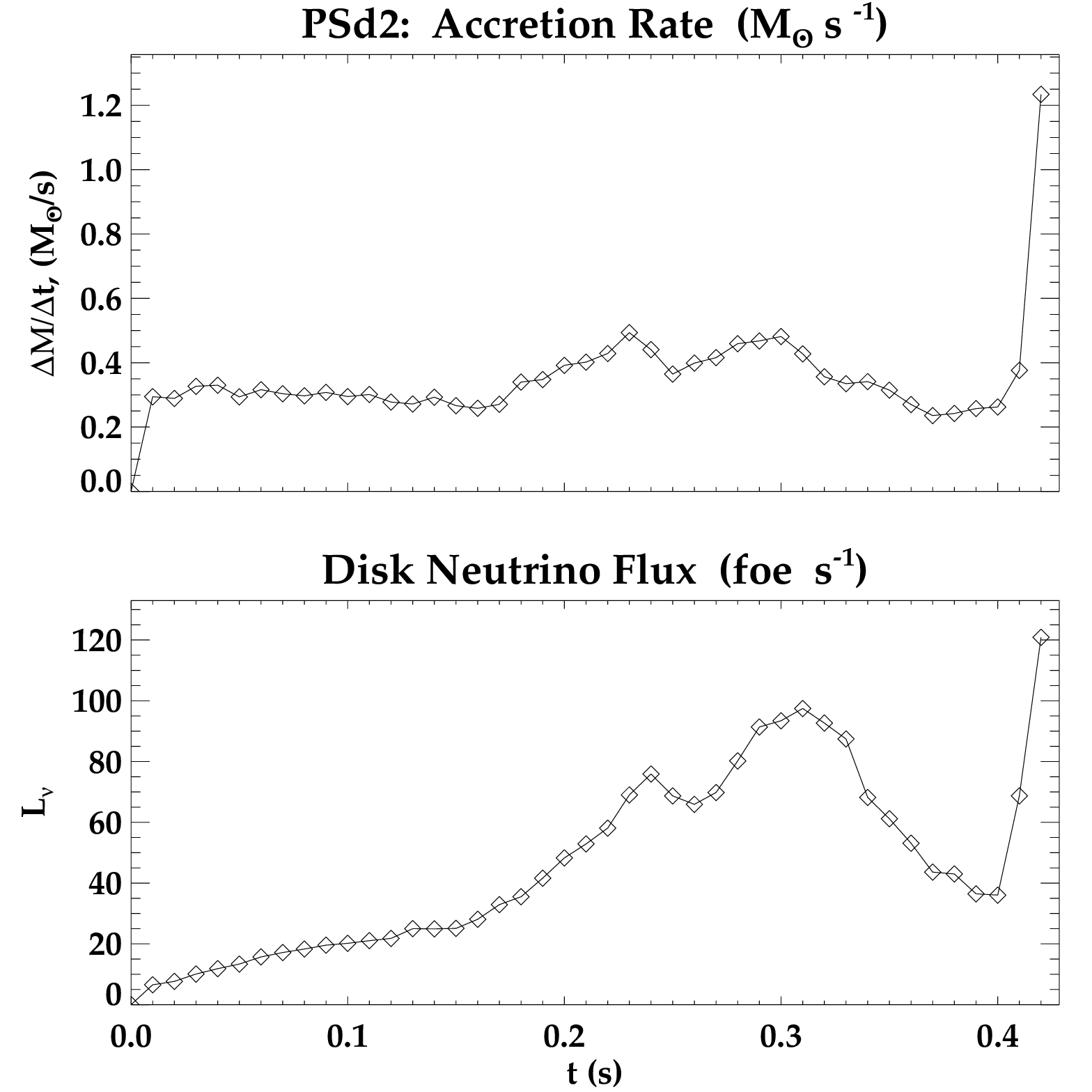}
    \caption{\label{fig:PSd2_mdot} Model PSd2: mass accretion rates
      and neutrino luminosity.}
    \end{center}
\end{figure}

For much of the disk evolution, the mass accretion rate is
consistently high (Fig.~\ref{fig:PSd2_mdot}).  After the influx of
low-$j$ material, non-axisymmetric structures form ($t>0.20$~s) and
disk accretion increases to $\dot{M}=0.4~\mps$, peaking at the high
rate of $\dot{M}\approx 1.2~\mps$.  This is reflected also in the
neutrino luminosity, which reaches $L_{\nu}\gtrsim
100~\foeps$. Therefore, in this model a successful central engine for
a GRB jet is produced by both mechanisms being examined in this study.
High accretion rates (with $a\approx0.5$ and assuming a large, central
$B$-field) lead to $\dot{E}_{\rm{B-Z}}\approx\dot{\it{E}}_{\rm{GRB}}$,
while the large neutrino production from the shock-heated disk leads
to $\enu>\dot{E}_{\rm{GRB}}$ (assuming a reasonable $\nu$-annihilation
efficiency).

As in the previous models, mainly light elements constitute the
disk. Although the inner regions here produce a much greater amount of
heavy elements, $0.048~\msun$ of $^{54}$Fe, $0.069~\msun$ of Fe,
$0.029~\msun$ of Co and $0.037~\msun$ of $^{56}$Ni, the latter
remains below that required to power typical HNe (though, as noted
previously, a repetition of accretion events may additively increase
abundances).  The outflow velocities in the equatorial region reach
$5\times10^8<v_{\rm{R}}\lesssim10^9~\cmps$.  Much of the material at
the outer edge of the disk has high specific entropy with
$\ye\sim0.45$, making case (i) outflows a possible site for non-NSE
nucleosynthesis.  While there is some vertical flow from the
equatorial plane, most material remains confined to the coronal
regions, limiting the possibility of case (ii) processes.

There is very little asymmetry observed in Fe and O, with production
of the latter slightly increased along the equatorial plane near the
coronal structures.

\subsection{PSd5, PSm2}

Neither model PSd5 nor PSm2 ($v_\phi$ decreased by a factor of 5 and
increased by a factor of 2, respectively) results in a successful LGRB
progenitor.  In model PSd5, the accretion onto the central object
remains quasi-spherical, with 45\% of $\dot{M}$ coming from the polar
direction even at t = 0.20~s.  The infall rate is quite steady, with
no outflows. This model has too little rotation to form a significant
disk structure or a potential LGRB, and it shows very little
polarisation in element abundances.

In PSm2, the Toomre parameter remains $\geq2$ at all times.  Since
PSm1, with 25\% of the kinetic energy of this model, was only locally
unstable, it is unsurprising that this disk appears to be globally
stable.  The mass of dense ($\rho>10^8 \gcc$) disk material is
significantly less than that of model PSm1, and by $t=1$~s the
central object has accreted approximately half as much mass.  In this
case, the distribution of elemental O is significantly polarised along
the equatorial plane.

Results for both models are given in Tables~\ref{tab:nsenucl} and
\ref{tab:modelsres}.

\section{Cylindrical rotation models}
We briefly discuss the cylindrical rotation profile models, which have
lower $E_{\rm{kin},\phi}$ but higher $J_0$ than the
shellular models.  In general, the results for respective scalings of
velocity for each series are qualitatively similar.

\subsection{PCm1}
\begin{figure}
  \begin{center}
    \includegraphics[width=74mm]{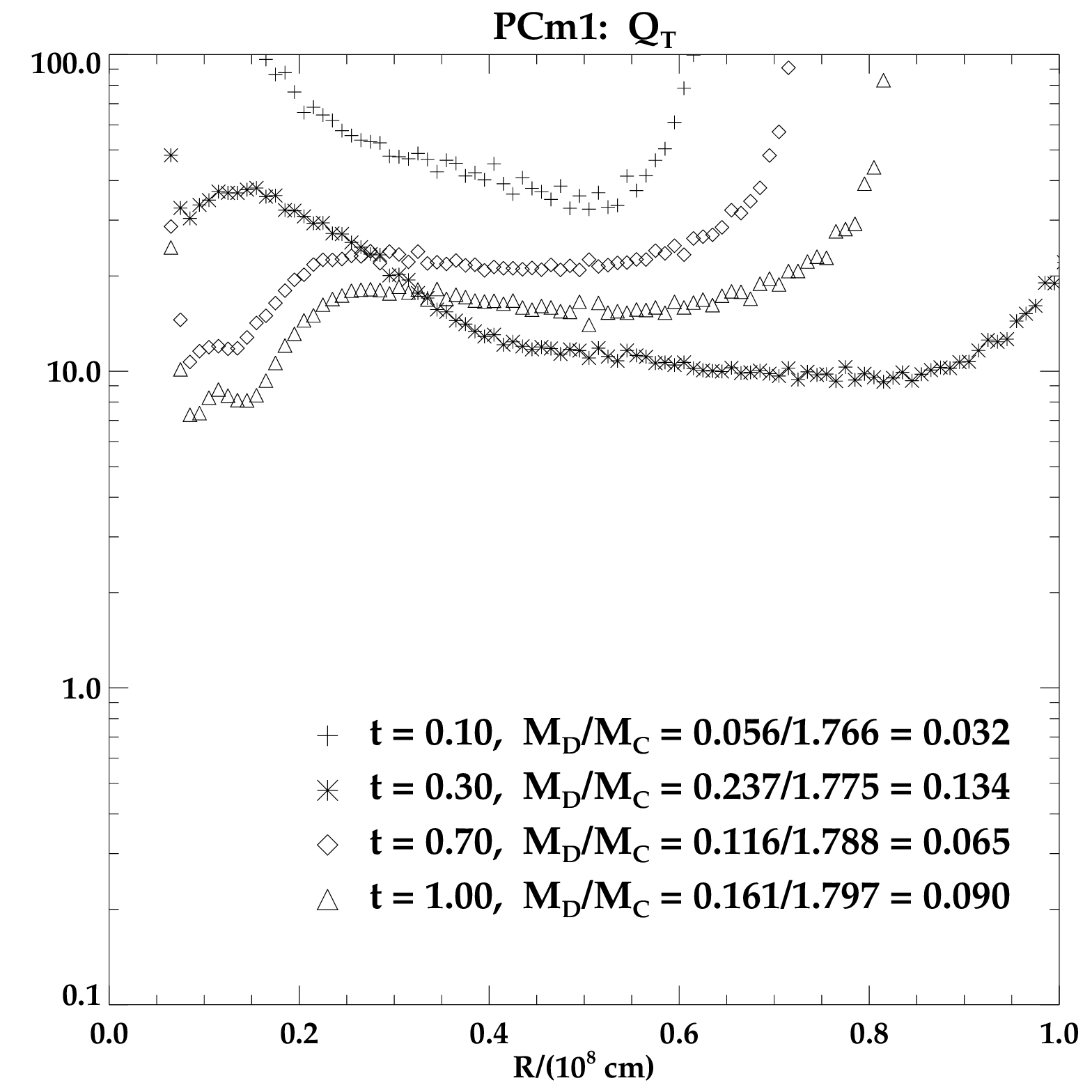}
    \caption{\label{fig:PCm1_qt} For PCm1, $Q_{\rm{T}}\gg 1$ for the
    entire evolution, and the disk remains stable.}
    \end{center}
\end{figure}

The evolution of $Q_{\rm{T}}$ and of the system's disk/central object
mass ratios are shown in Fig.~\ref{fig:PCm1_qt} for PCm1 (full
$v_{\phi}$).  In this model the collapsing material forms a disk
without any locally unstable region, even though the disk average
$Q_{\rm{T}}\gtrsim10$ is less than in PSm1.

\begin{figure}\vspace{-0.4cm}
  \begin{center}
    \includegraphics[width=78mm]{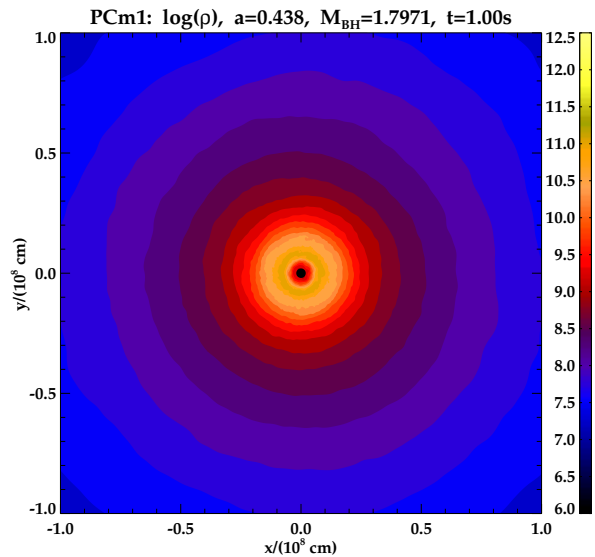}
    \caption{\label{fig:PCm1_den} Equatorial density profile for PCm1,
      which shows peak density less than PSm1.}
    \end{center}
\end{figure}

The densities in the inner disk region remain approximately an order
of magnitude less than in PSm1, with a slowly decreasing density
profile in the outer regions (Fig.~\ref{fig:PCm1_den}).  The
temperature and radial velocity profiles are similar, although $\ye$
remains above 0.35.  In general, the vertical structure and the
relative locations of Fe and O are similar to those of PSm1.  Roughly
a factor of five less $^{56}$Ni is produced in PCm1, which produces
$\approx60\%$ more Fe and $^{54}$Fe instead
(Table~\ref{tab:nsenucl}).

After a brief accretion period of low-$j$ material, $\dot{M}$ settles
to $\approx0.03$~M$_{\odot}~\textrm{s}^{-1}$.  Similarly, even with
100\% neutrino annihilation efficiency, $\enu$ remains below
$\dot{E}_{\rm{GRB}}$.  Therefore, this model, like PSm1, does not
appear to be a candidate for either a LGRB or a HN.

\subsection{PCdsq2}
\begin{figure}
  \begin{center}
    \includegraphics[width=74mm]{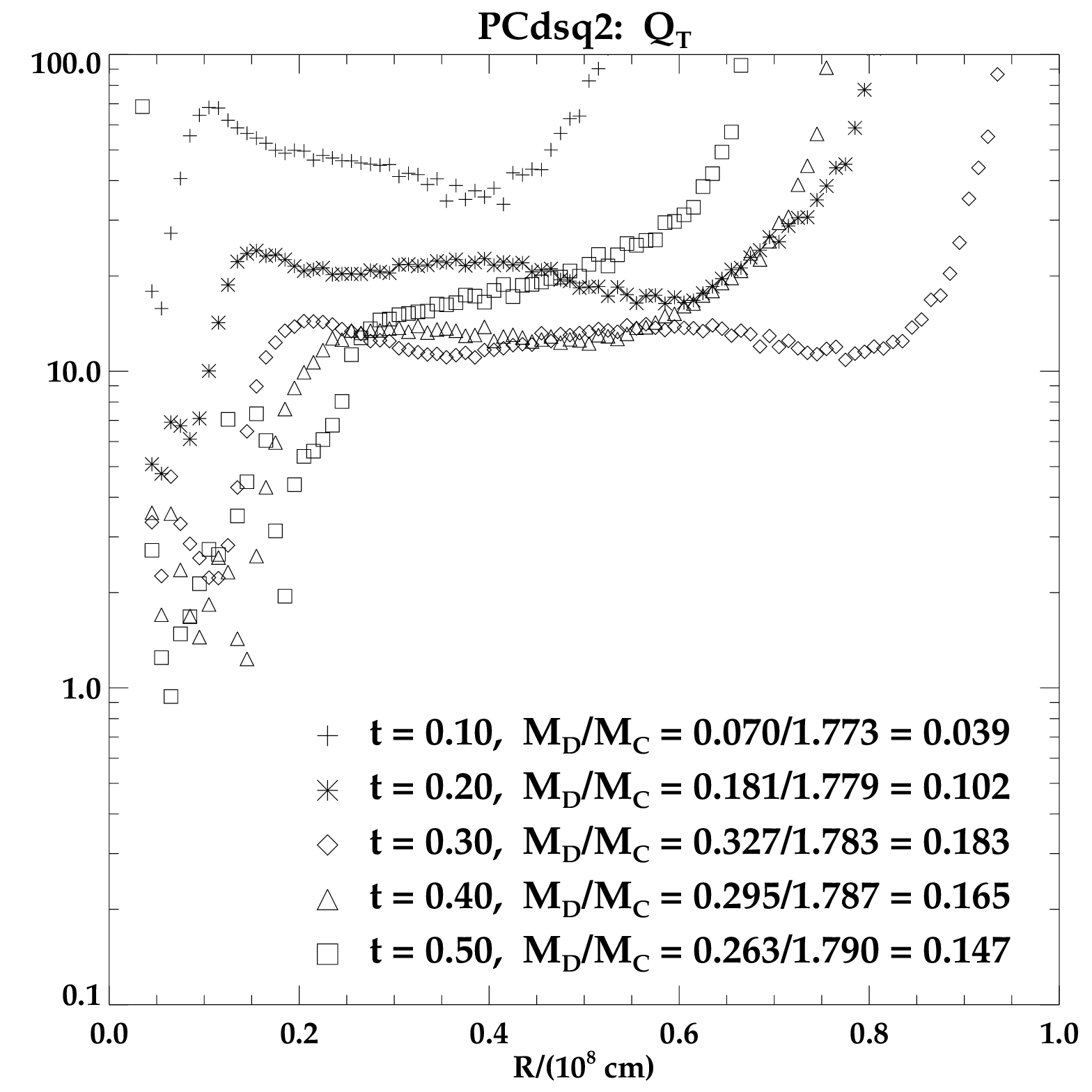}
    \caption{\label{fig:PCdsq2_qt} $\qt$ evolution for
      PCdsq2, which is unstable by $t=0.35$~s.}
    \end{center}
\end{figure}

\begin{figure*}\vspace{-0.4cm}
  \begin{center}
    \begin{tabular}{cc}
      \includegraphics[width=78mm]{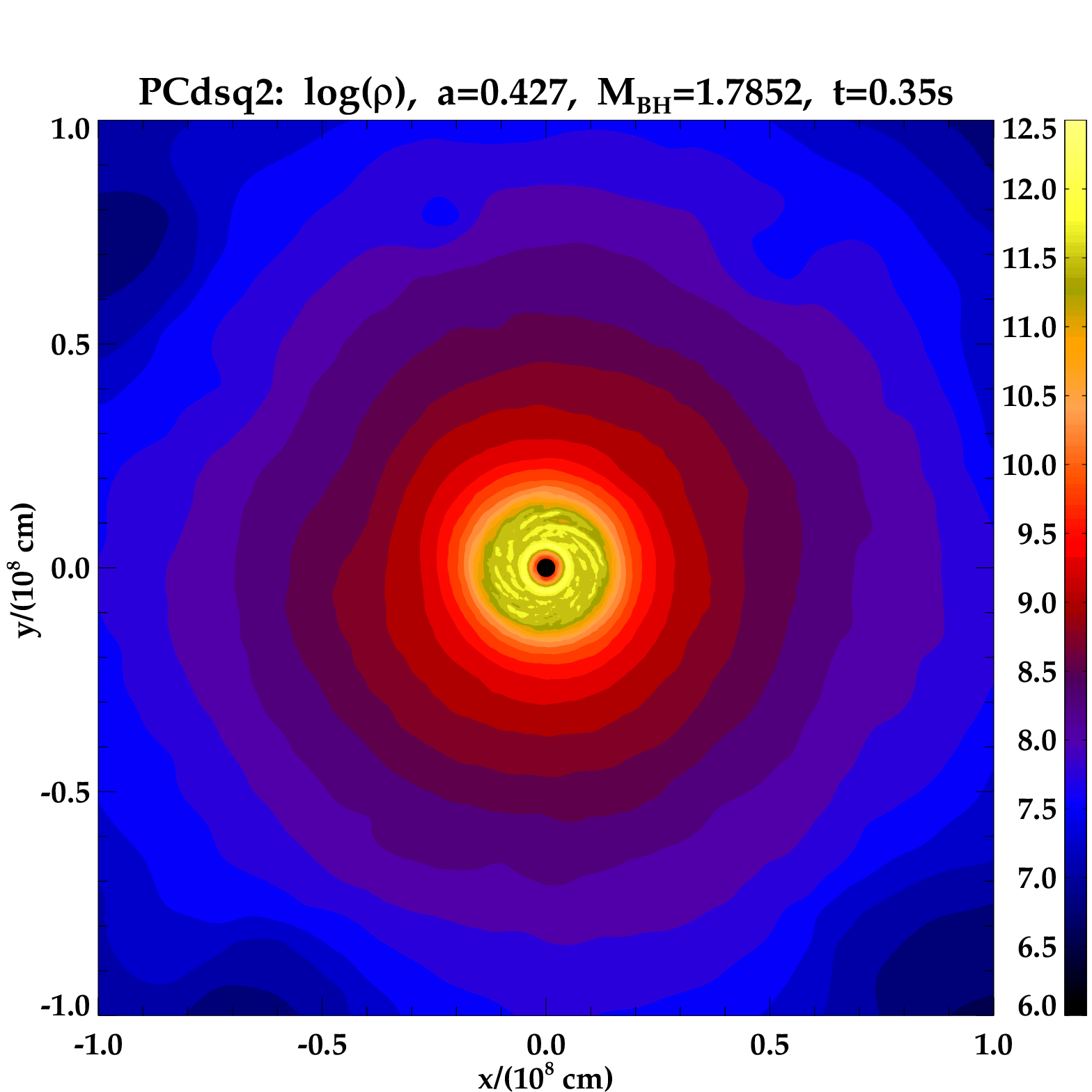}&
      \includegraphics[width=78mm]{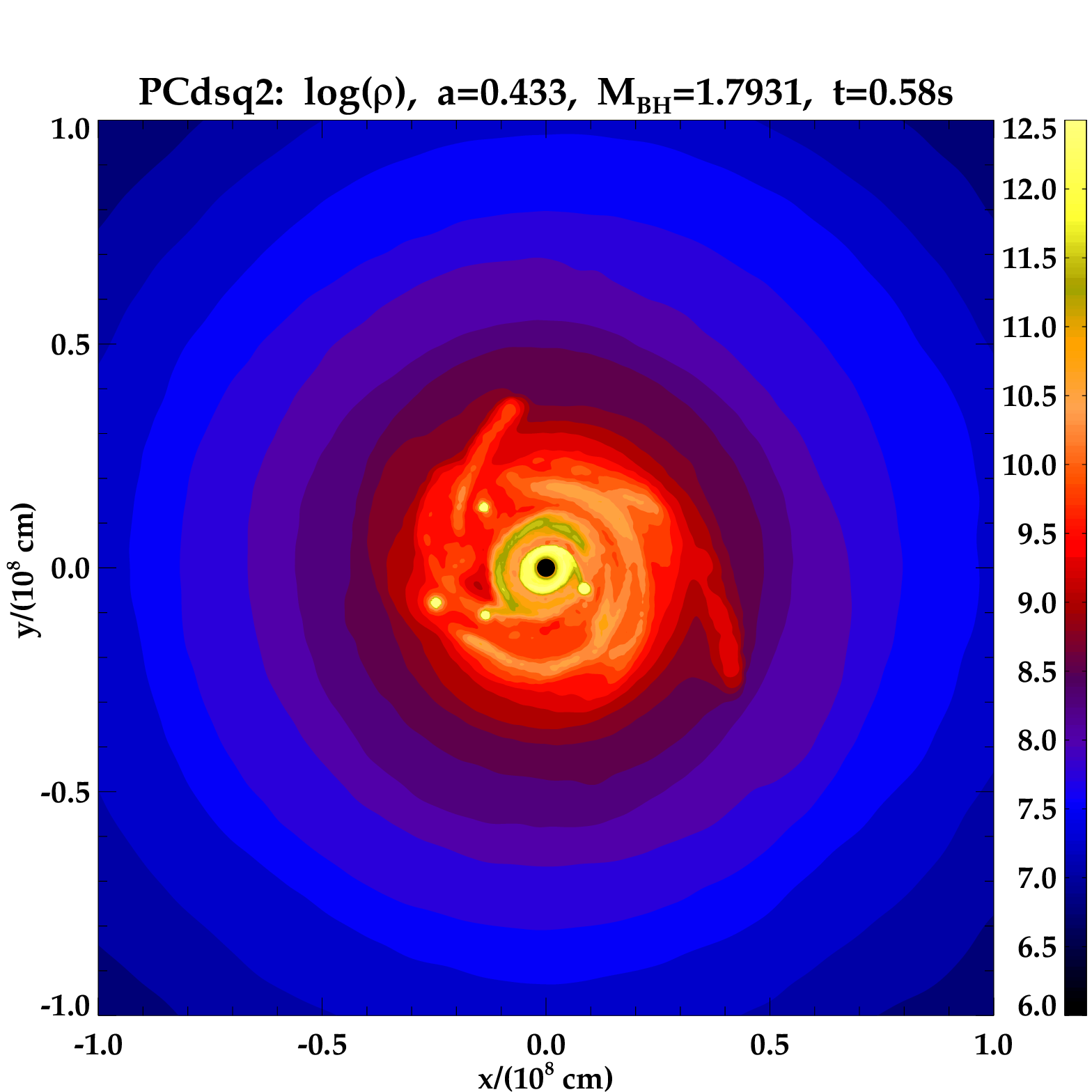}\\[-.2cm]
      \end{tabular}
      \caption{\label{fig:PCdsq2_den1} Evolution of density
        perturbations in PCdsq2.}
   \end{center}
\end{figure*}

The evolution of $\qt$ for PCdsq2 ($v_{\phi}$ decreased by a factor of
$\sqrt{2}$) is shown in Fig.~\ref{fig:PCdsq2_qt}.  The disk becomes
unstable by $t=0.35$~s, earlier than for PSdsq2 and with a larger
radial extent.  The resulting formation of spirals and an $m=2$
structure is shown in Fig.~\ref{fig:PCdsq2_den1}.  High velocity
radial outflows form, $v_{\rm{R}}\sim 10^9~\textrm{cm~s}^{-1}$,
carrying material with $\ye\approx0.43$ outward.  The values of
$\bar{A}$ increase much more rapidly with radius than in PSdsq2, with
much of the material in the outflow regions having $\bar{A}>4$.

\begin{figure}
  \begin{center}
    \includegraphics[width=78mm]{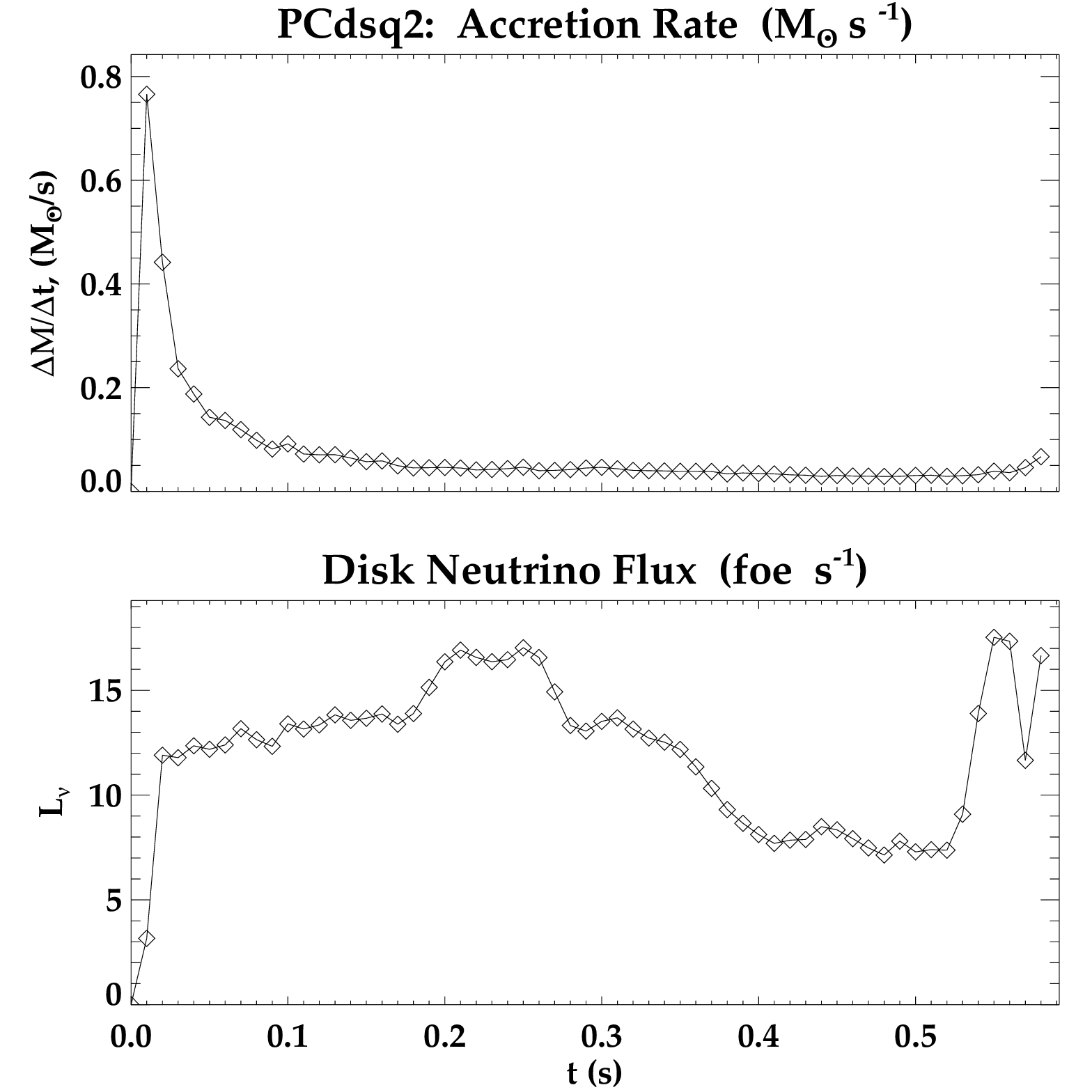}
    \caption{\label{fig:PCdsq2_mdot} Model PCdsq2: evolution of mass
      accretion rates and neutrino luminosity.}
    \end{center}
\end{figure}

In the vertical plane the coronal regions extend to greater height,
with a much larger amount of material with $\ye\approx0.43$.  Again,
the average nucleon number increases and the temperature drops more
quickly outside of the main disk/coronal area.  The outer corona again
contains O preferentially, with the Fe abundance dominating at larger
radii.  The amount of $^{56}$Ni ($0.096~\msun$) remains below HN
values, however, by a factor of 3-5.  This is similar to PSdsq2,
where O was more prevalent in all regions, and significantly smaller
amounts of Fe-group elements were produced.

After the initial infall of low-$j$ material, accretion rates (shown
in Fig.~\ref{fig:PCdsq2_mdot}) remain low,
$\dot{M}<0.05~\rm{M}_{\odot}~\textrm{s}^{-1}$, even after the
formation of spiral structure, although there is an upturn at late
times.  Neutrino luminosity is fairly high at $L_{\nu}\approx
10~\textrm{foe~s}^{-1}$, but $\enu$ remains below $\dot{E}_{\rm{GRB}}$
for reasonable annihilation efficiencies.

\subsection{PCd2}
\begin{figure}
  \begin{center}
    \includegraphics[width=74mm]{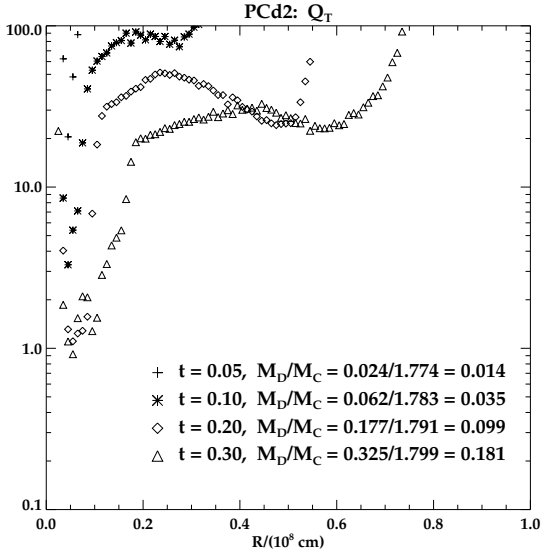}
    \caption{\label{fig:PCd2_qt} $\qt$ evolution for PCd2,
      which is unstable by $t\gtrsim0.20$~s.}
    \end{center}
\end{figure}

\begin{figure}\vspace{-0.4cm}
  \begin{center}
    \includegraphics[width=78mm]{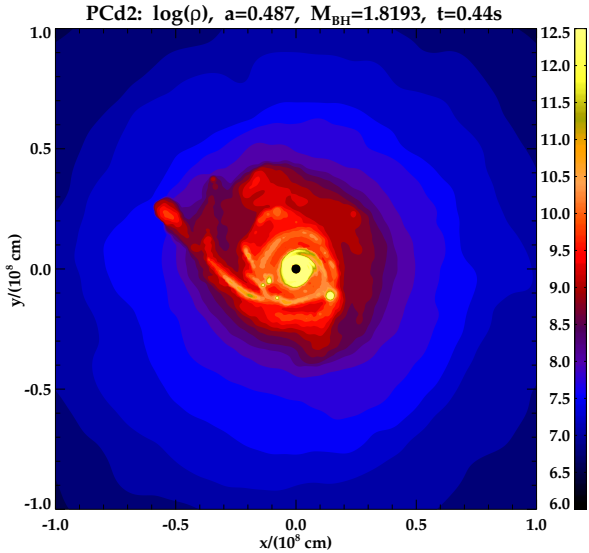}
    \caption{\label{fig:PCd2_den} Density structure in the equatorial
      plane for model PCd2 at $t=0.44$~s.}
    \end{center}
\end{figure}

As in the shellular case, PCd2 ($v_{\phi}$ decreased by a factor of 2)
becomes unstable to the formation of non-axisymmetric structure
quickly ($t<0.20$~s, Fig.~\ref{fig:PCd2_qt}).
Fig.~\ref{fig:PCd2_den} shows the spiral modes which create shocks in
the fluid flow.  The evolution of structure, with the development of
finer spirals followed by a global $m\approx 2$ mode, is very similar
to that of PSd2.  Eventually, moderate-$\ye$ material moves outward in
the equatorial plane due to the transfer of angular momentum in the
disk.

In the vertical plane again $\rho$, $s/A$ and $\ye$ decrease more
slowly with distance from the equatorial plane than in the equivalent
shellular model, PSd2.  $^{56}$Ni and Fe are produced in very similar
quantities to those of PSd2.  Typically, the outer corona contains O
preferentially, with Fe abundances being greater at larger radii.

During the spiral structure phase, accretion rates begin near $\dot{M}\lesssim
0.1~\textrm{M}~\textrm{s}^{-1}$, increasing to $\dot{M}\lesssim
0.4~\textrm{M}~\textrm{s}^{-1}$ at late times.  The neutrino
luminosity is fairly high during the spiral structure phase and
increases to $L_{\nu}=100-150~\textrm{foe~s}^{-1}$.  This model
therefore provides sufficient fuel for a LGRB central engine for both
the B-Z mechanism and $\nu$-annihilation, as did PSd2.

\section{Low temperature models}
The low temperature (shellular) models produced results very similar
to those with similar rotation profiles discussed above in
\S\ref{sec:shell}.  Comparisons between the relevant models are discussed
here.

\subsection{PSm1Kd5}
\begin{figure}
  \begin{center}
    \includegraphics[width=74mm]{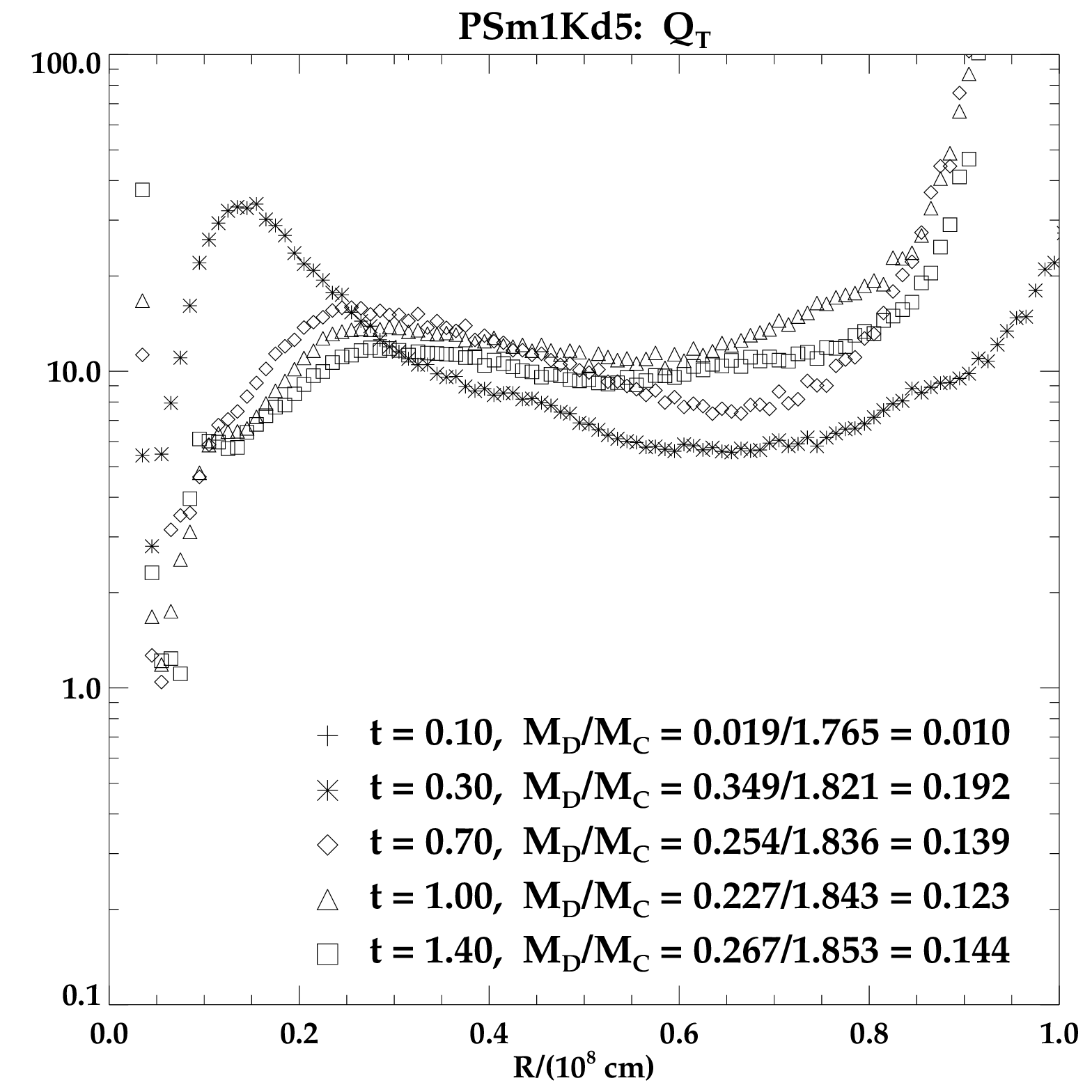}
    \caption{\label{fig:PSm1Kd5_qt} $\qt$ evolution for PSm1Kd5, which
      remains stable globally.}
    \end{center}
\end{figure}

\begin{figure}\vspace{-0.4cm}
  \begin{center}
    \includegraphics[width=78mm]{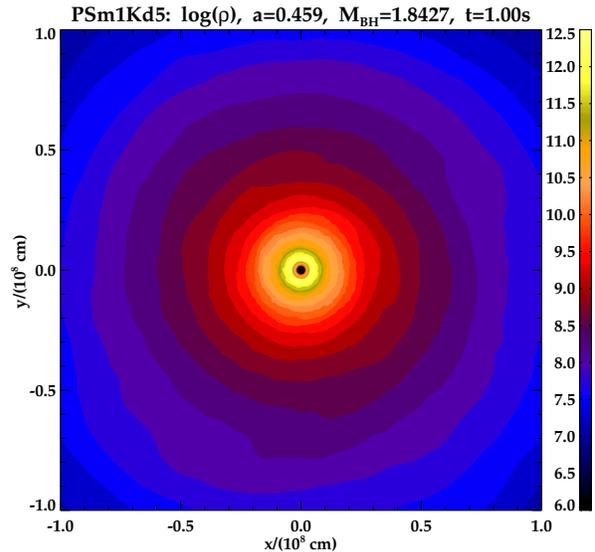}
    \caption{\label{fig:PSm1Kd5_den} Density in the equatorial plane
      for PSm1Kd5, which has a very similar profile to that of PSm1,
      excepting the local instability
      (cf. Fig.~\ref{fig:PSm1_eqplden}).}
    \end{center}
\end{figure}

The mass of the disk which forms in PSm1Kd5 (full $v_{\phi}$) is
greater that of PSm1.  The shapes of the $\qt$ profiles are similar,
with PSm1Kd5 tending to lower values, though its minimum value
increases after $t\approx 0.70$~s, and the disk remains globally
stable throughout its evolution
(Figs.~\ref{fig:PSm1Kd5_qt}\ref{fig:PSm1Kd5_den}).  Much smaller
regions of outflow are produced and with lower $v_{\rm{R}}$, due to
the greater stability of the disk.

Through the first $1.00$~s, PSm1 and PSm1Kd5 have similar temperature
profiles in the equatorial plane. In the latter, density decreases and
specific entropy increases more slowly with radius, and, while the
electron fraction is greater in the inner region ($\ye\approx0.30$),
$\ye$ increases slowly so that a larger region contains material with
$\ye\approx0.43-0.45$; $\bar{A}$ behaves similarly.

Vertically, in PSm1Kd5 density decreases rapidly with distance from
the equatorial plane, and a smaller corona surrounds the disk than in
PSm1.  In this region, there is a larger extent of low $s/A\leq1$
material, before the entropy increases rapidly.  The electron fraction
behaves similarly in the corona, with $\ye$ mainly $\approx0.43$. The
amount of $^{56}$Ni produced is smaller by an order of magnitude, as
are the total masses of the Fe-group elements calculated here in
general (Table~\ref{tab:nsenucl}).  The relative regional
distributions of O and Fe are similar to those of PSm1, with a
slightly smaller coronal extent of O.

Both $\dot{M}$ and $L_{\nu}$ closely resemble those for PSm1.  After
the initial accretion of shock-heated, low-$j$ material, accretion
rates settle to $\dot{M}\approx0.03~\textrm{M}_{\odot}~\rm{s}^{-1}$,
and neutrino luminosity goes to
$L_{\nu}\approx0.1\textrm{~foe~s}^{-1}$.  Therefore, model PSm1Kd5 is
also unable to produce a successful LGRB.

\subsection{PSd2Kd5}
\begin{figure}
  \begin{center}
    \includegraphics[width=74mm]{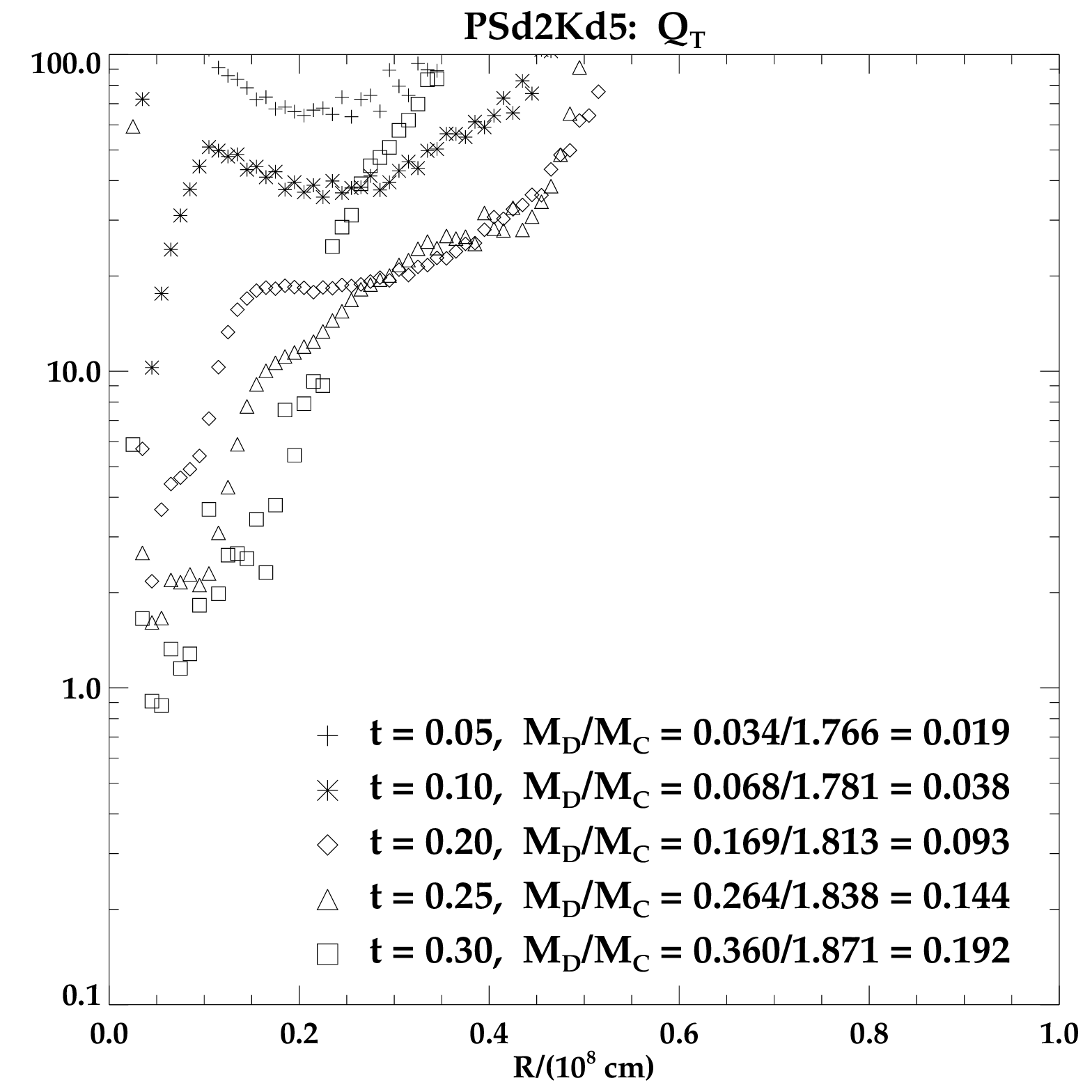}
    \caption{\label{fig:PSd2Kd5_qt} For PSd2Kd5, $Q_{\rm{T}}$ becomes
      unstable after $t=0.20$~s, slightly earlier than PSd2; the
      shapes of the curves are very similar, with PSd2Kd5 being
      $\approx 10\%$ less (cf. Fig.~\ref{fig:PSd2_qt}).}
    \end{center}
\end{figure}

\begin{figure}\vspace{-0.4cm}
  \begin{center}
    \includegraphics[width=78mm]{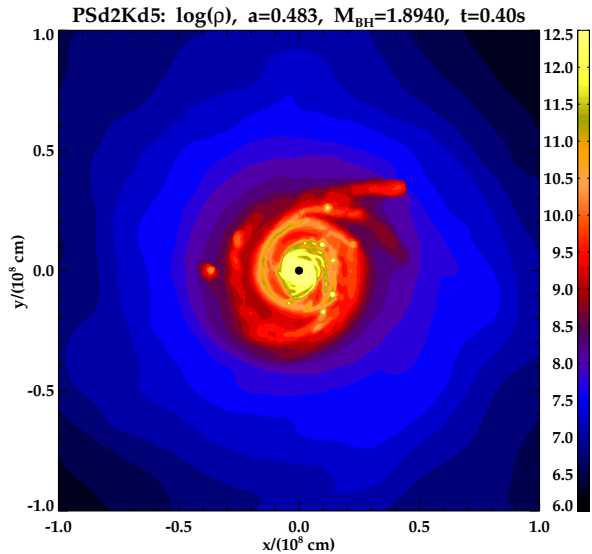}
    \caption{\label{fig:PSd2Kd5_den} Equatorial density profile for
      PSd2Kd5, which shows very similar late time structure to PSd2
      (cf. Fig.~\ref{fig:PSd2_rho}).}
    \end{center}
\end{figure}

The shapes of the $\qt$ curves for PSd2Kd5 ($v_{\phi}$ decreased by a
factor of 2) shown in Fig.~\ref{fig:PSd2Kd5_qt} are very similar to
those of PSd2, with the former having $\approx 10\%$ lower values.
For $t<0.20$~s, the disk masses are nearly identical as well.  Again,
tightly wound spirals develop into `finger-like' substructure in a
global $m=2$ mode (Fig.~\ref{fig:PSd2Kd5_den}).  It is perhaps
surprising that the thermal and specific entropy structures are also
nearly identical to those of PSd2, despite an initially large
difference in internal energy. Also, the $\bar{A}$ and $\ye$ values
are very similar for both models. However, the outflow velocities in
the equatorial plane are much greater in PSd2Kd5, reaching
$v_{\rm{R}}>0.1c$.

The vertical profiles of PSd2Kd5 closely resemble those of PSd2, but
the coronal regions are smaller in the former, and $\rho$ decreases
more rapidly with distance from the equatorial plane. The relative
locations of Fe and O production follow those in PSd2.  However, the
amounts of Fe-group elements produced are quite different
(Table~\ref{tab:nsenucl}).  Here, $0.10~\msun$ of $^{56}$Ni is made,
approximately three times the amount in PSd2 and only a factor of a
few below that which is estimated for some HNe.  The masses of the
competing end-products, $0.017~\msun$ of $^{54}$Fe and $0.029~\msun$ of
Fe, are a factor of three lower than in PSd2.

\begin{figure}
  \begin{center}
    \includegraphics[width=78mm]{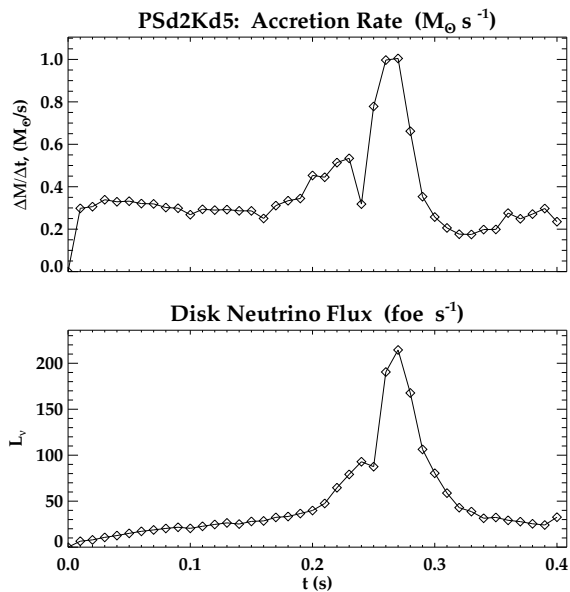}
    \caption{\label{fig:PSd2Kd5_mdot} Model PSd2Kd5: evolution of mass
      accretion rates and neutrino luminosity.}
    \end{center}
\end{figure}

In this model accretion reaches very high rates, $\dot{M}\approx
1~\textrm{M}_{\odot}~\textrm{s}^{-1}$ at peak, with an average of
$\dot{M}\approx 0.25-0.30~\textrm{M}_{\odot}~\textrm{s}^{-1}$
otherwise (Fig.~\ref{fig:PSd2Kd5_mdot}).  The neutrino flux reaches
the high values of $L_{\nu}> 200~\textrm{foe~s}^{-1}$, although it
remains below $50~\textrm{foe~s}^{-1}$ for much of the evolution.
Therefore, while at peak values both $\dot{E}_{\rm{B-Z}}$ and $\enu$
are $>\dot{E}_{\rm{GRB}}$, the B-Z mechanism appears to be
significantly better for providing a longer duration central engine.

\section{Simple jet test, PSdsq2J}
\label{sec:psdsq2j}
In order to test the behaviour of the collapsar flow in the presence
of a collimated outflow, we implemented a very simple `jet' structure
originating from the central object (starting with model PSdsq2).
This was not intended to be a physically realistic jet, but merely a
toy model to gain insight into affected regions of the system, i.e. in
disk structure, in the nature of any coronal material swept outward,
etc.

The prescription inserted into PSdsq2 at $t=1.10$~s was that a narrow
jet region with polar angular radius of $\theta=10^{\circ}$ around the
rotation axis developed and grew quickly.  An acceleration directed
along the envelope of the jet, away from the equatorial plane, was
applied to every fluid element which approached this region to within a
small fraction of its smoothing length.

\begin{figure}\vspace{-0.4cm}
  \begin{center}
    \includegraphics[width=78mm]{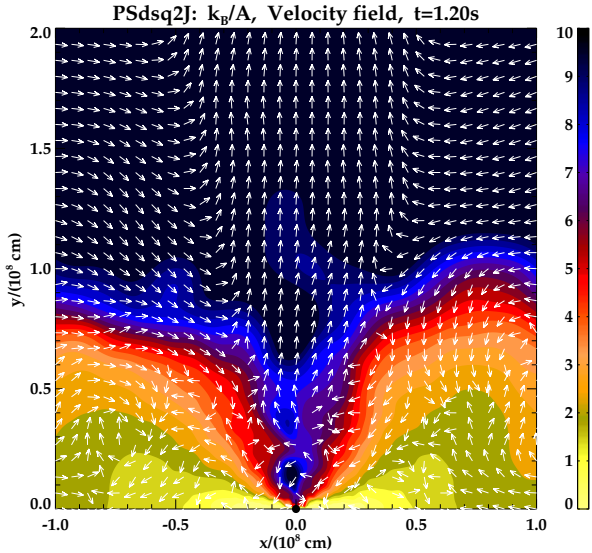}
    \caption{\label{fig:PSdsq2J_den} Specific entropy contours with
      normalised velocity vector field showing direction of flow in
      vertical slice half-plane (above equatorial plane) for
      comparison with Fig.~\ref{fig:PSdsq2_vel}.  Material in coronal
      regions retains small, turbulent velocities.  A strong vertical
      outflow is created by the jet.}
    \end{center}
\end{figure}

In Fig.~\ref{fig:PSdsq2J_den}, model PSdsq2J is shown 0.10~s after the
insertion of the jet.  A comparison with the same model without a jet
in Fig.~\ref{fig:PSdsq2_vel} shows that the vertical structures are
affected only slightly, but most importantly, high $s/A$ material
(with moderate $\ye$) is carried upwards from the corona and inner
edges of the disk at high velocities, $0.05$-$0.1c$.  These conditions
make the inner disk region a strong candidate for producing large
amounts of $^{56}$Ni in outflows.  We note, too, that the vertical
outflow material becomes more likely to form Fe-group elements,
assuming NSE is maintained; therefore, the polar direction becomes
more Fe-rich, while the equatorial plane remains O-rich, as in
polarised HNe.  Explosive nucleosynthesis in the jet itself may also
be a source of heavy elements such as $^{56}$Ni, though the abundances
which can be produced remain uncertain
\citep*[e.g.,][]{2006ApJ...647.1255N}.

\section{Comparing GR and Newtonian potentials}
A simple test to gauge the effects of including GR was made by
utilising a purely Newtonian central potential in (shellular)
simulations with full rotation (model NSm1) and with $v_{\phi}$
decreased by a factor of 2 (model NSd2).  In general, the results are
quite similar to those using the SEP pseudo-potential (PSm1 and PSd2,
respectively).

\begin{figure}
  \begin{center}
    \includegraphics[width=74mm]{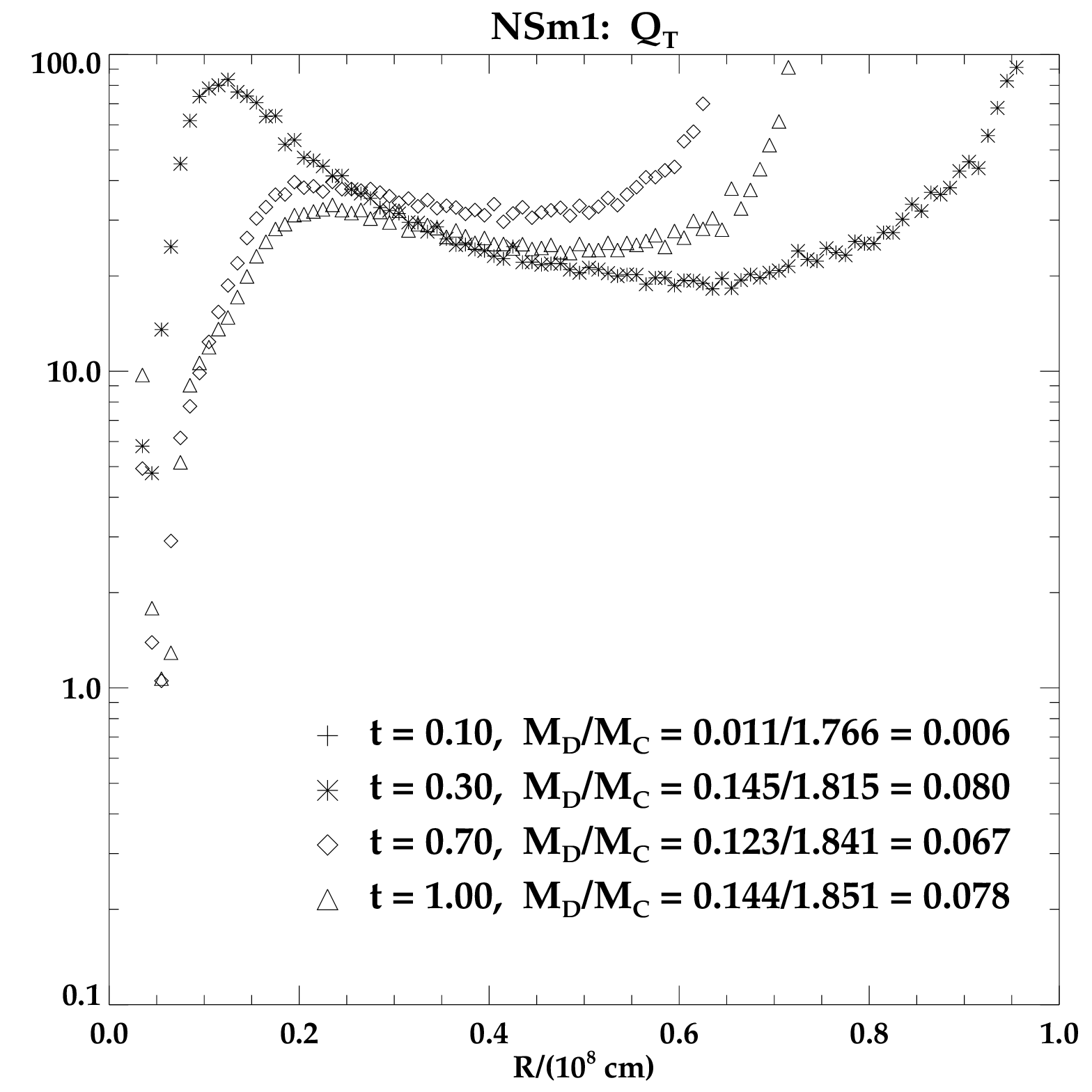}
    \caption{\label{fig:NSm1_qt} Evolution of $\qt$ for the Newtonian,
      shellular model, NSm1.}
    \end{center}
\end{figure}

Fig.~\ref{fig:NSm1_qt} shows $Q_{\rm{T}}$ for NSm1.  The greatest
difference occurs in the minimum-region which nears instability, which
is $\approx50\%$ wider in the pseudo-potential system.  The result is
that the NSm1 disk remains both globally and locally stable, without
the formation of non-axisymmetric structures which developed in the
PSm1 model. Due to this (temporary) lack of perturbation, the radial
outflow is much lower in the Newtonian case.

The profiles of hydrodynamic and microphysical quantities, in both
equatorial and polar directions, are very similar.  However, NSm1
produces orders of magnitude less $^{56}$Ni and a factor of five less
Fe than PSm1.  The $\dot{M}$ and $L_{\nu}$ values are typically
$\approx 15-20\%$ greater in NSm1.

\begin{figure}
  \begin{center}
    \includegraphics[width=74mm]{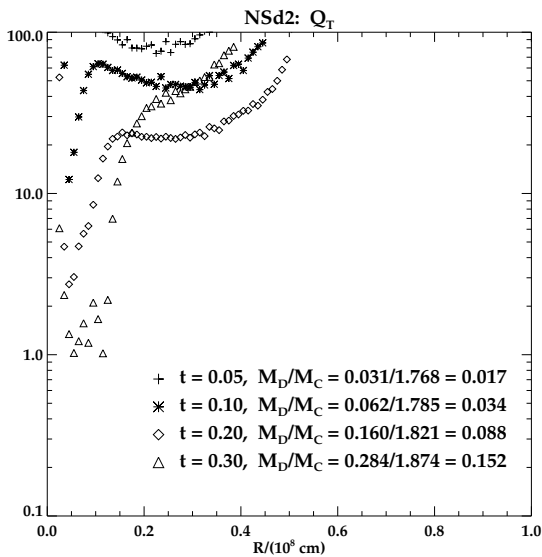}
    \caption{\label{fig:NSd2_qt} Evolution of $\qt$ for the Newtonian,
      shellular model with reduced rotation, NSd2.}
    \end{center}
\end{figure}

For the reduced velocity case, the $\qt$ curves for NSd2
(Fig.~\ref{fig:NSd2_qt}) closely resemble those of its
pseudo-potential analogue, PSd2.  Moreover, the structural,
hydrodynamic and nuclear constituent properties of the evolving disk
appear similar, as well.  In this case, the only difference in
Fe-group production appears in $^{56}$Ni, which is $25\%$ less than
the amount produced by PSd2.  Again, the $\dot{M}$ and $L_{\nu}$
curves for NSm1 are qualitatively similar to those of PSm1 but
with the values of each of the former being $\approx 20\%$ greater.

\section{Discussion and conclusions}
\label{sec:discusssim}

\begin{table*}
  \begin{center}
    \begin{minipage}{175mm}
      \caption{\label{tab:modelsres} Results for the various collapsar
        models, in terms of maximum accretion rates and energy
        production.  The values of $\dot{E}_{\rm{B-Z,max}}$ have been
        calculated using Eq.~(\ref{edot_bz}).  In general, scenarios
        of $\dot{M}_{\rm{max}}>0.1~\mps$ and
        $L_{\nu,\rm{max}}>100~\foeps$ produce successful LGRBs
        (although the profiles for each exhibit rapid fluctuations).}
\begin{tabular}{lccccccccl}
\hline
\mc{1}{c}{Model}& \mc{1}{c}{Vel.}& \mc{1}{c}{$v_{\phi}$}   & \mc{1}{c}{ $E_{\rm{th}}$}& \mc{1}{c}{$y_j$} & \mc{1}{c}{$\dot{M}_{\rm{max}}$  }&\mc{1}{c}{$\dot{E}_{\rm{B-Z,max}}$  }& \mc{1}{c}{ $L_{\nu,\rm{max}}$}& \mc{1}{c}{ $M(^{56}\rm{Ni})^{\dag}$}& \mc{1}{c}{Instability}  \\
&\mc{1}{c}{Profile}& \mc{1}{c}{scale }& \mc{1}{c}{scale}  & \mc{1}{c}{($\bar{j}_0/j_{\rm ISCO}$)}  & \mc{1}{c}{(M$_{\odot}$~s$^{-1}$)} & \mc{1}{c}{(foe~s$^{-1}$)}& \mc{1}{c}{(foe~s$^{-1}$)}& \mc{1}{c}{(M$_{\odot}$)} &  \mc{1}{c}{ (description) }\\
\hline
PSm2		& shell		& $\times$2	& full &65.5 &0.023	& 0.13	&  0.073 &$10^{-4}$	& none \\
PCm1		& cyl.		& full		& full &36.3  &0.035	& 0.20	&0.23	&0.005	& none \\
PSm1		& shell		& full		& full &32.8	&0.032	& 0.21 &0.41	&0.020	&  local perturbation (temp.) \\
PSm1Kd5		& shell		& full		& /5  & 32.8	&0.027	& 0.17	&0.12	&0.001	& none (slight non-axis.) \\
NSm1		& shell		& full		& full &32.8 	&0.035	& 0.241	&0.20	&$10^{-4.2}$	& none \\
PCdsq2		& cyl.		& /$\sqrt{2}$	& full  &25.7 & 0.067	& 0.38	&17.5	&0.010	& global spirals, late $m=2$ \\
PSdsq2		& shell		& /$\sqrt{2}$	& full &23.1 &0.67	& 8.33	&29.7   &$10^{-8.3}$	& global spirals, late $m=0$ \\
PCd2		& cyl.		& /2		& full  & 18.2  & 0.37 	& 2.51	&167	&0.032	& global spirals and $m=1,2$ \\
PSd2		& shell		& /2		& full  &16.4  & 1.23	& 8.82	&121	&0.038	&  global spirals and $m=2$ \\
PSd2Kd5		& shell		& /2		& /5    &16.4 &1.00  	& 6.49	&214	&0.101	& global spirals and $m=2$ \\
NSd2		& shell		& /2		& full  &16.4 & 0.63 	& 4.41	&121	&0.029	& global spirals and $m=2$ \\
PSd5		& shell		& /5		& full 	&~6.5 &-	& -	&-	&-	& none (no disk) \\
\hline
\end{tabular}
\medskip
{$^{\dag}\,$}\small{Estimated at late times in the evolution.}
\end{minipage}
\end{center}
\end{table*}

\subsection{Numerical timesteps}
It was noted in \S\ref{sec:psdsq2} that the timesteps in the more
dynamic simulations became too short to continue.  In Gadget-2
standard hydrodynamical criteria assign the length of time between
`kicks' for each SPH particle: the Courant-Friedrich-Levy (CFL)
condition, which requires small timesteps for dense material ($\Delta
t_{\rm{step}} \propto \rho^{-1}$); and a dynamic condition, which
requires small timesteps for particles with rapidly changing velocity
($\Delta t_{\rm{step}} \propto |\textbf{a}|^{-1/2})$.

In the case of collapsars, densities often reach high values, much
greater than neutron drip.  Similarly, dense matter in the inner disk
is shocked by spiral arms, and infalling matter is shocked to high $T$
at the edge of the corona.  Finally, particles are removed from the
simulation when they are accreted, and during periods of rapid
accretion, the computer must keep searching for and recalculating
neighbours for nearby un-accreted particles; the material close to the
accretion boundary is also typically the densest, meaning that when
small amounts accrete, neighbour lists must be recalculated for many
particles. For some simulations in this study, the conjunction of
these conditions caused timesteps to become quite small, and the
computational expense per timestep, quite large.

In order to prevent some of these difficulties, particularly due to
neighbour searches when accreting, one may utilise a method of
replacing many SPH particles in a high-density region with a single
particle, as in \citet*{1995MNRAS.277..362B}. A simplified version of
this procedure was tested in some models with only limited success.
To avoid these difficulties in future work, more refined numerical
treatments of this process and of the accretion will be required.

\subsection{Hydrodynamic Instability}
In all of the models, the global stability of the collapsar disks
consistently depended on the scaling of azimuthal velocity in the
progenitor.  The dimensionless parameter, $y_j=\bar{j}_0/j_{\rm
  ISCO}$, was introduced to characterise the progenitor in terms of
relevant quantities; here, systems with values of, very roughly,
$10-15<y_j\lesssim30-32$ (with the upperbound seen more directly) were
hydrodynamically unstable; for large $y_j$, angular momentum
stabilises the disk, and for small values, no significant disk forms.
The initial velocity profile shape and even the scaling of
$E_{\rm{th}}$ by half an order of magnitude had comparatively very
little influence.  Results from the simulations are summarised in
Table~\ref{tab:modelsres} (for reference, the scalings of the various
model progenitors are included as well).  We note that in future
studies, the quantity $y_j$ may be particularly useful in predicting
the behaviour of a collapsar-candidate progenitor.

The full rotation (and $v_{\phi}\times 2$) models proved to be
globally stable against forming non-axisymmetric structures, with PSm1
temporarily developing a locally unstable ring.  By scaling $v_{\phi}$
in the progenitor by $1/\sqrt{2}$, the resulting disk became unstable
to thin spirals which became tightly wound and shocked surrounding
fluid.  These spirals eventually led to the development of larger,
radial oscillations ($m=0$ modes) for PSdsq2 and to global
non-axisymmetric structure for PCdsq2.  In all cases, scaling
$v_{\phi}$ by 1/2 led to the rapid formation of tightly wound spirals
and also to the appearance of global $m=1~$or$~2$ modes which
dominated the subsequent disk evolution.

As a general trend, instability set in a number of dynamical timesteps
earlier for lower $E_{\rm{kin},\phi}$, with cylindrical models
(prefixed with `PC') tending to become unstable slightly more quickly
than the shellular ones.  The spirals typically formed in the disk
within the radial interval of $4-8\times 10^6$~cm (of order $20
\times GM/c^2$), with lower rotation models becoming unstable at the
smaller end of the range.  We note that for several models, disks
formed localised density clumps as well, with signficant repercussions
for the accretion rate profile and energetics.

We note that it is not surprising that the $v_\phi$ profile of each of
the models which formed global structures and non-axisymmetric modes
had been scaled downward from that of the original progenitor. The
total angular momentum of a binary remnant depends heavily on the
redistribution mechanism of the merger process, which remains an area
of active research. For the basic progenitor used in this study, the
merger prescription had maximised the possible $J$ retained by a
remnant, one of three pictures presented by
\citet{2005ApJ...623..302F}.  Therefore, it should not be worrying for
the general consistency of binary mergers as collapsar progenitors
that only the models with scaled-down $E_{\rm{kin},\phi}$ produced
globally unstable disks.

\begin{figure}
\includegraphics[width=85mm]{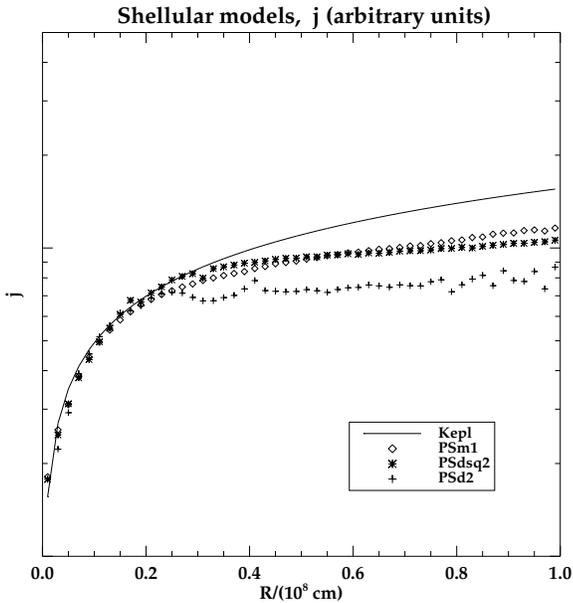}
\caption{ \label{fig:all_jspan} A relative comparison of specific
  angular momentum profiles for the shellular models PSm1, PSdsq2 and
  PSd2 at $t=1.00, 1.25, 0.42$~s, respectively. Analogous behaviour
  occured in the cylindrical and thermally-scaled models. All values
  are scaled in arbitrary units to the Newtonian, Keplerian profile
  (line).}
\end{figure}

\subsection{Accretion rates, $\nu$ luminosity}
Accretion rates were nearly constant (after accreting low-$j$
material) for the full-rotation models which did not form spirals,
settling to be $\approx0.03~\textrm{M}_{\odot}~\rm{s}^{-1}$, even in
the model with local instability.  As shown in
Table~\ref{tab:modelsres}, these rates are too low by a factor of 5-8
to produce a burst via the B-Z mechanism, and the accompanying $\enu$
is also several orders of magnitude below the LGRB limit.  Moreover,
these models would provide no central engine mechanism for the large
variations in observed lightcurves which occur on the order of ms
timescales (though other phases of evolution may contribute).

In the $v_{\phi}/\sqrt{2}$ models, accretion rates only increased
slightly after the initial onset of the spiral phase and before the
formation of global modes; in the PSdsq2 model, the accretion rate
then briefly peaked at
$\dot{M}>0.65~\textrm{M}_{\odot}~\rm{s}^{-1}$.  The neutrino
luminosity was $\approx 10~\textrm{foe~s}^{-1}$ during the early
spiral phase, increasing by a factor of two in PCdsq2 and by a factor
of three in PSdsq2 as global modes appeared.  While some of the
accretion rates reach LGRB energetic requirements, the $\enu$
typically remained too low with the annihilation efficiency taken into
account.  Model PSdsq2J showed no significant change in disk structure
due to the inclusion of a polar jet.

In the $v_{\phi}/2$ models accretion rates were consistently
$>0.30~\mps$ and $L_{\nu}>100-150 ~\textrm{foe~s}^{-1}$ during the
spiral phase, with peak magnitudes occuring in the presence of global
modes.  Both shellular models (full and reduced $E_{\rm{th}}$) had a
higher average accretion rate than the cylindrical model, which
averaged and peaked at $\dot{M}\approx0.1$ and $0.4~\mps$,
respectively.  These models all produced accretion rates and neutrino
luminosities high enough to fuel LGRBs.

In cases with rapid accretion, variations of up to half an order of
magnitude in $\dot{M}$ and $L_{\nu}$ occured as rapidly as the
snapshot interval of 0.02~s.  Such fluctuations have been observed in
the spectra of LGRBs, although several factors in the system may
combine to produce these, such as jet structure, environment, etc.
Here, the bursts of accretion are due in large part to the presence of
dense clumps which have been created by the spirals and fragmentation.

We also note that while $\dot{M}$ and $L_{\nu}$ show strong (positive)
correlation, no direct scaling appeared between the quantities.  In
general, the rates of the former appear to make the B-Z mechanism a
significantly more likely scenario than $\nu$-annihilation for
producing LGRBs, keeping in mind the simplifying assumptions made in
calculating both.  The neutrino rate was probably too low much of the
time, as $L_{\nu}$ only reached above 100~foe$\textrm{~s}^{-1}$ at
brief intervals for accretion rates typically of $>0.5~\mps$.

The effects of general relativity in the system were non-negligible.
During the evolution of the Newtonian models, typical (though not
peak) accretion rates and neutrino fluxes were roughly $15-20\%$
higher than in the pseudo-potential models.  At the intermediate
central object spin rates present in these collapsar simulations
($a\approx0.5$), the overall difference in potential wells was quite
small ($\approx1\%$, see caption in Table~\ref{tab:models}).  The
deviations in disk temperature and structure appear to be the
determining factor in the energetics.  This stresses the importance of
including GR in as realistic a manner as possible in simulations, and
further work on this task is being done.

\subsection{Nucleosynthesis}
In studying NSE-produced materials in the collapsar, it was found that
light elements dominated the disk, and that, during the early
evolution, the masses of $^{56}$Ni produced in the corona and beyond
were too small to power a HN.  Typically the amount of $^{56}$Ni
present in the system remained constant after the early phases of the
disk ($t\gtrsim0.30$~s), when the corona had reached an approximate
state of equilibrium.  More $^{56}$Ni was produced in the lower
$v_{\phi}$ models, with the greatest amount in the low-$T$ model,
$0.10~\msun$.  In investigating areas for further production of
radioactive Ni, the $v_{\phi}/2$ models produced the strongest
equatorial outflows, even though those disks were the smallest;
PSd2Kd5 produced the highest outward velocity.

It was also found that, by the inclusion of a simple collimated jet,
rapid polar outflows were produced (here, with $v\sim 0.1c$, but this
may be strongly model-dependent), which carried high entropy material
with moderate and low electron fraction outward.  This material from
above the disk and the coronal edges which was swept outward may be
suitable for forming $^{56}$Ni, and in some cases, for the $r$-process
to manufacture other heavy (neutron-rich) metals. Due to the
necessarily high velocity of any outflow, one might expect that, once
begun, a relatively large amount of $^{56}$Ni and Fe-group metals
could be deposited outside the disk in a short period of time.

Studying the relative locations of Fe and O did not provide
overwhelming evidence for or against the presence of polarised
asymmetry, as has been observed and predicted in HNe.  In general, O
is prevalent in the outer corona, and Fe, outside that region.
However, in the presence of the jet, material from the corona would be
carried up along the polar axis, with a strong likelihood of heavy
elements being manufactured.

\subsection{Collapsar implications}
\label{sec:resresres}

While not spanning the lifetime of a typical LGRB, all models in this
study ran for a large number of dynamical timescales, and they provide
a basis for postulating continued behaviour for long-duration
phenomena. It is worth considering the likely behaviour of our models
continuing after the end of our simulations, particularly in
connection with its implications for both LGRBs and HNe.

Regarding the former, the long-term stability and behaviour of the
collapsar disk must be considered.  In the scaled $v_\phi/2$ models,
high accretion rates occured, and the size of the spiral structures
grew globally; in the $v_\phi/\sqrt{2}$ models, the densest material
($\rho>\rho_{\rm drip}$) tended to be accreted, most likely requiring
reformation of the disk to permit high accretion rates.  In all of
these systems, however, the $\qt$ values generally remained low at
late times, and a high degree of non-axisymmetric structure remained
even in the catastrophic cases, along with dense clumps of material.
Moreover, the flattening of specific angular momentum profiles (shown
for shellular profiles in Fig.~\ref{fig:all_jspan}) for models with
decreased $v_{\phi}$ also suggests continued instability for the
disks.  With sustained infall, the longterm disk behaviour can then be
broadly classified into two possible scenarios: alternating
destruction and reformation, and a continual presence with more
`balanced' mass loss/gain.  Again, these may be evidenced in and also
explain some of the variety of observed LGRB lightcurves.  While short
term fluctuations would be due to clumps and local density differences
in the flow, longer term ones may arise from the oscillating behaviour
of the disk as a whole.

The details of this behavior may strongly affect HNe as well, in
considering the effects of time-integration on heavy element
production.  The yields from the models in this study must be
considered as minima, as the inclusion of non-NSE nucleosynthetic
processes will certainly lead to increased amounts of both $^{56}$Ni,
which powers any HN lightcurve, and other Fe-group products, which
characterize the non-spherical geometry of the event.  In the presence
of near-jet outflows, even NSE yields may be enhanced as heavy metals
are produced and ejected.  Disk evolution, whether proceeding by
destruction and reformation or by balanced mass gain and loss, will
influence whether metal production is quasi-continual or -discrete, or
whether potential Fe-group material is accreted rapidly, delaying any
possible HN event.

Within the scope of these trends, there is also, importantly, a wide
range of properties obtained from this related sample of LGRB
candidates.  For example, very different structures of the $\dot{M}$
and $L_\nu$ curves were obtained; equatorial outflows of various
velocity ranges were produced; the amount of heavy metals
(particularly $^{56}$Ni and Fe), synthesised in the disk and corona
varied by orders of magnitude, even for models of similar total
angular momentum.  Such diversity exists within the observed LGRB/HN
population. For example, a system with high accretion rates, low
$^{56}$Ni content in the corona and small outflows in the equatorial
plane might produce a bright LGRB with a weak SN explosion (due to low
radioactive Ni content).  Conversely, a system with low accretion
rates and a large mass of $^{56}$Ni may result in a strong HN with a
weak LGRB or with a lower energy XRF, or possibly only a bright,
polarised SN. We conclude that {\em dynamically unstable collapsar
  disks provide a unified and simple explanation for the large variety
  of these events}.

Among the family of potential collapsar progenitors investigated here,
the principal discriminating factor appeared to be the overall amount
of rotation initially present.  Simulation results led to the
classification of LGRB candidates by a simple $y_j=\bar{j}_0/j_{\rm
  ISCO}$ parametrisation with a window of values for systems that
evolved hydrodynamically unstable disks able to fuel a central engine.
Importantly, this $y_j$ prescription is quite general. Given the
diversity of observed LGRBs, the heterogeneity of the GRB-SN/HN
connection and the probable relations with phenomena such as XRFs, it
is likely that a variety of progenitor scenarios will require
examination, including both binary and single-star systems, to fully
understand LGRBs. Further considerations may refine progenitor
characterisation, but we expect the trends of this study to be broadly
applicable across a wide range of possible collapsars.

Additional work should be done to further these results, in particular
by including non-NSE nucleosynthetic processes and a more realistic
equation of state as well as a physically-motivated jet.  These
physical inputs are necessary for understanding not only LGRBs
themselves but also their particular connection with HNe.  Finally,
while known to be essential for driving burst jets, magnetic fields
were not directly included in this study, and their full role during
collapse and disk evolution must be investigated further.

\section*{Acknowledgments}

PT would like to thank Chris Fryer, Alexander Heger, Shazrene Mohamed,
Orazio Nicotra, S\'ebastien Peirani, Stephan Rosswog and Adrianne Slyz
for informative discussions and for useful datasets with regards to
initial conditions and analysis, as well as Frank Timmes in particular
for the NSE calculations which have been reproduced herein.  The
authors would also like to thank Lorne Nelson for making available the
cluster at the Centre de Calcul Scientifique de l'Universit\'e de
Sherbrooke, with which many of the computations reported in this work
were performed. This work was partly supported by CompStar, a Research
Networking Programme of the European Science Foundation.

\bibliography{mnras_sph_grb}

\appendix

\section[A]{Dissipation of MRI, B-field conditions}
\label{sec:nobfields}
In consideration of the onset of MRI in an accretion disk,
\citet{1991ApJ...376..214B} estimated conditions for non-ideal MHD
effects to damp the growth of magnetic modes from dimensional
analysis.  Specifically, the local dynamical timescale was compared to
the diffusion timescale of modes due to finite resistivity and thermal
conductivity.  This resulted in the following criterion for damping to
be negligible (for a $z$-component):
\begin{equation}\label{finitemhdcrit}
\frac{3~\Omega_{\rm{K}}\, \chi}{{\it v}^2_{\rm{Az}}} \ll 1,
\end{equation}
where $\Omega_{\rm{K}}$ is the orbital frequency assuming Keplerian
rotation; $v_{\rm{Az}}$, the poloidal Alfv\'{e}n velocity; and $\chi$,
a general diffusion coefficient.  They stated that the minimum $B$-field
strength for damping to be ignored, and therefore for the MRI to grow
nonlinearly, was extremely small for most accretion disks; as a
result, a magnetic field, internal or external, of any size would lead
to the development of global modes in the disk on dynamical
timescales.  We investigate the same criteria in the realm of the
collapsar disk structure, which has already been noted to be of an
extreme nature.

In considering finite electrical resistivity, plasma theory leads to
an expression \citep{1965pfig.book.....S} for the coefficient in
Eq.~(\ref{finitemhdcrit}):
\begin{eqnarray}
\chi_{\rm{er}}&=& \frac{3.8\times10^{12}\,{\it Z}\,\ln(\Lambda)}{\gamma_{\rm{E}}\,T^{\rm{3/2}}}~~~\rm{cm}^2~\rm{s}^{-1}~,\\
\Lambda&=& \frac{3}{2} \frac{(k_{\rm{B}} {\it T})^{\rm{3/2}}}{{\it Z}\,e^3(\pi n_{\rm{e}})^{1/2}}=\frac{4.31\times10^{20}~{\it T}^{3/2}}{(\ye \rho)^{1/2}}~, \nonumber
\end{eqnarray}
where $Z$ is ionic charge; $\gamma_{\rm{E}}\approx1$, the ratio of
conductivity to that of an ideal Lorentz gas; and $\ln(\Lambda)$, the
Coulomb logarithm.  Substituting this coefficient into
Eq.~(\ref{finitemhdcrit}) for a $Z=1$ (fully ionised) gas, the
general condition on the local poloidal $B$-field magnitude for
dissipation due to electrical resistivity not to damp out the MRI is:
\begin{equation}
B_{\rm{z,er}}^2~\gg~\frac{2.83\times10^{27}\ln(\Lambda)~\rho}{(RT)^{3/2}}
\left(\frac{M}{\msun}\right)^{1/2}~~~\rm{G^2},
\end{equation}
where $M$ is the mass of the central BH.

The analogous diffusion coefficient representing the thermal
conductivity of the plasma is \citep{1991ApJ...376..214B}:
\begin{equation}
  \chi_{\rm{tc}}= \frac{1.84\times10^{-5}}{\ln(\Lambda)}T^{5/2}~~~~\cmps,
\end{equation}
and the corresponding condition on the $B$-field is
\begin{equation}
 B_{\rm{z},tc}^2\gg \frac{57.8~T^{5/2}}{\ln(\Lambda)~R^{3/2}}
  \left(\frac{M}{\msun}\right)^{1/2}~~~\rm{G^2}.
\end{equation}

As an example, we check the requirements on the magnetic field in the
near-inner regions of the collapsar disk, inserting reasonable values
of physical quantities ($R=10^7~\rm{cm}$, $\rho=10^{11}~\gcc$,
$\ye=0.4$, $T=10^{10}~\rm{K}$, $M/\msun=3$). The conditions with
electrical resistivity and thermal conductivity require that
$B_{\rm{z}}\gg3.3\times10^7~\rm{G}$ and
$B_{\rm{z}}\gg2.2\times10^7~\rm{G}$, respectively, in order that
dissipation is unable to damp out MRI growth.  Therefore, the minimal
necessary magnetic field strength for nonlinear growth
($B\sim10^{9}$~G), while not unphysically high, is certainly not
negligible, as it is generally for lower density disks.  

\section[B]{Relativistic pseudo-potentials}

The generic task of a pseudo-Newtonian potential\footnote{The same
  notations as in \S\ref{sec:grsr} are used here.} is to capture
important relativistic aspects without the computational demands of
full GR.  The most well-known example of such an approximation for a
rotation-dominated flow is the Paczy\'{n}ski-Wiita potential for a
non-rotating BH, $\Phi_{\rm{PW}}(R)=-M(R-2r_{\rm{g}})^{-1}$
\citep{1980A&A....88...23P}, and there are also similar potentials for
motion in the Kerr metric.  The accuracy of such potentials can be
tested by the comparison of output to exact GR solutions, such as for
the location of the innermost stable circular orbit ($R_{\rm{ISCO}}$);
the Keplerian orbital velocity ($\Omega_{\rm{K}}$) and the epicyclic
frequency ($\kappa$), which are no longer equal, as they are in the
Newtonian regime; and the structural and temporal properties of
evolving accretion disks, such as specific energy and dissipation.

We note that this involves identifying the Newtonian $R$ in all of
these terms with the Boyer-Lindquist (B-L) coordinate, $R_{\rm{B-L}}$.
To quantify the resulting deviation involved in this, we compare the
proper circumference at $R_{\rm{B-L}}=r_{\rm{in}}$ to the
corresponding `Newtonian' value, $2\pi r_{\rm{in}}$, in the equatorial
plane.  Examples of ($a$, $C_{\rm{B-L}}/C_{\rm{N}}$) are (0,1),
(0.5,1.01) and (0.9,1.13).  The ratio quickly and monotonically
approaches 1 with increasing radius, e.g. for $a=0.9$, it deviates by
only 1\% at $R=3r_{\rm{in}}$.

For the cases considered here (corotating orbits and not extremely
fast BH spin, $0.4 \leq a \leq 0.9$), most pseudo-potentials are
essentially Newtonian for radii $\geq20-25~r_{\rm{g}} \sim
3-4\times10^{6}~$cm, while also reproducing the correct locations of
$R_{\rm{ISCO}}$, which is used as the BH accretion boundary here.
However, no single potential is able to reproduce all of the
above-noted properties simultaneously to errors within even 20\% in
the relativistic regimes, and therefore one must choose to focus on
certain properties over others depending on the system.

We have considered four different pseudo-potentials for corotating
motion around BHs, which we implement as acceleration terms.  The
$\rm{a_{SEP}}$ acceleration which is implemented in this study was
derived by \citet{2003ApJ...582..347M}, and is given in
Eq.~(\ref{SEPequ}) above.  As shown here, this form reproduced most
exact GR properties within tolerable limits.

\begin{figure}
\includegraphics[width=81mm]{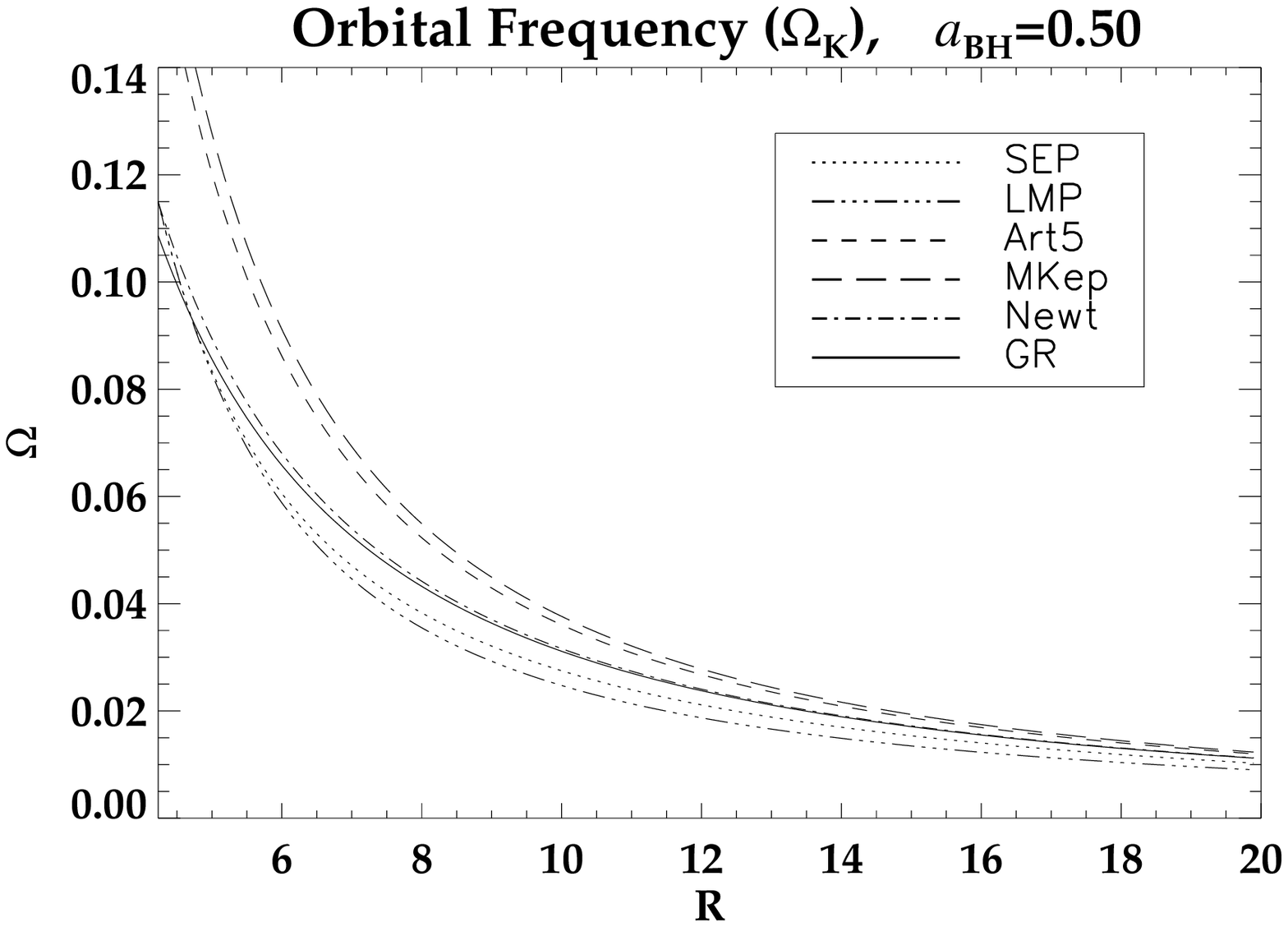}
\caption{A comparison of orbital velocities for different
pseudo-Newtonian accelerations, including the exact GR value, for
$a$=0.5. ($R$ is measured in units of $M$.)}
\label{fig_om_F}
\end{figure}

\begin{figure}
\includegraphics[width=81mm]{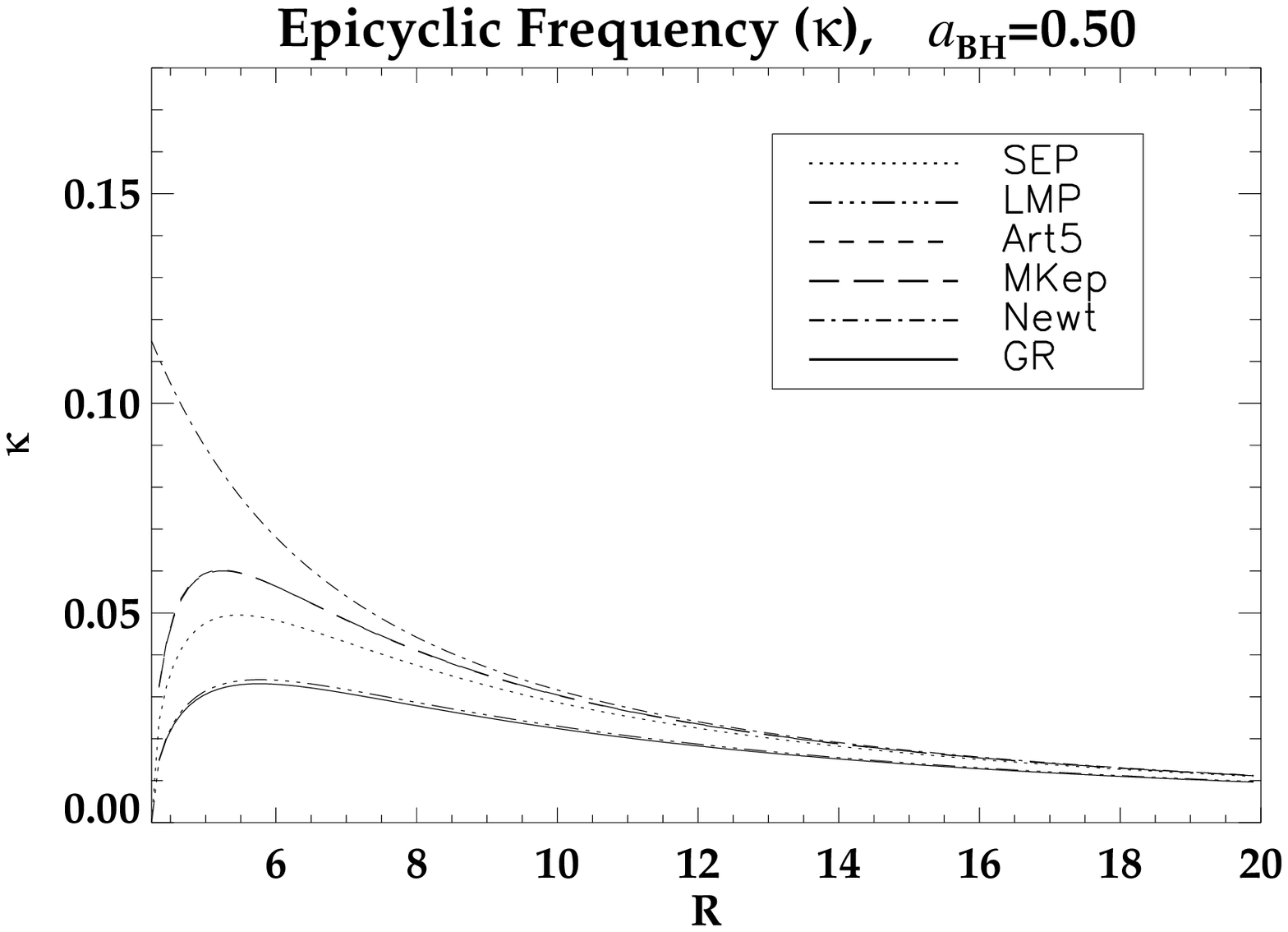}
\caption{A comparison of elliptical frequencies for different
pseudo-Newtonian accelerations, including the exact GR value, for
$a$=0.5. Note: $\rm{a_{Art5}}$ and $\rm{a_{MKep}}$ produce nearly the
same frequencies for this BH spin.}
\label{fig_el_F}
\end{figure}

From \citet*{1996ApJ...461..565A}, the accelerations $F_5$ and $F_6$
(their Eqs. (13)-(18)) are practically identical, and so we
implemented only the former:
\begin{eqnarray}
 \rm{a_{Art5}}(\textit{R}) &=& \frac{M}{\textit{R}^{2-\beta}(\textit{R}-\textit{r}_1)^{\beta}}\,, \\[0.2cm]
 r_1  &= & 1 + \sqrt{1-a^2}\,,\nonumber \\[0.2cm]
 \beta  &= & \frac{r_{\rm{in}}}{r_1} - 1\,, \nonumber
\end{eqnarray}
where $r_1$ denotes the location of the event horizon (with radius and
spin in units of $M$).  Figs. 4-6 of \citet{1996ApJ...461..565A} show
the use of this term in reasonably approximating an optically thick
$\alpha$-disk with a low accretion rate by comparing optical depth
($\propto$ surface density) and temperature to a fully relativistic
case for different spins.  However, a comparison of angular and
particularly epicyclic frequencies in Fig.~\ref{fig_om_F} and
Fig.~\ref{fig_el_F}, respectively, shows great disparities with the
exact GR results for $a=0.5$, and these increase with $a$.

\begin{figure}
\includegraphics[width=81mm]{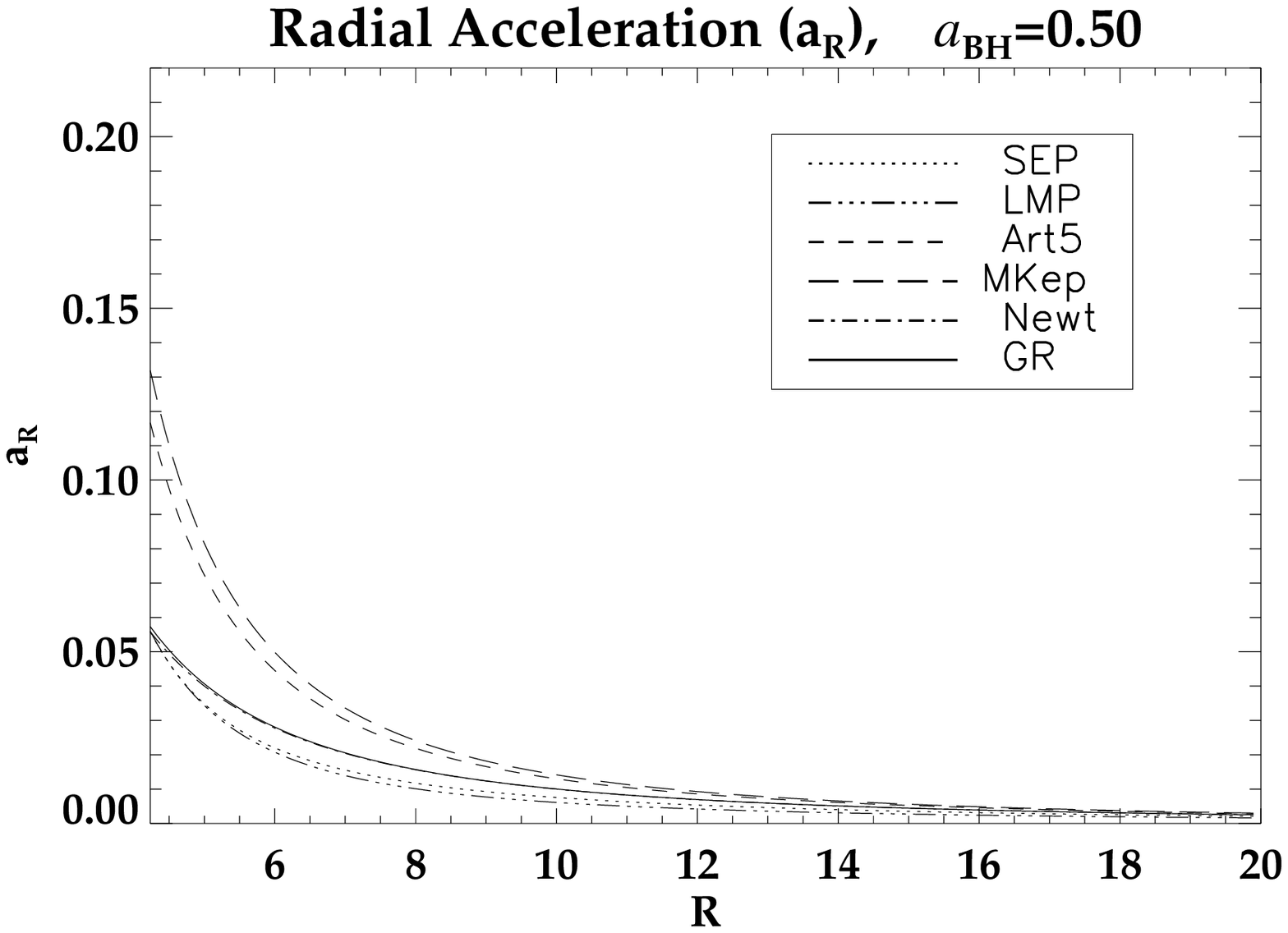}
\caption{A comparison of pseudo-Newtonian accelerations, for
$a$=0.5. In these inner regions, $\rm{a_{Art5}}$ and $\rm{a_{MKep}}$
have noticeably larger accelerations than the other potentials, of a
factor 2-3 at $R_{\rm{ISCO}}$.}
\label{fig_ps_F}
\end{figure}

Similar results are produced with the acceleration given by
\citet{2002ApJ...581..427M}, derived from exact GR solutions for
specific energy and angular momentum.  His Eq. 12,
\begin{equation}
\rm{a_{MKep}}(\textit{R}) = \frac{\textit{M}(\textit{R}^2 - 2\textit{a}\sqrt{\textit{R}} + \textit{a}^2)^2}{\textit{R}^3 (\textit{R}^{3/2} - 2\sqrt{\textit{R}} + \textit{a}^2)^2 }\,,
\end{equation}
reproduces GR energy dissipation rates calculated very well (his
Fig. 1), but does not reproduce well the rotational kinematics, as seen
again in Figs.~\ref{fig_om_F}-\ref{fig_el_F}.

In order to obtain more accurate epicyclic frequencies,
\citet{2003ApJ...582..347M} integrated a function approximating the GR
value of $\kappa$, yielding their `Logarithmically Modified Potential'
(LMP), which has a radial acceleration:
\begin{eqnarray}
\rm{a_{LMP}}(\textit{R}) &=& \frac{M\textit{r}_{\rm{in}}}{\textit{R}^3} \left( \frac{9}{20}(\textit{r}_{\rm{in}}-1) - \frac{3}{2}\,\rm{ln} \left[\frac{\textit{R}}{(3\textit{R}-\textit{r}_{\rm{in}})^{2/9}}
\right]
\right) \nonumber \\ 
& ~&+\,\frac{\textit{M}}{\textit{R}^2}.
\end{eqnarray}
Figs.~\ref{fig_om_F}-\ref{fig_el_F} show that the kinematic properties
of these orbits are fairly accurate; however, Fig. 5 of
\citet{2003ApJ...582..347M} shows that the specific energy of these
orbits is much greater ($>30\%$) than the GR values.  In contrast, the
chosen potential in this study, SEP from the same authors, varies only
slightly from the GR specific energy ($<5\%$), and also from the
angular frequency as shown in Fig.~\ref{fig_om_F} above.  It deviates
the most from the GR epicyclic frequencies, up to $65\%$ as shown in
Fig.~\ref{fig_el_F}.

In the case of the collapsar system, we are studying global
hydrodynamic stability which depends on local values of $\kappa$, as
reflected by the form of $\qt$.  The structural energetics are
important globally, as stability is determined also by cooling rates,
determined by temperature and density.  Furthermore, LGRBs powered by
neutrino-annihilation require large amounts of neutrinos produced in
the innermost regions of the disk, where density and temperatures are
highest.

The relative values of the radial accelerations themselves vary quite
a lot, by more than a factor of 2 at the $R_{\rm{ISCO}}$, as seen in
Fig.~\ref{fig_ps_F}.  The inner boundary condition provided at this
point by $\rm{a_R}(\textit{R})$ is also key to the accretion rate
which drives the LGRB.  The GR value is calculated in the equatorial
plane for a particle at rest in the Kerr metric, and is well
approximated by the LMP and SEP accelerations for $a=0.5$.  However,
it is interesting that for $a=0.9$, $\rm{a_{GR}}(R_{\rm{ISCO}})$ has
increased much more rapidly than these, and is best approximated by
the Art5 formulation to within 15\%, being now a factor of two greater
than the LMP/SEP values.

However, even at this high value of $a$, the SEP produces a very
accurate value of $\Omega_{\rm{K}}$ throughout most of the inner
relativistic region, unlike Art5 and MKep.  The SEP (Art5/MKep)
potential gives a peak value for $\kappa$ roughly a factor of 2 (3)
greater than the GR one, and even the LMP value is $\approx$ 50\%
above the GR value.

However, over this whole range of BH spins, the SEP provides accurate
values of specific energy (and of $\Omega_{\rm{K}}$).  Even though we
are explicitly focused on the hydrodynamic stability of the LGRB disks
and the value of the parameter $\qt$, we have chosen a potential whose
greatest weakness is in reproducing the $\kappa_{\rm{GR}}$.  However,
we note that, through the early evolution of the disk, the minimum
values of $\qt$, denoting those regions approaching hydrodynamic
instability, are at radii of order $5-10\times10^6$~cm ($\sim20M$),
where the Newtonian approximation is quite good.  Also, we reiterate
the importance of the inner, relativistic regions in representing
accurate structure and energetics of the disk.  We therefore use this
SEP approximation in this study, whilst acknowledging its limitations.

\section[C]{NSE abundances}

\begin{figure}
  \includegraphics[width=84mm]{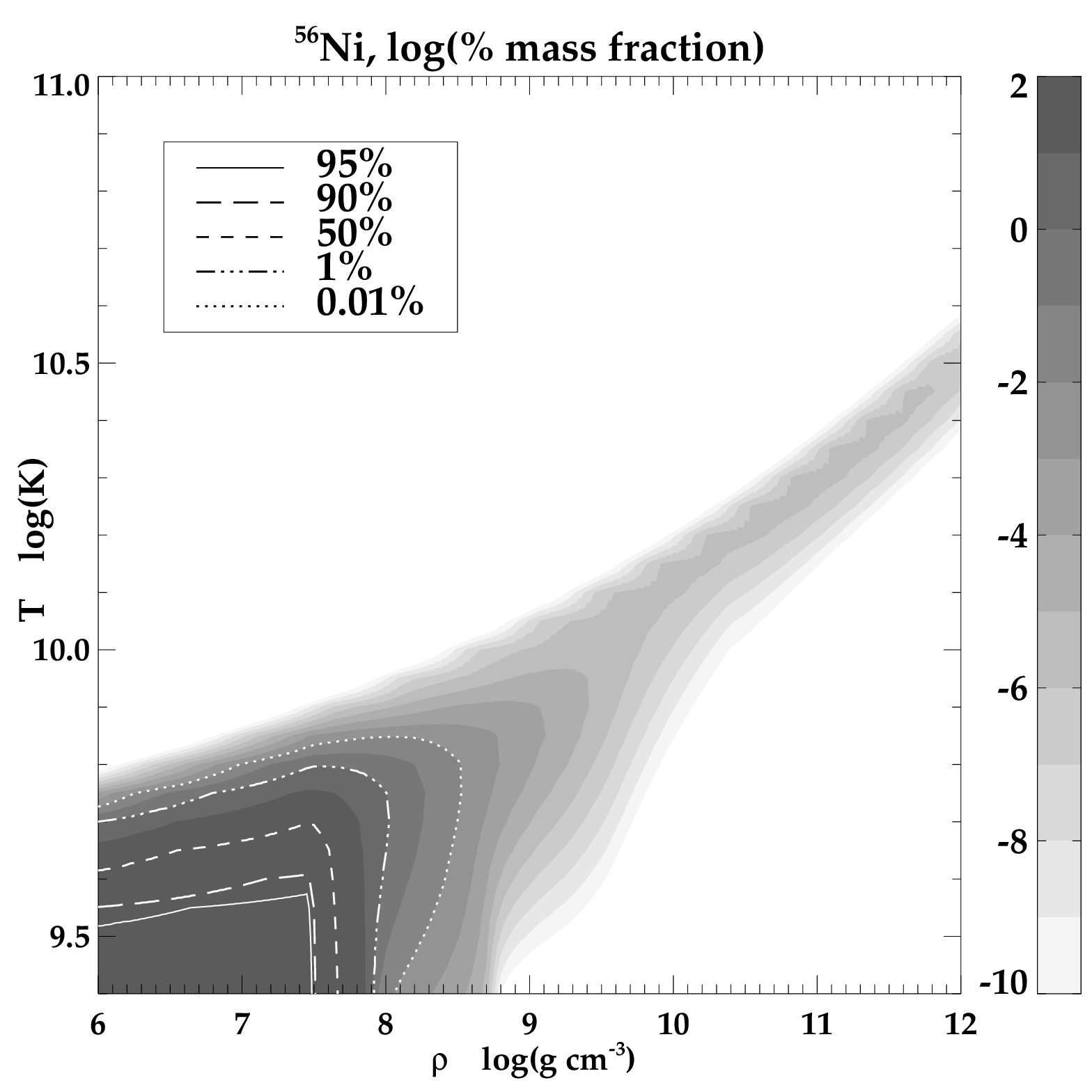}
  \caption{\label{abund_ni56} $^{56}$Ni abundances in NSE, as
    percentages of relative mass fraction in the relevant $(\rho, T)$
    phase space, with successive shades representing order of
    magnitude intervals in log(\%).}
\end{figure}

\begin{figure}
\includegraphics[width=84mm]{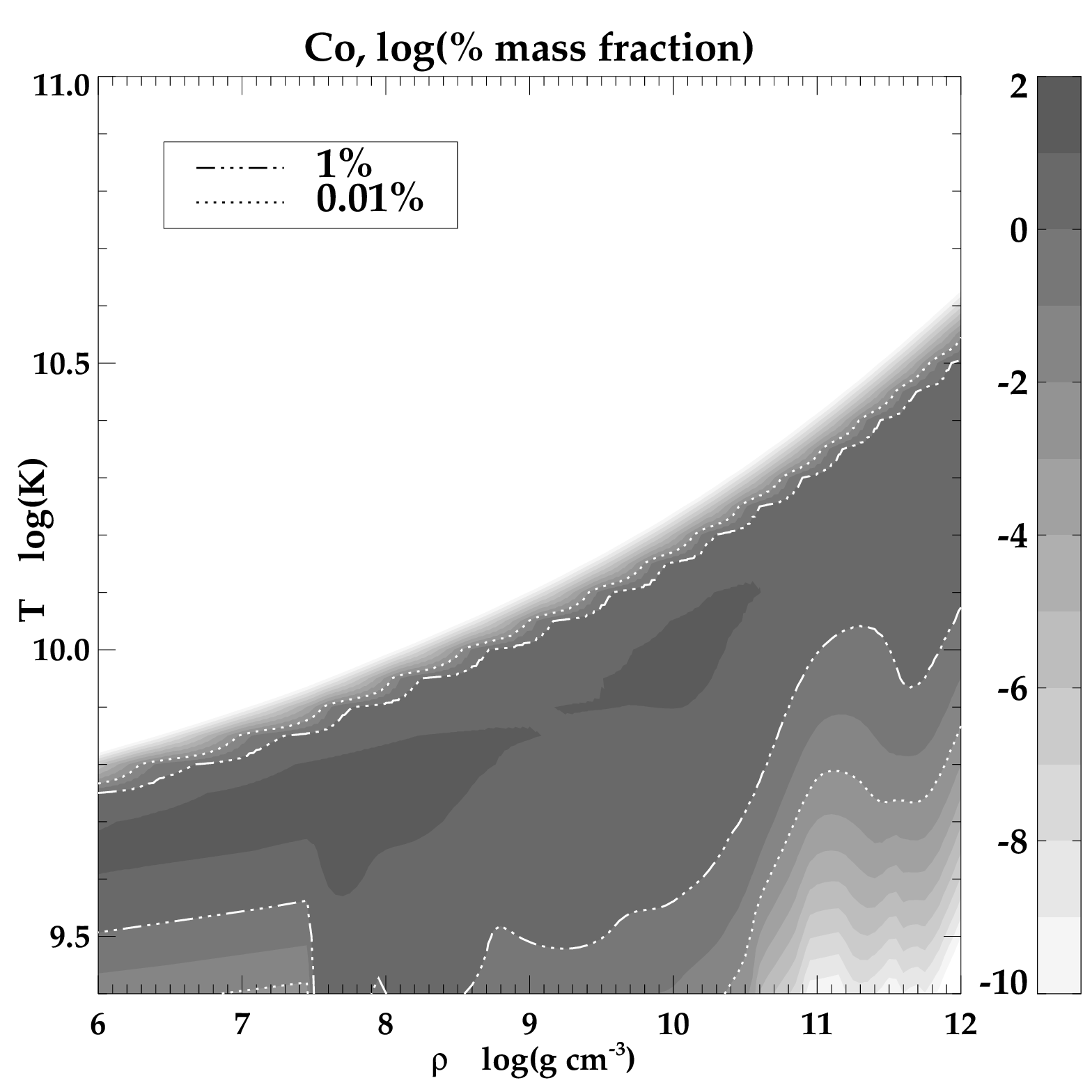}
\caption{Co abundances (and see Fig.~\ref{abund_ni56}).}
\label{abund_cob}
\end{figure}

\begin{figure}
\includegraphics[width=84mm]{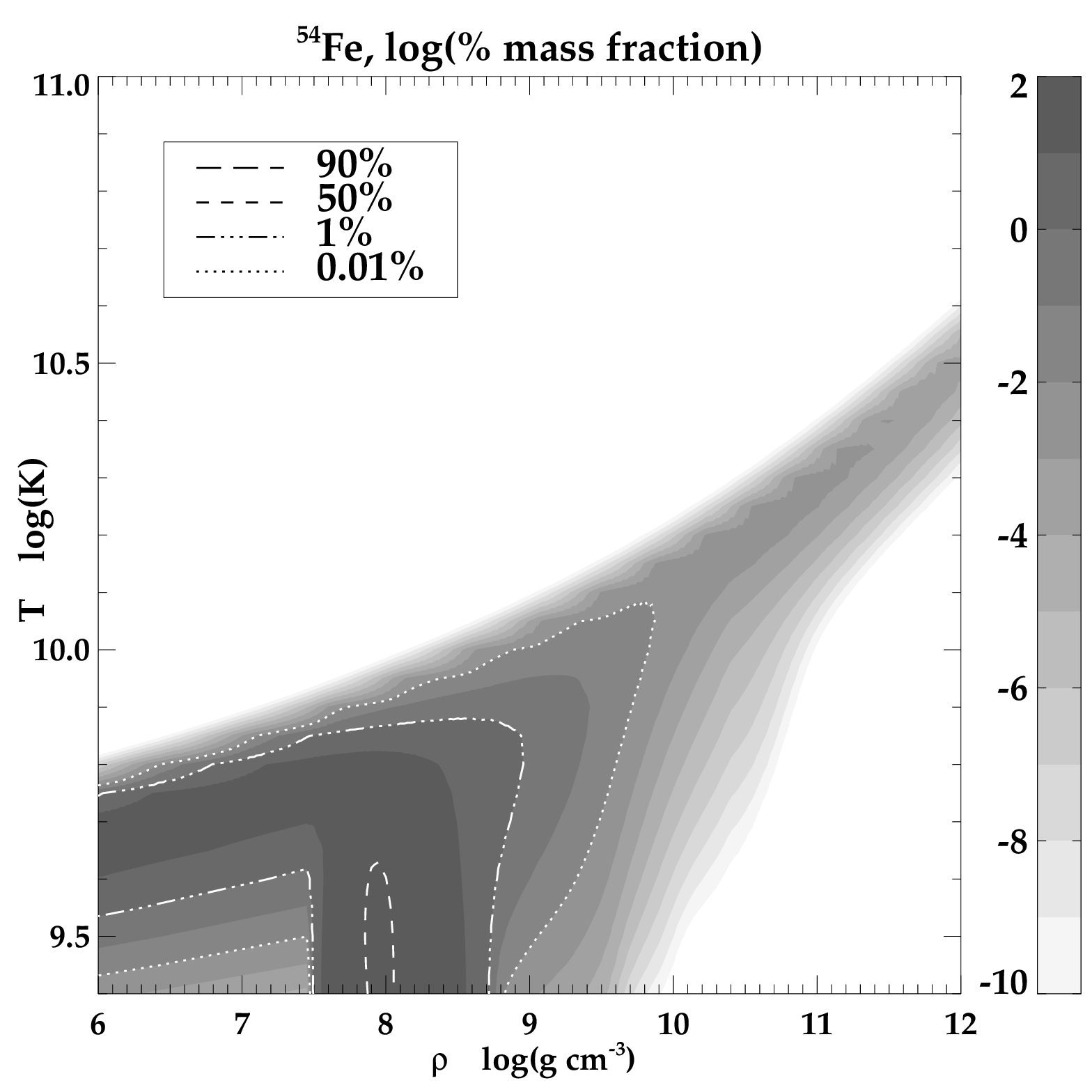}
\caption{$^{54}$Fe abundances (and see Fig.~\ref{abund_ni56}).}
\label{abund_fe54}
\end{figure}

\begin{figure}
\includegraphics[width=84mm]{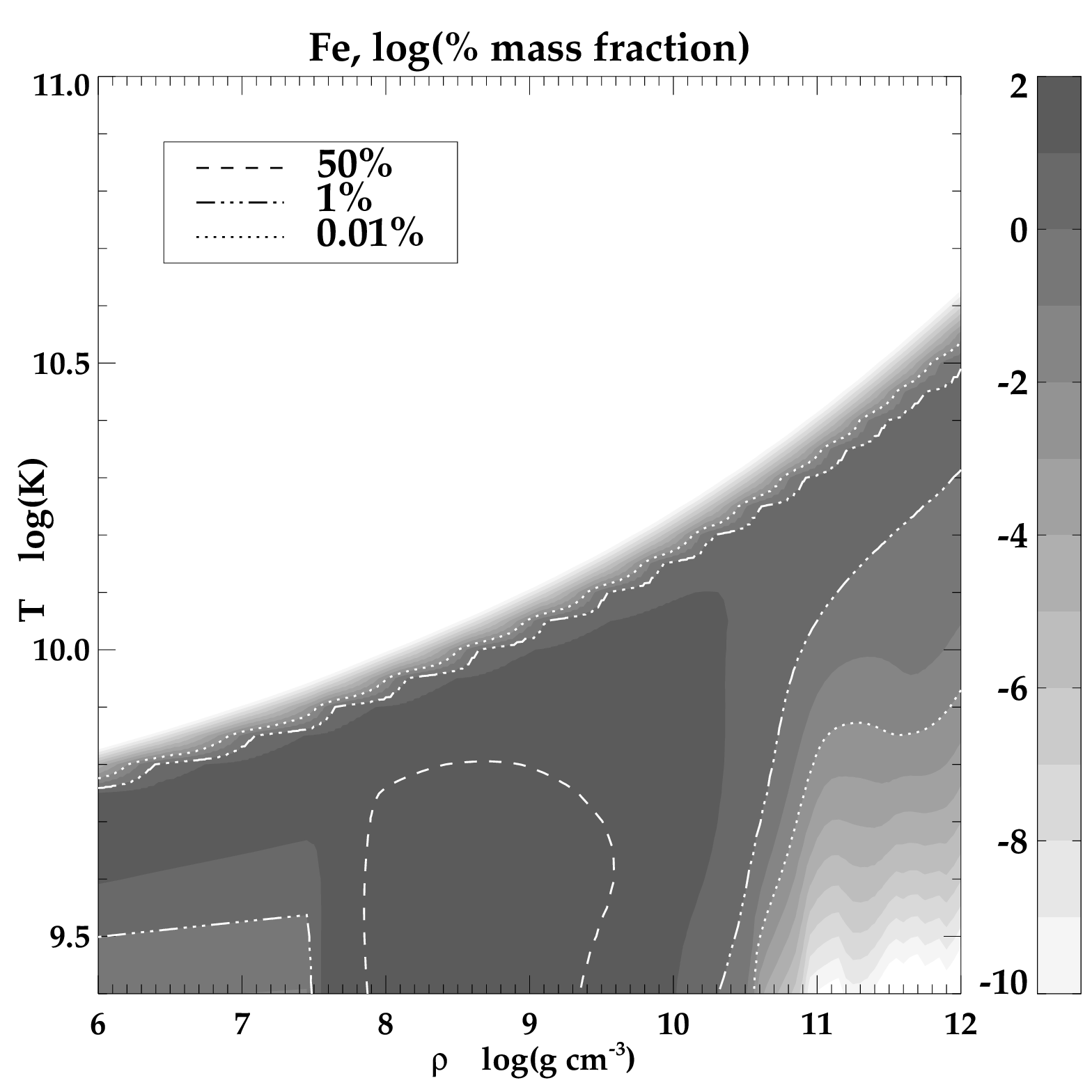}
\caption{Fe abundances (and see Fig.~\ref{abund_ni56}).}
\label{abund_iron}
\end{figure}

\begin{figure}
\includegraphics[width=84mm]{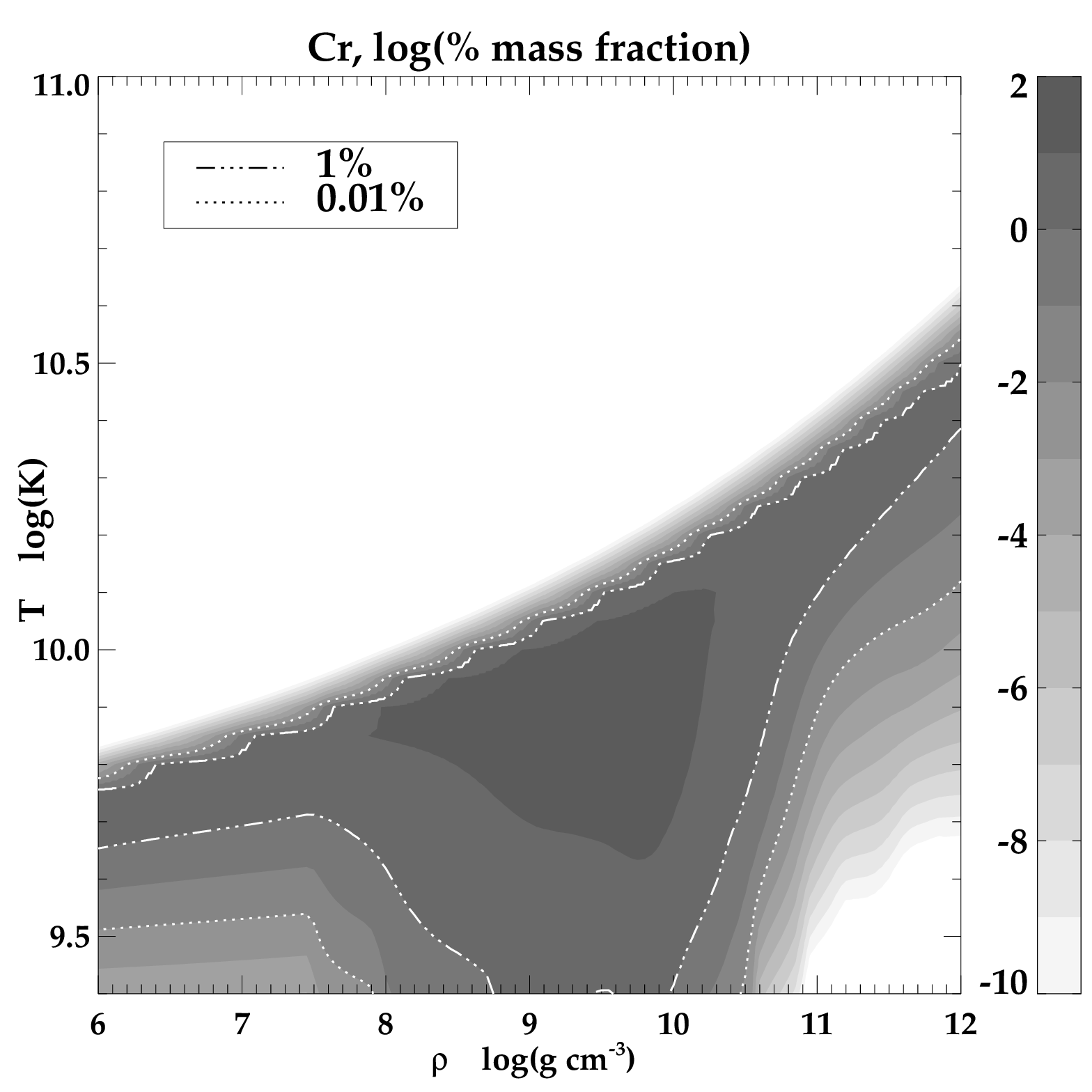}
\caption{Cr abundances (and see Fig.~\ref{abund_ni56}).}
\label{abund_chrom}
\end{figure}

\begin{figure}
\includegraphics[width=84mm]{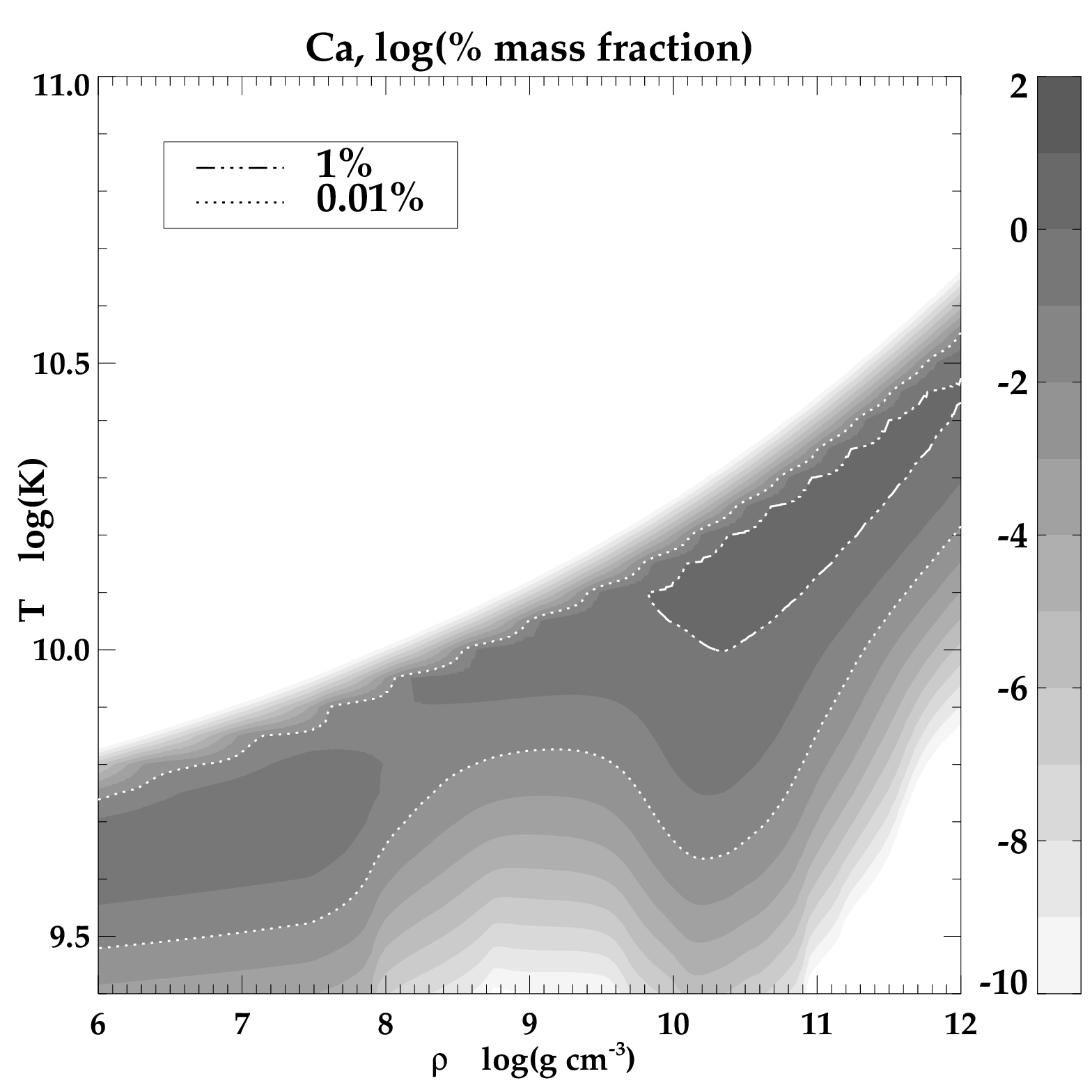}
\caption{Ca abundances (and see Fig.~\ref{abund_ni56}).}
\label{abund_calc}
\end{figure}

\begin{figure}
\includegraphics[width=84mm]{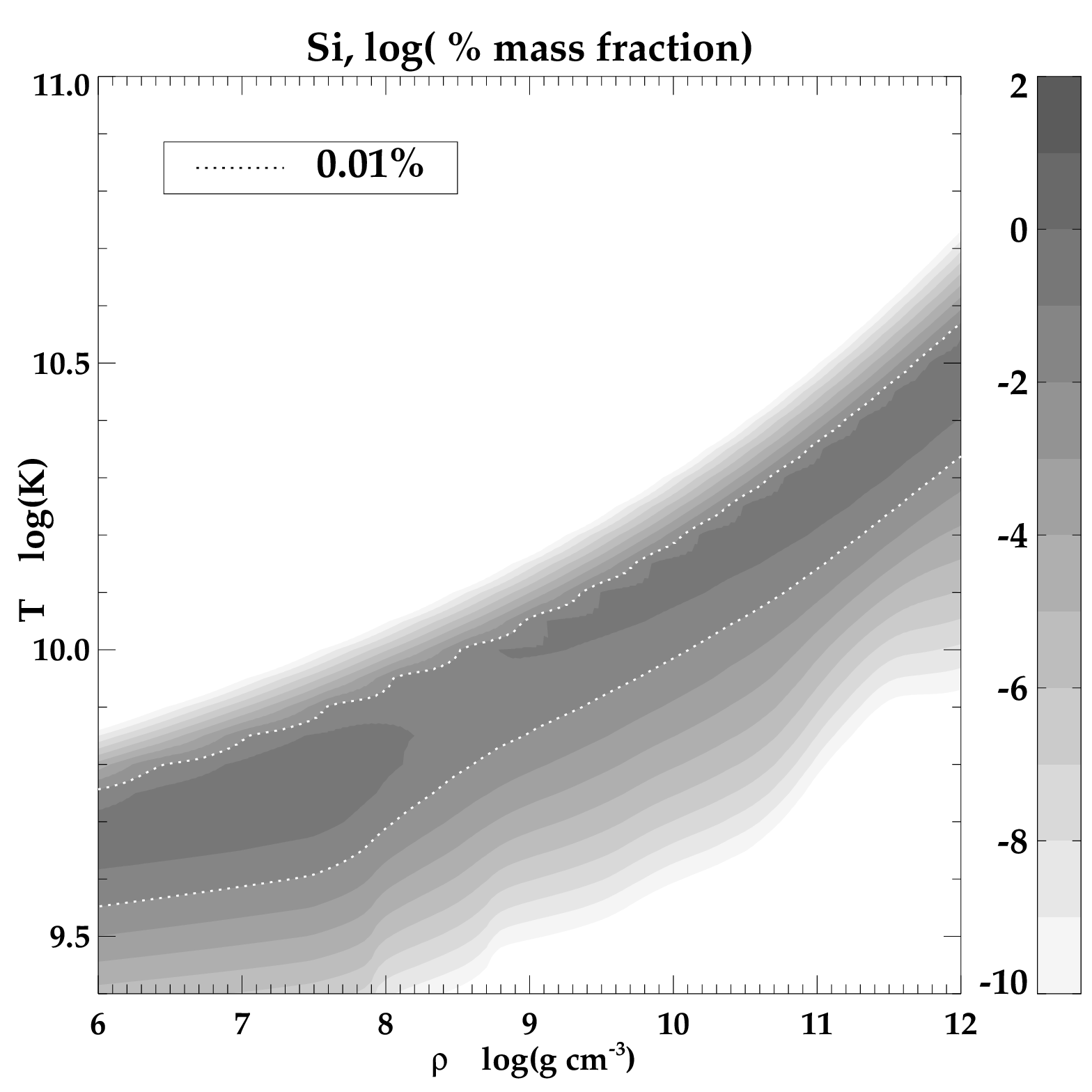}
\caption{Si abundances (and see Fig.~\ref{abund_ni56}).}
\label{abund_sili}
\end{figure}

\begin{figure}
\includegraphics[width=84mm]{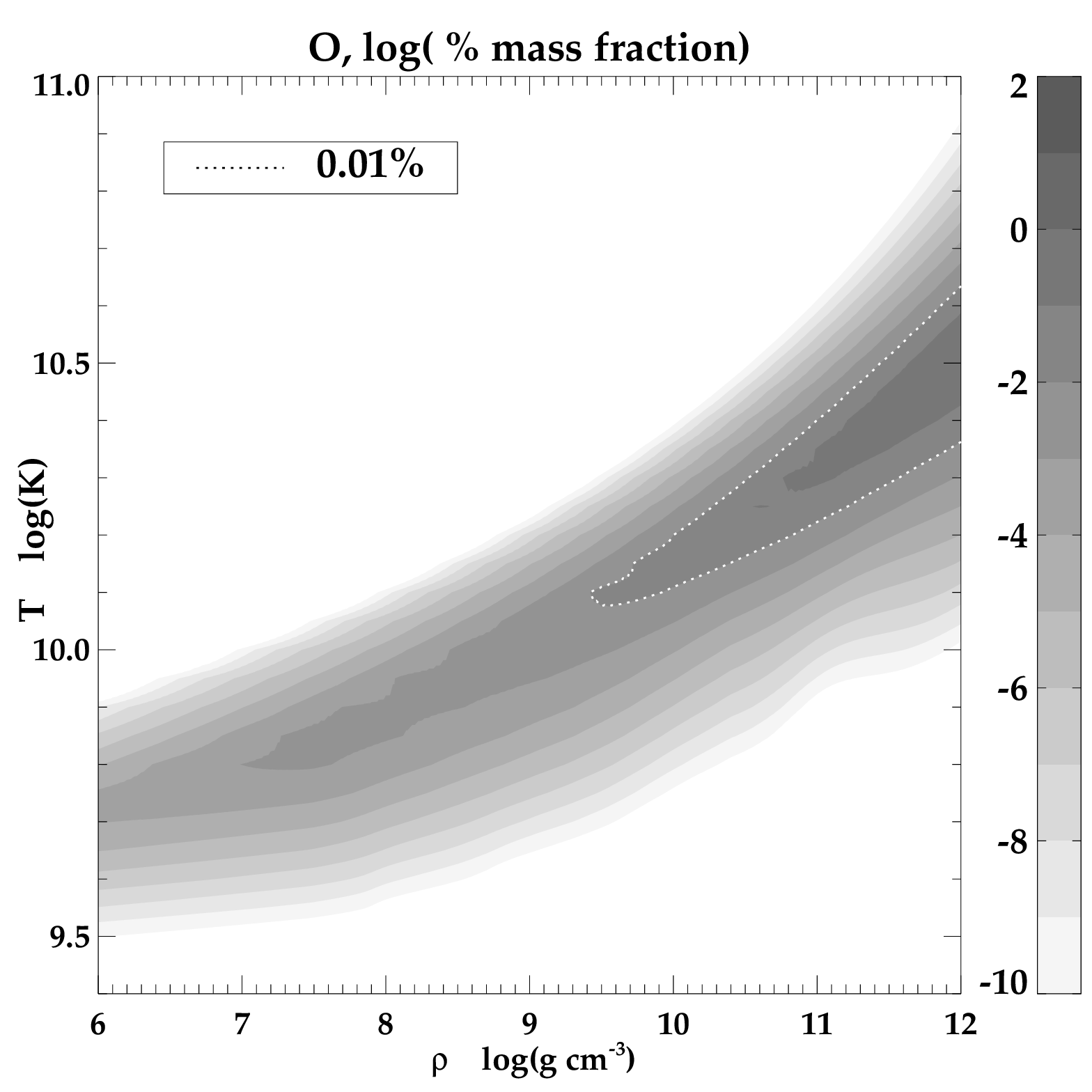}
\caption{O abundances (and see Fig.~\ref{abund_ni56}).}
\label{abund_oxy}
\end{figure}

\begin{figure}
\includegraphics[width=84mm]{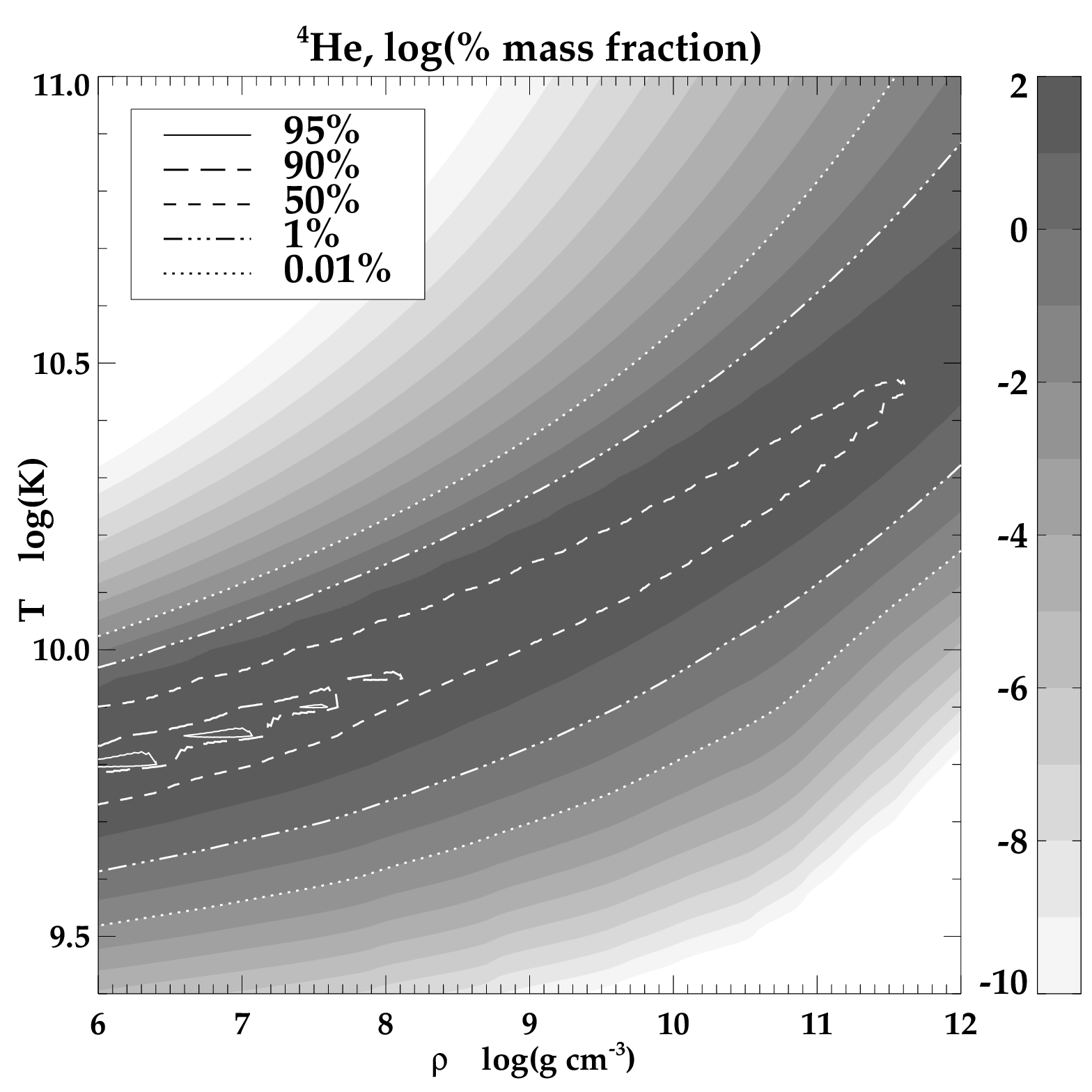}
\caption{$^4$He abundances (and see Fig.~\ref{abund_ni56}).}
\label{abund_he4}
\end{figure}

\begin{figure}
\includegraphics[width=84mm]{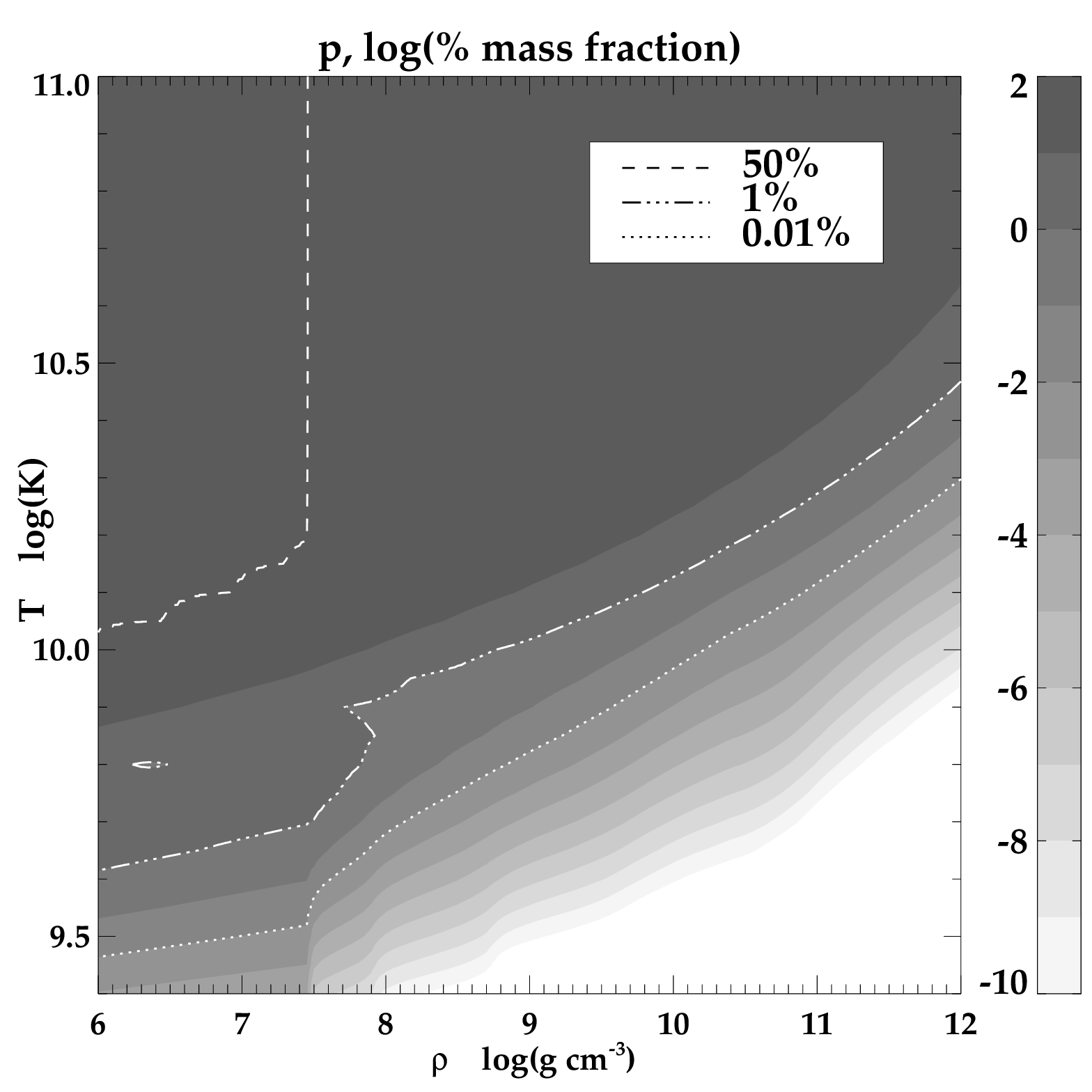}
\caption{Proton abundances (and see Fig.~\ref{abund_ni56}).}
\label{abund_prot}
\end{figure}

\begin{figure}
\includegraphics[width=84mm]{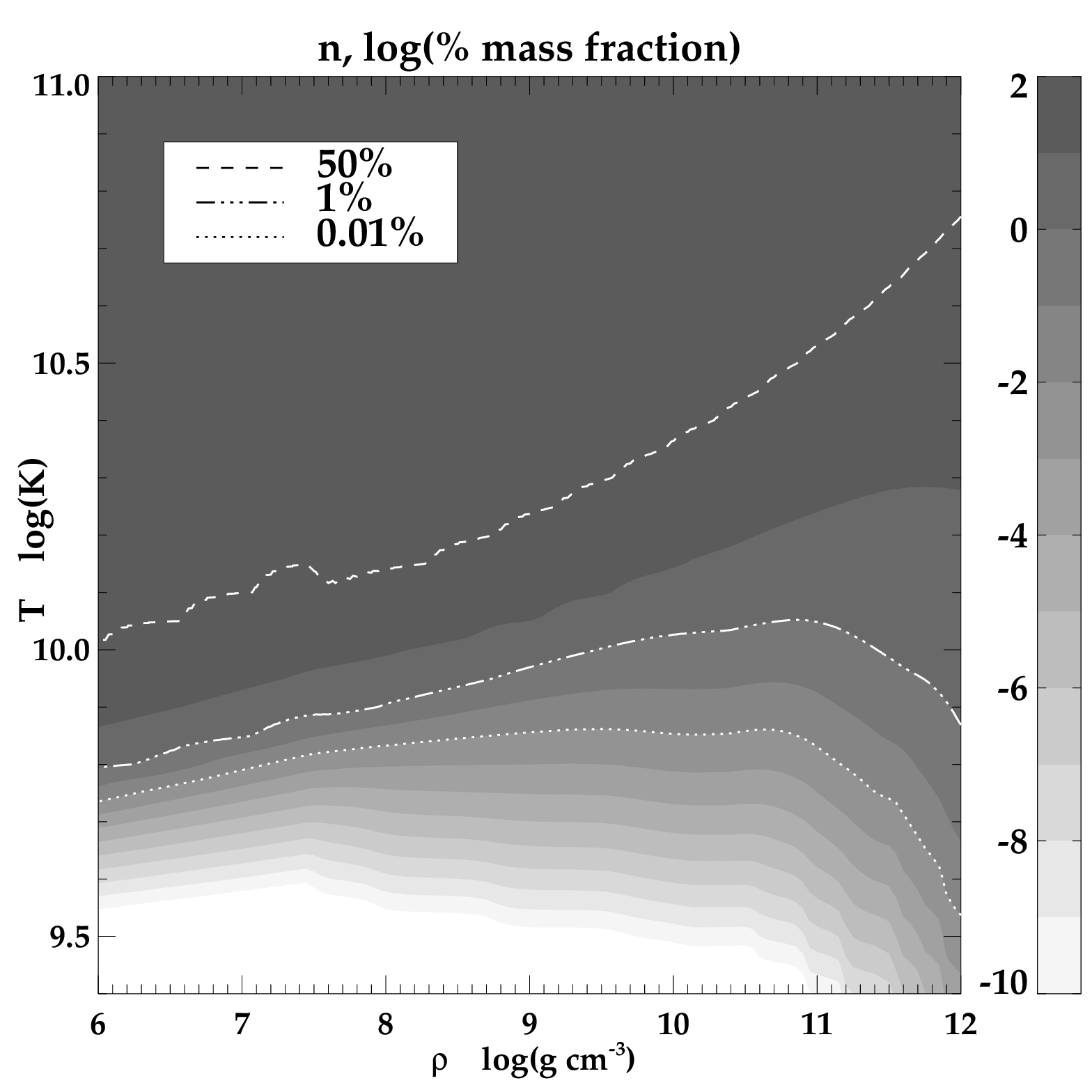}
\caption{Neutron abundances (and see Fig.~\ref{abund_ni56}).}
\label{abund_neut}
\end{figure}

All NSE calculations of $\abar$ and element and isotope abundances
used in this study have been made by F. Timmes using a $3000+$ isotope
reaction network.  The code used is related to the publicly available
NSE code at \href{http://cococubed.asu.edu/code\_pages/nse.shtml}{http://cococubed.asu.edu/code\_pages/nse.shtml},
but including a larger network of isotopes and more detailed physics,
such as Coulomb corrections.  Figs.~\ref{abund_ni56}-\ref{abund_neut}
show the mass fraction abundances of the NSE material analysed in this
work, in the relevant ($\rho, T$) phase space.

\bsp
\label{lastpage}

\end{document}